\pdfoutput=1

\documentclass[acmsmall]{acmart}

\AtBeginDocument{
  \providecommand\BibTeX{{
    \normalfont B\kern-0.5em{\scshape i\kern-0.25em b}\kern-0.8em\TeX}}}

\setcopyright{acmcopyright}
\copyrightyear{2023}
\acmYear{2023}
\acmDOI{XXXXXXX.XXXXXXX}

\acmJournal{FACMP}
\acmMonth{1}

\usepackage{multirow}

\usepackage{amssymb}
\usepackage{xspace}
\usepackage{paralist}
\usepackage{hyperref}
\usepackage{natbib}
\usepackage{fancybox}
\usepackage{color, colortbl}
\usepackage{float}
\usepackage{subcaption}
\usepackage{bbding}

\newcommand{\rom}[1]{\uppercase\expandafter{\romannumeral#1}}

\newcommand{\Apollo}{\textsc{Apollo}\xspace}
\newcommand{\Autoware}{\textsc{Autoware}\xspace}
\newcommand{\OpenPilot}{\textsc{OpenPilot}\xspace}
\newcommand{\Pylot}{\textsc{Pylot}\xspace}
\newcommand{\carla}{\textsc{Carla}\xspace}
\newcommand{\LGSVL}{\textsc{LGSVL}\xspace}
\newcommand{\BeamNG}{\textsc{BeamNG}\xspace}
\newcommand{\SUMO}{\textsc{SUMO}\xspace}
\newcommand{\PreScan}{\textsc{PreScan}\xspace}
\newcommand{\CarSim}{\textsc{CarSim}\xspace}
\newcommand{\CarMaker}{\textsc{CarMaker}\xspace}
\newcommand{\matlabSimulink}{\textsc{Matlab/Simulink}\xspace}
\newcommand{\VTD}{\textsc{VTD}\xspace}
\newcommand{\VISSIM}{\textsc{VISSIM}\xspace}
\newcommand{\Webots}{\textsc{Webots}\xspace}
\newcommand{\Gazebo}{\textsc{Gazebo}\xspace}
\newcommand{\Unity}{\textsc{Unity}\xspace}
\newcommand{\AirSim}{\textsc{AirSim}\xspace}
\newcommand{\rFpro}{\textsc{rFpro}\xspace}
\newcommand{\Cognata}{\textsc{Cognata}\xspace}
\newcommand{\NVIDIADriveSim}{\textsc{NVIDIA Drive Sim}\xspace}
\newcommand{\ADAMS}{\textsc{ADAMS}\xspace}
\newcommand{\ProSiVIC}{\textsc{Pro-SiVIC}\xspace}
\newcommand{\SCANeRStudio}{\textsc{SCANeR Studio}\xspace}
\newcommand{\LBC}{\textsc{LBC}\xspace}
\newcommand{\NvidiaCNNLaneFollower}{\textsc{Nvidia CNN Lane Follower}\xspace}
\newcommand{\UdacityDNNModels}{\textsc{Udacity DNN Models}\xspace}
\newcommand{\Chauffeur}{\textsc{Chauffeur}\xspace}
\newcommand{\Epoch}{\textsc{Epoch}\xspace}
\newcommand{\DeepDriving}{\textsc{DeepDriving}\xspace}
\newcommand{\BeamNGAI}{\textsc{BeamNG.AI}\xspace}
\newcommand{\carlapid}{\textsc{Carla PID}\xspace}
\newcommand{\myparagraph}[1]{\medskip\noindent{\bf #1.\xspace}}
\newcommand{\DonkeyCar}{\textsc{Donkey Car}\xspace}

\begin{document}
\title{A Survey on Automated Driving System Testing: Landscapes and Trends}

\author{Shuncheng Tang}
\email{scttt@mail.ustc.edu.cn}

\author{Zhenya Zhang}
\email{zhang@ait.kyushu-u.ac.jp}

\author{Yi Zhang}
\email{zhangyi2122@mail.ustc.edu.cn}

\author{Jixiang Zhou}
\email{zjxmail@mail.ustc.edu.cn}

\author{Yan Guo}
\email{guoyan@ustc.edu.cn}

\author{Shuang Liu}
\email{Shuang.liu@tju.edu.cn}

\author{Shengjian Guo}
\email{sjguo@baidu.com}

\author{Yan-Fu Li}
\email{liyanfu@tsinghua.edu.cn}

\author{Lei Ma}
\email{ma.lei@acm.org}

\author{Yinxing Xue}
\email{yxxue@ustc.edu.cn}

\author{Yang Liu}
\email{yangliu@ntu.edu.sg}

\authorsaddresses{
Authors’ addresses: S.~Tang, Y.~Zhang, J.~Zhou, Y.~Guo and Y.~Xue are with University of Science and Technology of China, China, Z.~Zhang is with Kyushu University, Japan, S.~Liu is with College of Intelligence and Computing, Tianjin University, China, S.~Guo is with Baidu Security, USA, Y.~Li is with Department of Industrial Engineering, Tsinghua University, China, L.~Ma is with University of Alberta, Alberta Machine Intelligence Institute, Canada, and Kyushu University, Japan and Y.~Liu is with School of Computer Science and Engineering, Nanyang Technological University, Singapore.\\
Y.~Xue is corresponding author, email: yxxue@ustc.edu.cn.
}

\renewcommand{\shortauthors}{S. Tang et al.}

\begin{abstract}
\emph{Automated Driving Systems} (\emph{ADS}) have made great achievements in recent years thanks to the efforts from both academia and industry. A typical ADS is composed of multiple modules, including sensing, perception, planning, and control, which brings together the latest advances in different domains. Despite these achievements, safety assurance of ADS is of great significance, since unsafe behavior of ADS can bring catastrophic consequences.
Testing has been recognized as an important system validation approach that aims to expose unsafe system behavior; however, in the context of ADS, it is extremely challenging to devise effective testing techniques, due to the high complexity and multidisciplinarity of the systems.
There has been great much literature that focuses on the testing of ADS, and a number of surveys have also emerged to summarize the technical advances. Most of the surveys focus on the system-level testing performed within software simulators, and they thereby ignore the distinct features of different modules. 
In this paper, we provide a comprehensive survey on the existing ADS testing literature, which takes into account both module-level and system-level testing.
Specifically, we make the following contributions: 
\begin{inparaenum}
    \item we survey the module-level testing techniques for ADS and highlight the technical differences affected by the features of different modules;
    \item we also survey the system-level testing techniques, with focuses on the empirical studies that summarize the issues occurring in system development or deployment, the problems due to the collaborations between different modules, and the gap between ADS testing in simulators and the real world;
    \item we identify the challenges and opportunities in ADS testing, which pave the path to the future research in this field.
\end{inparaenum}
\end{abstract}



\maketitle

\section{Introduction}
\label{sec:introduction}
With the aim of bringing convenient driving experience, increasing driving safety and reducing traffic congestion, \emph{automated driving systems} (\emph{ADS}, a.k.a. \emph{self-driving cars}) have attracted significant attention from both academia and industry. According to the statistics from a recent report~\cite{dollar}, the autonomous car market was valued for more than 22 billion dollars in 2021. However, the state-of-the-practice ADS are still vulnerable to numerous safety and security threats, due to either complicated external environments or deliberate attacks from various sources. These threats may lead to system failure, which could bring catastrophic consequences and unacceptable losses~\cite{cui2019review}.
Despite the rapid progress that has been made so far, safety assurance of ADS is still a major challenge to their full-scale industrialization. Some recent news, e.g., the report of Tesla's fatal accident~\cite{accident}, further highlights the importance of research in the safety assurance of automated driving.

In general, an ADS is composed of several modules for the functionalities of sensing, perception, planning and control. The sensing module collects and preprocesses the environmental data using a number of intelligent sensors, such as camera, radar, and LiDAR. The perception module extracts information from the sensors to understand the environmental conditions, including road conditions, obstacles, and traffic signs. Based on the output of the perception module, the planning module generates the optimal driving trajectories which are expected to be followed by the ADS. Lastly, the control module sends the lateral and longitudinal control signals to drive the ADS along the planned trajectories. In particular, some ADS adopt a special \emph{end-to-end} design that integrates the perception, planning and control functionalities in a single module.
These modules collaborate with each other and jointly decide the behavior of the ADS~\cite{pendleton2017perception}; the abnormal function of any module can lead to system failures, which severely threatens the safety and security of the ADS.

Testing has been an effective approach to exposing potential problems and ensuring the safety of systems. However, the testing of ADS is known to be extremely challenging, due to the complexity and multidisciplinarity of those systems. In recent years, there have been a surge of studies that focus on ADS testing. These published papers span over mainstream venues in various domains, such as transportation venues (e.g., ITSC, IV), software engineering venues (e.g., ICSE, ASE), artificial intelligence venues (e.g., CVPR, AAAI), and security venues (e.g., CCS, USENIX Security), which tackle the challenges in ADS testing from various perspectives (see the detailed statistics and analysis in~\S{}\ref{sec:collectionResults}). Numerous testing approaches have been proposed for solving different problems, and numerous bugs and vulnerabilities have been reported to facilitate the system reengineering that repairs the existing problems and ensures the system safety. 

To better understand the landscape of ADS testing, there have been a number of surveys~\cite{grigorescu2020survey, rosique2019systematic,zhang2022finding,zhong2021survey,jahangirova2021quality} that summarize the recent advances in this field.
Grigorescu et al.~\cite{grigorescu2020survey} investigate the deep learning techniques for different modules of ADS and discuss the safety risks of these techniques.
Rosique et al.~\cite{rosique2019systematic} analyze the characteristics of the common sensors used for perception, as well as the performance of different simulators for the simulation of perception systems.
Zhang et al.~\cite{zhang2022finding} present a literature review on the techniques for identification of critical scenarios, in which they point out the necessity of combining different scenario identification methods for safety assurance of ADS.
Zhong et al.~\cite{zhong2021survey} review the works about scenario-based testing in high-fidelity simulators, and discuss the gap between the virtual environment and the real-world environment.
Jahangirova et al.~\cite{jahangirova2021quality} propose a set of driving quality metrics and oracles for ADS testing, and demonstrate the effectiveness of combining the 26 best metrics as the functional oracles.

Most of the existing surveys view the ADS under study as a whole and investigate the methodologies of ADS testing from the perspective of the system level. In that case, as a typical problem setting, ADS testing consists in generating critical scenarios that can lead to system failures, such as collisions with obstacles. In addition, because of the high cost of testing ADS in the real world, most of the studies in these surveys adopt software simulators as the testing environments. 
While these surveys are useful, they are not sufficient to show the comprehensive landscape of ADS testing. Indeed, since ADS are complex and composed of multiple modules that differ from each other in technical design, their testing should capture the features of  different modules and address the challenges in different domains. 
Moreover, at the system level, the testing should concern the problems arising from the collaborations between different modules, and highlight the gap between simulation-based testing and real-world testing.

\myparagraph{Contributions} To bridge this gap, we conduct a survey on ADS testing that focuses on both module-level testing and system-level testing. Specifically, at the module level, we reveal the distinction of the testing techniques for different modules due to their different features; at the system level, we focus on the challenges introduced by the cooperation between different modules and also discuss the different levels of realism of the testing environments. 
In particular, we answer the following research questions in this survey: 
\begin{compactitem}[$\bullet$]
    \item \textbf{RQ1:} What are the techniques adopted for testing different modules of an ADS?
    \item \textbf{RQ2:} What are the techniques adopted for system-level testing of ADS?
    \item \textbf{RQ3:} What are the challenges and opportunities in the field of ADS testing?
\end{compactitem}

\smallskip
In order to answer these questions, we make the following contributions in this paper:
\begin{compactitem}[$\bullet$]
\item To answer \textbf{RQ1}, we survey the testing techniques for the different modules of ADS, and in particular, we highlight the technical differences in these testing techniques due to the characteristics of different modules;
\item To answer \textbf{RQ2}, we survey the system-level testing techniques, with a focus on the following:
\begin{compactitem}
    \item First, we review the existing empirical studies on the issues/bugs in public reports and repositories. These studies reveal the system issues occurring in their development or deployment, and show a bird's-eye view on the potential system problems without running them;
    \item Second, we study the existing investigations on the safety problems at the system level, when different modules collaborate and interact with each other during the running of the systems; 
    \item Third, we focus on the gap between simulation-based testing and real-world testing, which is an emerging topic of great importance, in order to understand the quality of testing.
\end{compactitem}
\item To answer \textbf{RQ3}, we identify the challenges and potential research opportunities for ADS testing, based on our survey results.
\end{compactitem}

To the best of our knowledge, our work is the first one that unveils the intrinsic differences and challenges in ADS testing w.r.t. different modules; meanwhile, we give a specific emphasis on the comparison between the currently popular simulation-based testing and real-world testing. Moreover, our analysis and discussion on the challenges and opportunities exhibit the landscapes, and stimulate future research in this important field.

\myparagraph{Paper organization} The rest of the paper is organized as follows: \S{}\ref{sec:preliminaries} overviews the background of the ADS;  \S{}\ref{sec:paper_collection} describes the survey methodology, including the detailed scope, collection process, and collection results. The main results of this survey are in~\S{}\ref{sec:static_analysis}, \S{}\ref{sec:module} and \S{}\ref{sec:system}. In \S{}\ref{sec:static_analysis}, we survey the literature of empirical study on ADS testing; in \S{}\ref{sec:module}, we survey the literature of techniques on module-level ADS testing and answer RQ1; in \S{}\ref{sec:system}, we survey the literature of techniques on system-level ADS testing and answer RQ2.
We then show the statistics and analysis of the works in~\S{}\ref{sec:analysis}. We summarize the challenges and potential research directions in~\S{}\ref{sec:challenges&opportunities} and answer RQ3. Lastly, we conclude this survey in~\S{}\ref{sec:conclusion}.

\section{preliminaries}\label{sec:preliminaries}
Nowadays, autonomous systems have been deployed in various application domains, such as transportation, robotics, and healthcare, and they have made huge differences to our daily lives.
In this work, we pay particular attention to \emph{automated driving systems} (\emph{ADS}), i.e., \emph{self-driving cars}, as a typical example to exemplify the concerns in the quality aspects of those systems.
In this section, we first provide an overview of the categorization of ADS according to the levels of automation, from L0 to L5; then we show the general architecture of ADS; lastly, we showcase four open-source ADS and one simulation platform.

\subsection{Overview of Automated Driving Systems}
According to the complexity and variety of the ADS, the \emph{society of automotive engineers }(\emph{SAE}) proposed the taxonomy and definitions of driving automation systems, known as \emph{SAE J3016}\footnote{\url{https://www.sae.org/standards/content/j3016_202104/}}, which has become a classification standard in recent years. 
This standard categorizes driving automation systems into six levels, ranging from \emph{no driving automation} (Level 0) to \emph{full driving automation} (Level 5). These levels are usually referred to as L0 to L5. 

\begin{table}[!tb]
  \small
  \centering
  \caption{Automation Levels and Definitions by SAE}
  \label{tab:autoLevel} 
    \resizebox{\textwidth}{!}{
    \renewcommand{\arraystretch}{1.2}
    \begin{tabular}{p{0.05\textwidth}|p{0.25\textwidth}|p{0.4\textwidth}|p{0.2\textwidth} }
    \toprule
    \textbf{Level} & \textbf{Name} & \textbf{Description} & \textbf{Example}\\
    \midrule
     0     
     & No Driving Automation 
     & Drivers perform all of the DDT
     & LDW \\
    \hline
    1    
    & Driver Assistance
    & The system performs part of the DDT: either steering or acceleration/deceleration
    & ALC or ACC \\ 
    \hline
    2     
    & Partial Driving Automation 
    & The system performs part of the DDT: steering and acceleration/deceleration
    & ALC and ACC \\ 
    \midrule
    3
    & Conditional Driving Automation
    & Drivers or fallback-ready users need to be receptive to ADS-issued requests
    &   Traffic jam chauffeur\\
    \hline
    4     
    & High Driving Automation
    & The system performs all of the DDT and DDT fallback within a specified ODD
    & Local driverless taxis \\
    \hline
    5     & Full Driving Automation
    & The system performs all of the DDT and DDT fallback without ODD limitation
    & Full autonomous vehicles\\
    \bottomrule
    \end{tabular}}
\end{table}

The definitions of the systems from L0 to L5 are as follows: 
\begin{inparaenum}
\item L0 systems only perform warnings and momentary interventions, such as \emph{Lane Departure Warning} (\emph{LDW}) and \emph{Automated Emergency Braking} (\emph{AEB}), and the drivers need to perform all of the \emph{dynamic driving tasks} (\emph{DDT});
\item L1 systems support steering or acceleration/deceleration for drivers; example features include \emph{Automated Lane Centering} (\emph{ALC}) and \emph{Adaptive Cruise Control} (\emph{ACC});
\item L2 systems perform steering and acceleration/deceleration at the same time, and a typical L2 system should support both ALC and ACC;
\item L3 systems can execute responses to driving conditions within \emph{Operational Design Domain} (\emph{ODD}), which is an operational restriction imposed to the ADS at the design stage, but these systems require fallback-ready people to handle system failures; an example is a traffic jam chauffeur;
\item L4 systems can further support the system fallback, and an example is a local driverless taxi;
\item L5 systems can handle all driving conditions.
\end{inparaenum}
These are shown in Table~\ref{tab:autoLevel}.
      
A system in L0 to L2 is also known as an \emph{advanced driver assistance system} (\emph{ADAS}), since it is only in charge of a part of the DDT, such as lateral control or longitudinal control, and the safety of the whole vehicle still relies on drivers. 
In contrast, a system in L3-L5 performs all of the DDT and drivers are not expected to interfere during the driving process, so it realizes the real \emph{automated driving}. 

Note that there exist other identified synonyms of \emph{Automated Driving}, e.g., \emph{autonomous driving}, \emph{self-driving}, but in this paper, we follow the relevant terminology from \emph{SAE J3016}, in which the term ``ADS'' refers to \emph{Automated Driving System}. In literature~\cite{zhong2021survey}, an ADAS is usually referred to as a system that belongs to L0-L2, while an ADS is  referred to as a system that belongs to L3-L5. In this survey, since many testing techniques are independent of the automation levels of the systems under test, we sometimes mix the use of the terms and, by ADS, we refer to the systems over all of the levels of driving automation. 

\subsection{Architecture of ADS}\label{sec:ADS_architecture}
\begin{figure}[!tb]
  \centering
  \includegraphics[width=0.9\textwidth]{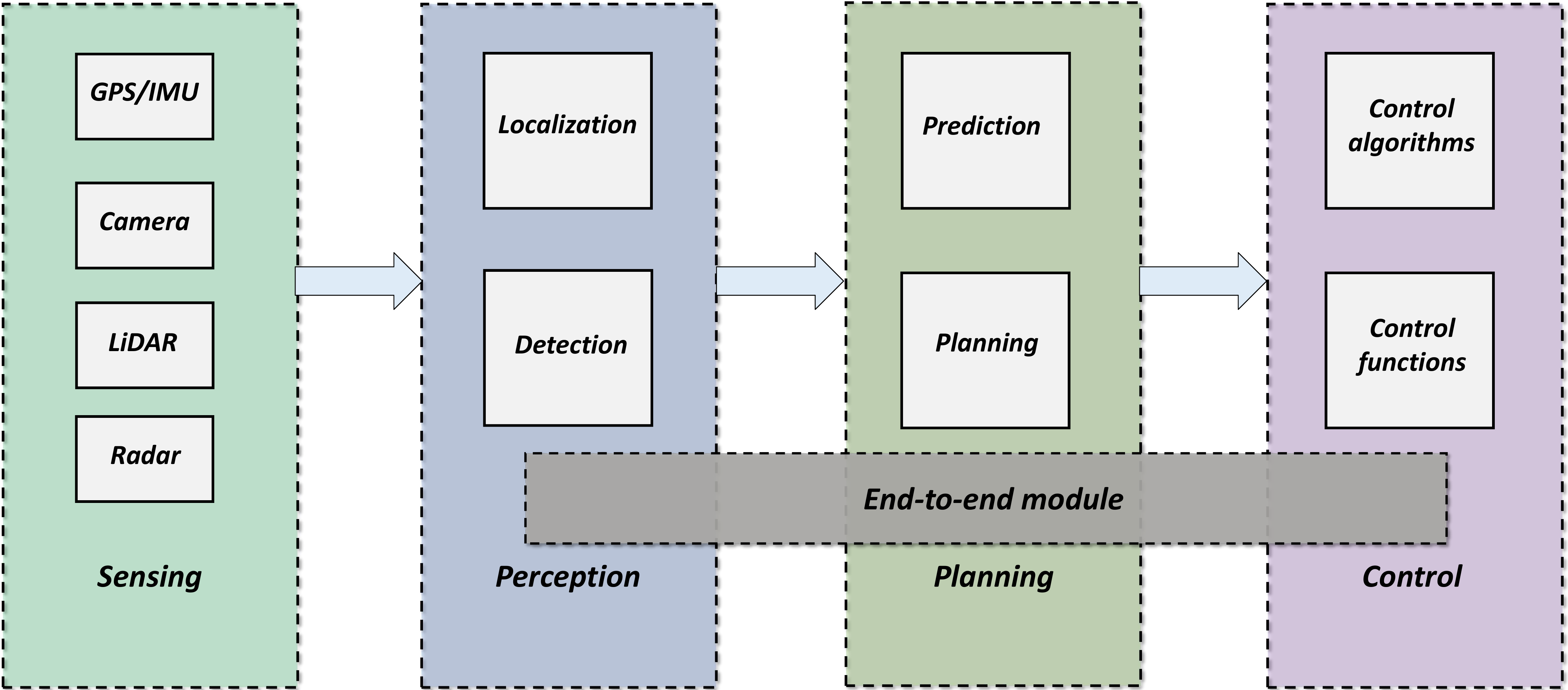}
  \caption{The typical architecture of an ADS}
  \label{fig:ADS_architecture}
\end{figure}

A common ADS is composed of four functional modules, namely, \emph{the sensing module}, \emph{the perception module}, \emph{the planning module} and \emph{the control module}, as shown in Fig.~\ref{fig:ADS_architecture}. In the sensing module, intelligent sensors (e.g., camera, radar and LiDAR) are used to collect the driving context from the physical world. The perception module extracts useful environmental information from the sensor data, and sends it to the planning module for motion planning. Based on the information, the planning module generates the optimal driving trajectory. Lastly, the control module outputs the control commands to drive the vehicle along the trajectory. Moreover, some modern ADS adopt a special design named \emph{end-to-end module}. In the remainder of this section, we elaborate on the functionalities of these modules in the typical architecture of an ADS.

\myparagraph{Sensing module}
By adopting various physical sensors, the sensing module takes charge of collecting and preprocessing driving environmental information from the physical world. The common sensors used by an ADS include \emph{Global Positioning Systems} (\emph{GPS}), \emph{inertial measurement units} (\emph{IMU}), cameras, \emph{radio detection and ranging} (\emph{radar}), and \emph{light detection and ranging} (\emph{LiDAR}). Specifically, GPS provides the absolute position data (e.g., latitude, longitude and heading angle) while IMU provides the temporal data (e.g., acceleration and angular velocity). The combination of these two sensors can provide more accurate real-time positioning of the autonomous vehicles. Cameras are used to record and capture visual information on the driving road for the perception module, and radar is used to detect obstacles by radio waves. Nowadays, LiDAR has become an indispensable component in many leading ADS (e.g., \Apollo and \Autoware), since it can collect 3D point cloud data and process it with higher measurement accuracy. In comparison with camera sensors that are sensitive to light conditions (e.g., shadows and bright sunlight), LiDAR sensors are more robust under these environments, and the generated 3D point cloud can be further utilized to build 3D models that better characterize the surrounding objects.

\myparagraph{Perception module}
With the help of deep learning techniques, the perception module processes sensor data (e.g., pictures and 3D point cloud) from the sensing module to accomplish a series of perception tasks, such as localization, detection, and prediction.
\begin{compactitem}[$\bullet$]
    \item \emph{Localization} provides the real-time location of the ADS during the driving process. Furthermore, localization is mostly implemented by fusing the data from GPS, IMU and LiDAR.
    Specifically, the 3D point cloud data of LiDAR are used to match the features stored in a \emph{High-Definition} (\emph{HD}) \emph{Map}, in order to determine the most likely location.
    \item \emph{Detection} includes lane detection, traffic light detection, and object detection. The data of camera are often used for lane detection and traffic light detection, while the data of camera, radar and LiDAR are often fused by several algorithms (e.g., extended \emph{Kalman filters}~\cite{julier1997new}) for object detection. These detection tasks are mostly implemented by using \emph{deep neuron networks} (\emph{DNNs}), such as faster RCNN~\cite{ren2015faster} and Yolov3~\cite{redmon2018yolov3}.
\end{compactitem}
The prediction task also benefits from the perception module and is mainly used for trajectory planning. We leave the introduction of this task below in the planning module.

\myparagraph{Planning module} 
By using DNNs and planning algorithms, the planning module takes perception data as input and makes decisions for the control module to control the vehicle. It has two submodules, namely, the \emph{prediction submodule} and the \emph{planning submodule}.
\begin{compactitem}[$\bullet$]
    \item \emph{The prediction submodule} estimates the future trajectories of the moving objects (e.g., vehicles and pedestrians) detected by the perception module. For a given moving object, the possibility of its path is often evaluated by machine learning (ML) algorithms, e.g., LSTM, RNN.
    \item \emph{The planning submodule} generates the optimal driving trajectory for ego vehicle based on the prediction results. Specifically, this module is responsible for three tasks, namely, \emph{route planning}, \emph{behavior planning} and \emph{motion planning}. 
    \begin{compactitem}
    \item Route planning selects the optimal path for the vehicle by using path algorithms, such as \emph{Dijkstra} and \emph{A*};
    \item Behavior planning makes decisions for the actions taken by the ADS, such as lane changing and car following, based on the system requirements and traffic rules;
    \item Motion planning generates velocity and steering angle plans which are locally optimal, by considering several factors, including safety, efficiency, and comfort.
    \end{compactitem}
\end{compactitem}

\myparagraph{Control module}
Based on the trajectories planned by the planning module, the control module finally takes charge of the longitudinal and lateral control of the vehicle. By using control algorithms (e.g., \emph{proportional integral derivative} (\emph{PID}) \emph{control}~\cite{pid} and \emph{model predictive control} (\emph{MPC})~\cite{mpc}), this module generates appropriate control commands (e.g., steering and braking) and sends them to the related hardware, i.e., the \emph{electronic control unit} (\emph{ECU}), via the protocol of \emph{controller area network} (\emph{CAN}) \emph{bus}. Note that, this module is critical for several functionalities provided by the ADS, including ACC, AEB, and \emph{Lane Keeping Assistance} (\emph{LKA}).

\myparagraph{End-to-end module}
As shown in Fig.~\ref{fig:ADS_architecture}, besides the common modules mentioned above, there exists another \emph{end-to-end} design that combines the perception, planning, and control processes in one module. 
To be specific, this module mainly consists of special deep learning models, which
are trained by labeled data that maps information from sensors directly to the corresponding control commands. Consequently, these models
could output the control commands based on the current driving environment.

\subsection{Open-Source Systems and Tool Stacks}\label{subsec:open-source_toolsets}
In this section, we introduce four open-source ADS, namely, \Apollo, \Autoware, \OpenPilot and \Pylot, and one simulation platform called \BeamNG~\cite{BeamNG.tech}. The first three ADS have been widely adopted for commercial usage in practice~\cite{apolloCommercial, autowareCommercial, openpilotCommercial} and the last ADS is from academia.

\myparagraph{\Apollo} \Apollo has been a popular open-source ADS developed by \emph{Baidu} since 2017; as of Dec. 2021, it has been updated to version 7.0.0. The hardware platform of \Apollo includes camera, LiDAR, \emph{millimeter wave radar}, and \emph{Human-Machine Interface} (\emph{HMI}) device, and currently the communications over different components are managed by CyberRT~\cite{cyberRT}. 
The functionalities of \Apollo include cruising, urban obstacle avoidance, and lane changing. 

\myparagraph{\Autoware} \Autoware is another open-source L4 ADS developed by the research group of \emph{Nagoya University} in 2015. Though it is mainly applicable to urban roads, it also suits highways and other road conditions. By using the sensors introduced in~\S{}\ref{sec:ADS_architecture}, it supports a series of functionalities including connected navigation, traffic light recognition, and object tracking. Unlike \Apollo which uses CyberRT, \Autoware adopts ROS~\cite{ros} for communications over different components.

\myparagraph{\OpenPilot} \OpenPilot is a popular open-source L2 ADAS developed by \emph{Comma.ai} and it has been updated to version 0.8.12 as of Dec. 2021. \OpenPilot supports common L2 features, such as ACC, ALC, and \emph{Forward Collision Warning} (\emph{FCW}).
Unlike other L2 ADAS,  \OpenPilot has a high degree of \emph{portability}---it can be compatible with more than 120 types of vehicle models by using related hardware set (e.g., \emph{Car Harness}~\cite{carharness} and \emph{Comma Two}~\cite{comma2}).

\myparagraph{\Pylot} 
\Pylot~\cite{gog2021pylot} is a modular and open-source autonomous driving platform developed by \emph{UC Berkeley} in 2021. For achieving the trade-off between latency and accuracy, it is built on a deterministic dataflow system called \emph{ERDOS}~\cite{gog2022d3}.
\Pylot also has other built-in features such as modularity, portability, and debuggability, which allow researchers to implement or test ADS functions with higher efficiency.

\myparagraph{\BeamNG} 
\BeamNG~\cite{beamng} is a popular image-generating simulation platform, which has been widely used in the \emph{Search-Based Software Testing competition} (\emph{SBST})~\cite{panichella2021sbst}. Specifically, it is based on a physically-accurate engine that can support customized vehicle models and realistic damage. For example, different components of a vehicle can have different degrees of deformation after a collision. In addition, \BeamNG also contains a driving agent called \BeamNGAI~\cite{BeamNG.AI}, which could take over one or more vehicles and drive in several different modes.

\section{Paper Collection Methodology and Result}\label{sec:paper_collection}
In this section, we introduce our paper collection methodology in~\S{}\ref{subsec:paperCollectionMethodology} and present the statistics and analysis of the results in~\S{}\ref{sec:collectionResults}.

\begin{figure}[!tb]
  \centering
  \includegraphics[width=0.9\textwidth]{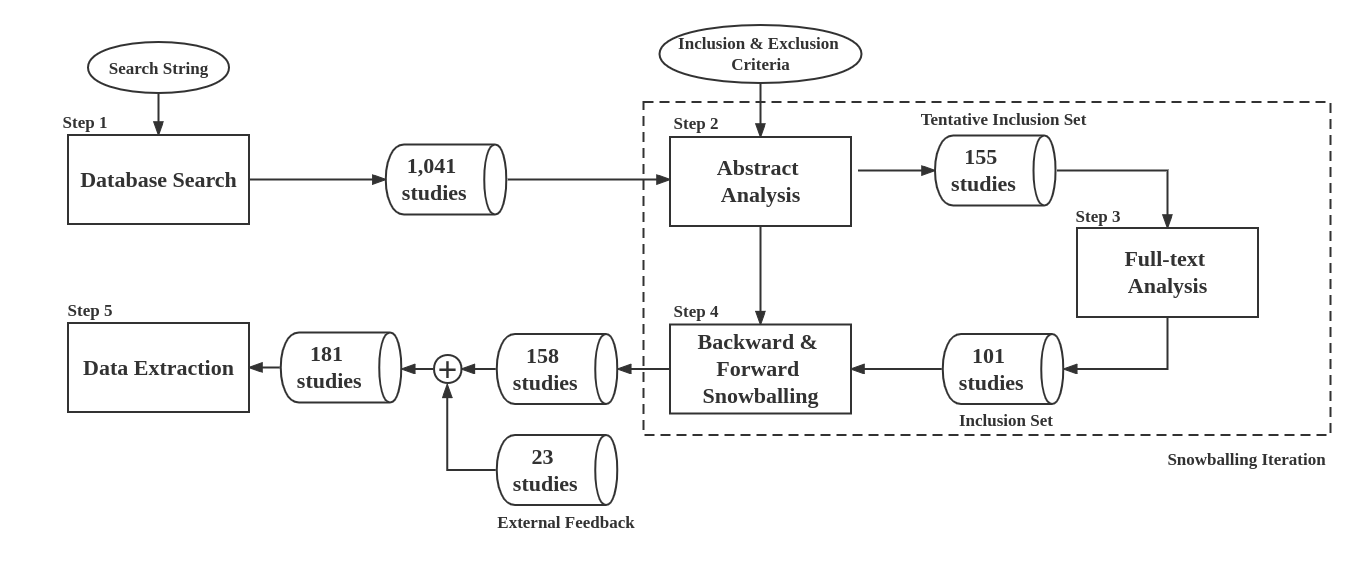}
  \caption{Overview of the paper collection methodology}
  \label{fig:paperCollection}
\end{figure}

\subsection{Paper Collection Methodology}
\label{subsec:paperCollectionMethodology}
This section introduces the methodology adopted in our paper collection process, which is illustrated in Fig.~\ref{fig:paperCollection}. The overall process consists of five main steps, namely, \emph{database search}, \emph{abstract analysis}, \emph{full-text analysis}, \emph{backward \& forward snowballing}, and \emph{data extraction}. The intermediate results of each step during the process are reported in our supplementary website\footnote{\url{ https://sites.google.com/view/ads-testing-survey}} and are also available in \emph{Zenodo}~\cite{tang_shuncheng_2022_7304210}. We now describe the details of each step, as follows: 
\subsubsection{Database Search}
\label{subsubsec:databaseSearch}
This step aims to find the potentially relevant papers by searching in electronic databases. Specifically, we select \emph{DBLP}\footnote{\url{https://dblp.org/}} as our database, which is a popular bibliography database containing a comprehensive list of research venues in computer science. Moreover, our search targets the titles of the papers, since the title often conveys the theme of a paper. We optimize the search string in an iterative manner, in order to collect as many related papers as possible. The final search string used during our search process is shown as follows:
 
\smallskip
\setlength{\fboxsep}{8pt}
\begin{center}
\Ovalbox{
\begin{minipage}{0.9\textwidth}
\small ((``automated vehicle'' OR ``automated driving'' OR ``autonomous car'' OR ``autonomous vehicle'' OR ``autonomous driving'' OR ``self-driving'' OR ``driver assistance system'' OR ``intelligent system'' OR ``intelligent vehicle'' OR ``intelligent agent'') \\
AND \\
(``test'' OR ``attack'' OR ``validation'' OR ``evaluation'' OR ``quality assurance'' OR ``quality assessment'' OR ``oracle'' OR ``mutation'' OR ``fuzzing''))
\end{minipage}
}
\end{center}
\smallskip

The first group of terms (above ``AND'') represents the identified synonyms of automated driving, which contains the terms such as ``autonomous driving'' and ``self-driving''; the second group of terms (below ``AND'') contains the common phases in the process of quality assessment of software systems (e.g., ``test'' and ``validation''), along with popular testing approaches (e.g., ``mutation'' and ``fuzzing'') and a keyword ``oracle'', which is a significant concept in software testing. The terms in each group are connected with \emph{OR} operator, while the two groups are connected with \emph{AND} operator, which means that a relevant paper should cover the characteristics of both groups. Overall, the application of the above search string on \emph{DBLP} retrieves 1185 papers. After removing 144 duplicates, the final number of papers we collected is 1041.

\subsubsection{Abstract Analysis}
\label{subsubsec:abstractAnalysis}
To determine whether each primary candidate paper is relevant to ADS testing, we perform a manual analysis on the abstracts of 1041 papers obtained from database search in~\S{}\ref{subsubsec:databaseSearch}. This process is conducted by two assessors, i.e., the first two authors, following the inclusion and exclusion criteria formulated as follows: 

\begin{compactitem}[$\bullet$]
\item Inclusion Criteria:
    \begin{compactenum}[\textbf{IC}1]
    \item papers that propose a method for testing the modules of ADS or the whole system;
    \item papers that introduce metrics as test oracles or adequacy criteria for testing the modules of ADS or the whole system;
    \item papers published between January 2015 and June 2022.
    \end{compactenum}
\item Exclusion Criteria:
    \begin{compactenum}[\textbf{EC}1]
    \item preprint papers or non-peer-reviewed papers;
    \item early results or preliminary studies;
    \item papers that do not target ADS;
    \item survey papers or summary papers;
    \item papers that do not focus on assessing quality aspects of ADS or its components;
    \item papers that focus on other quality aspects such as Human-Machine Interface (HMI), cyber security, and adversarial defense.
    \end{compactenum}
\end{compactitem}

Specifically, for \textbf{IC2}, test oracle refers to the metrics that measure whether an ADS or its components misbehave, and test adequacy refers to the criteria that judge whether a test suite has been sufficient for testing; for \textbf{EC3}, papers related to other intelligent systems, e.g., \emph{unmanned aerial vehicle} (\emph{UAV}), are excluded since we mainly focus on automated driving systems;
for \textbf{EC4}, such relevant papers are discussed in~\S{}\ref{sec:introduction} for a comparison with our survey; for \textbf{EC5}, papers that do not report novel techniques or metrics for ADS testing are excluded, e.g., the papers that focus on the engineering implementation of testbeds;
for \textbf{EC6}, only the studies that assess quality aspects, e.g., safety and security of the ADS or its modules, are considered.

As a result of manual analysis of the abstracts, the two assessors fully agree on the inclusion of 101 papers, and have divergent opinions on 54 papers, i.e., those included by one assessor but excluded by the other.
To cover more relevant studies, in this step, papers included by either one assessor or both assessors are all added into a \emph{tentative inclusion set}, which contains 155 papers.

\subsubsection{Full-text Analysis}
\label{subsubsec:full-textAnalysis}
In this step, we download the papers in the \emph{tentative inclusion set} and conduct a full-text analysis. For those papers included by both assessors, we further analyze the introduction, conclusion, or other parts to determine whether a certain paper proposes an approach or a metric for ADS testing. If the assessors' decisions conflict, the two assessors will first review the inclusion and exclusion criteria defined in~\S{}\ref{subsubsec:abstractAnalysis}, and have a discussion. In cases where the conflict still exists, a senior researcher will join the discussion and resolve the dispute. After an agreement on removing 54 irrelevant papers, the number of the inclusion set is 101.

\subsubsection{Backward \& Forward Snowballing}
In order to reduce the risk of missing relevant papers, we perform both backward snowballing and forward snowballing~\cite{snowballing} on the 101 papers in the \emph{inclusion set}, and the process is assigned to the two assessors. In backward snowballing, they check the reference list in the existing studies to obtain candidate papers, while in forward snowballing, they use \emph{Google Scholar\footnote{\url{https://scholar.google.com/}}} to access the papers that cite the existing studies. For those candidate papers produced by snowballing, the two assessors apply the inclusion criteria and exclusion criteria defined in~\S{}\ref{subsubsec:abstractAnalysis}, and conduct both \emph{Abstract Analysis in~\S{}\ref{subsubsec:abstractAnalysis}} and \emph{Full-text Analysis in~\S{}\ref{subsubsec:full-textAnalysis}} to identify the relevant papers that could be added into the \emph{inclusion set}. As a result of performing snowballing for one iteration, we identify 57 new papers that are relevant to ADS testing. Hence, the number of papers in the inclusion set after snowballing is 158. To avoid missing relevant papers that may not be obtained through our collection process, we also ask for feedback from domain experts and collect 23 papers as a result. Finally, we collect 181 papers for data extraction.

\subsubsection{Data Extraction}
In this step, all the resulting 181 papers are thoroughly read by the authors. Specifically, the authors need to identify the testing target (ADS module or system) and the proposed method or metrics for ADS testing. The identified information is then extracted into a \emph{data extraction form}.
Since the data extraction process requires careful reading of each paper, this task is conducted by three authors as the assessors to share the overall workload. Each assessor is assigned more than 50 papers, and to ensure accuracy, the extracted information from the three assessors is all reviewed in parallel by another author. All conflicting decisions are resolved in the discussion at this stage.

\subsection{Paper Collection Results}\label{sec:collectionResults}

\begin{figure}[!tb]
\centering
\begin{subfigure}[b]{0.46\textwidth}
\centering
\includegraphics[width=\textwidth]{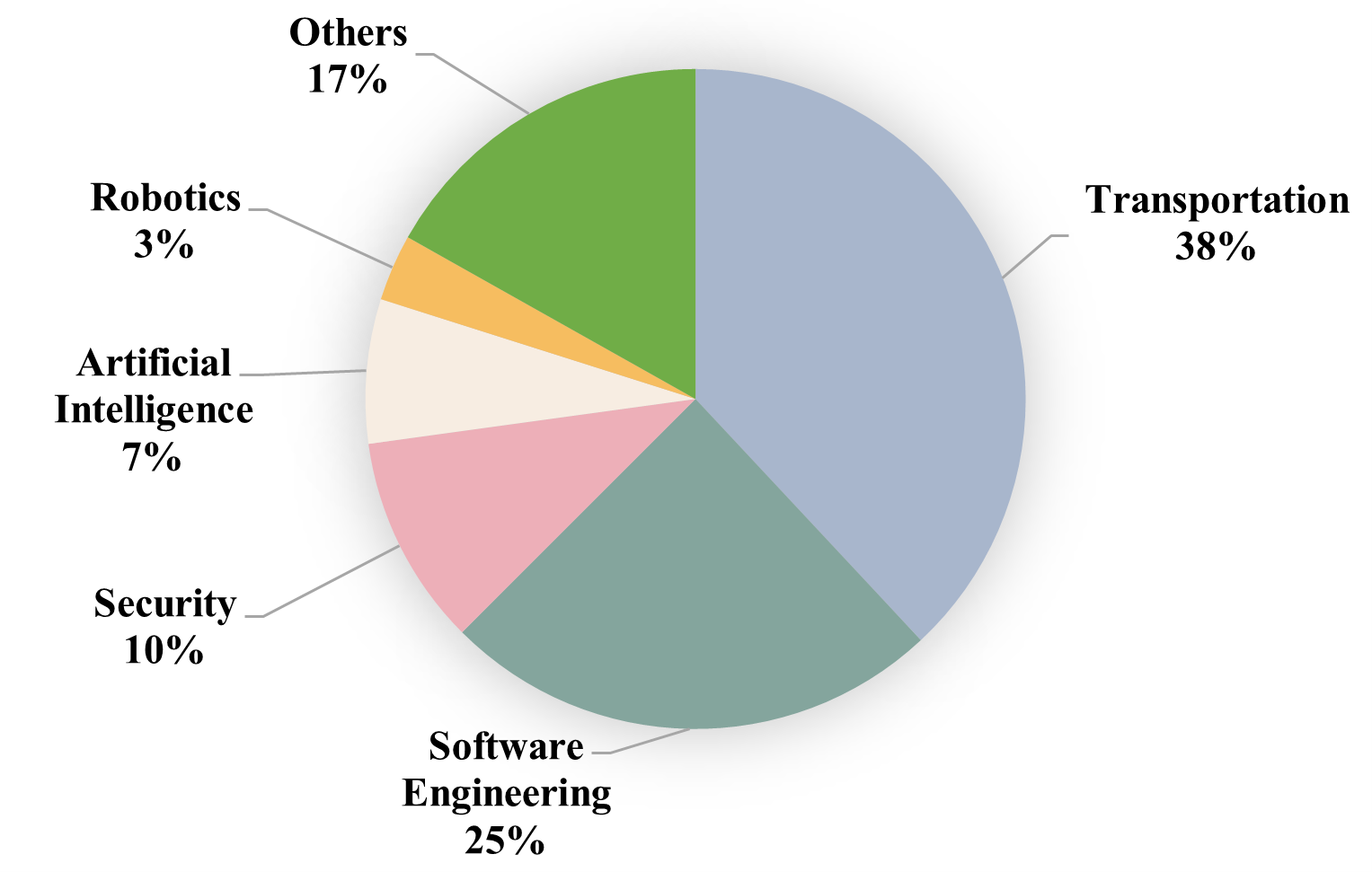}
\caption{Distribution of the publication venues}
\label{fig:publicationVenue}
\vspace{10pt}
\end{subfigure}
\hfill
\begin{subfigure}[b]{0.46\textwidth}
\centering
\includegraphics[width=\textwidth]{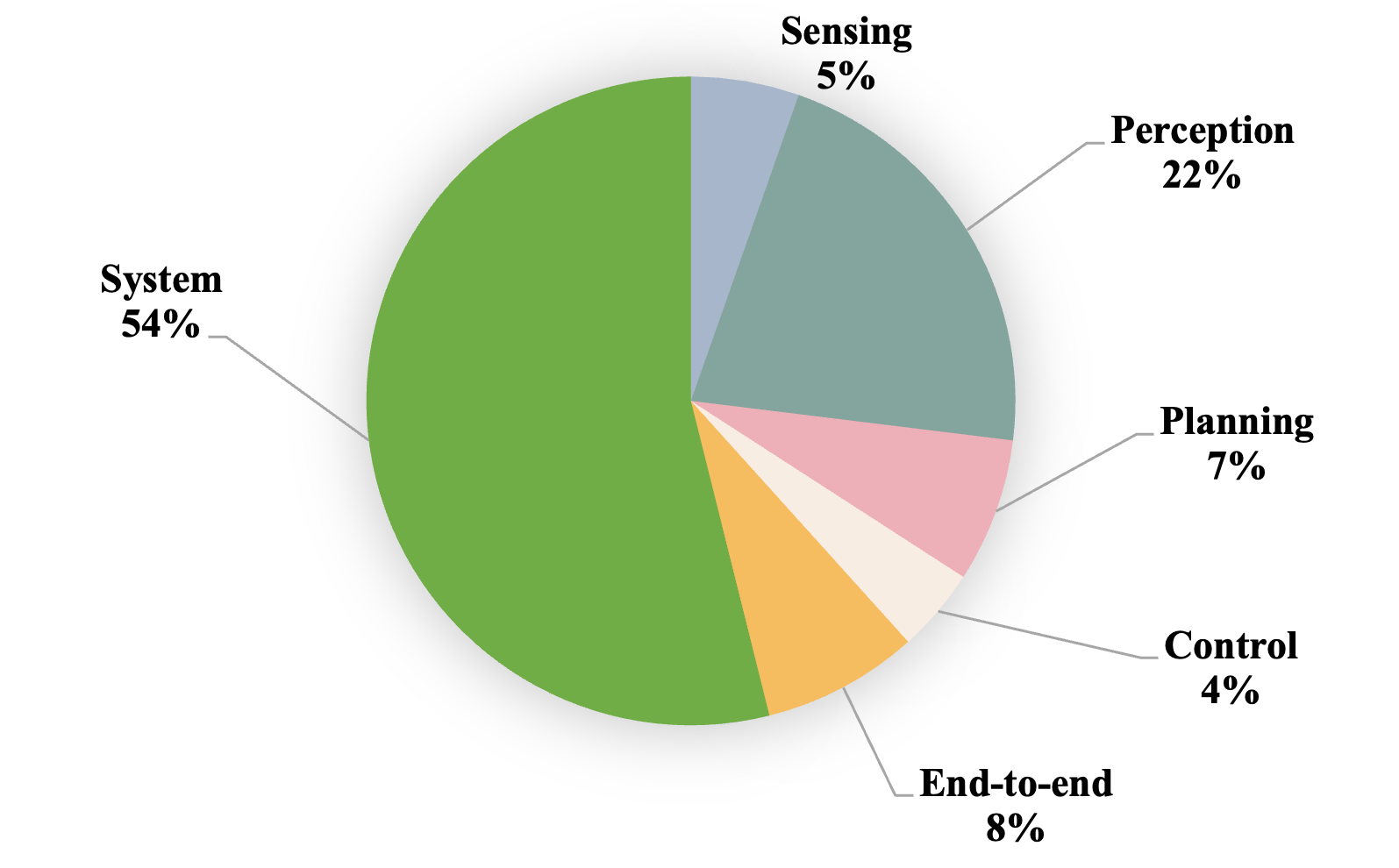}
\caption{Distribution of the testing targets}
\label{fig:publicationModule}
\vspace{10pt}
\end{subfigure}
\begin{subfigure}[b]{\textwidth}
\centering
    \includegraphics[scale=0.6]{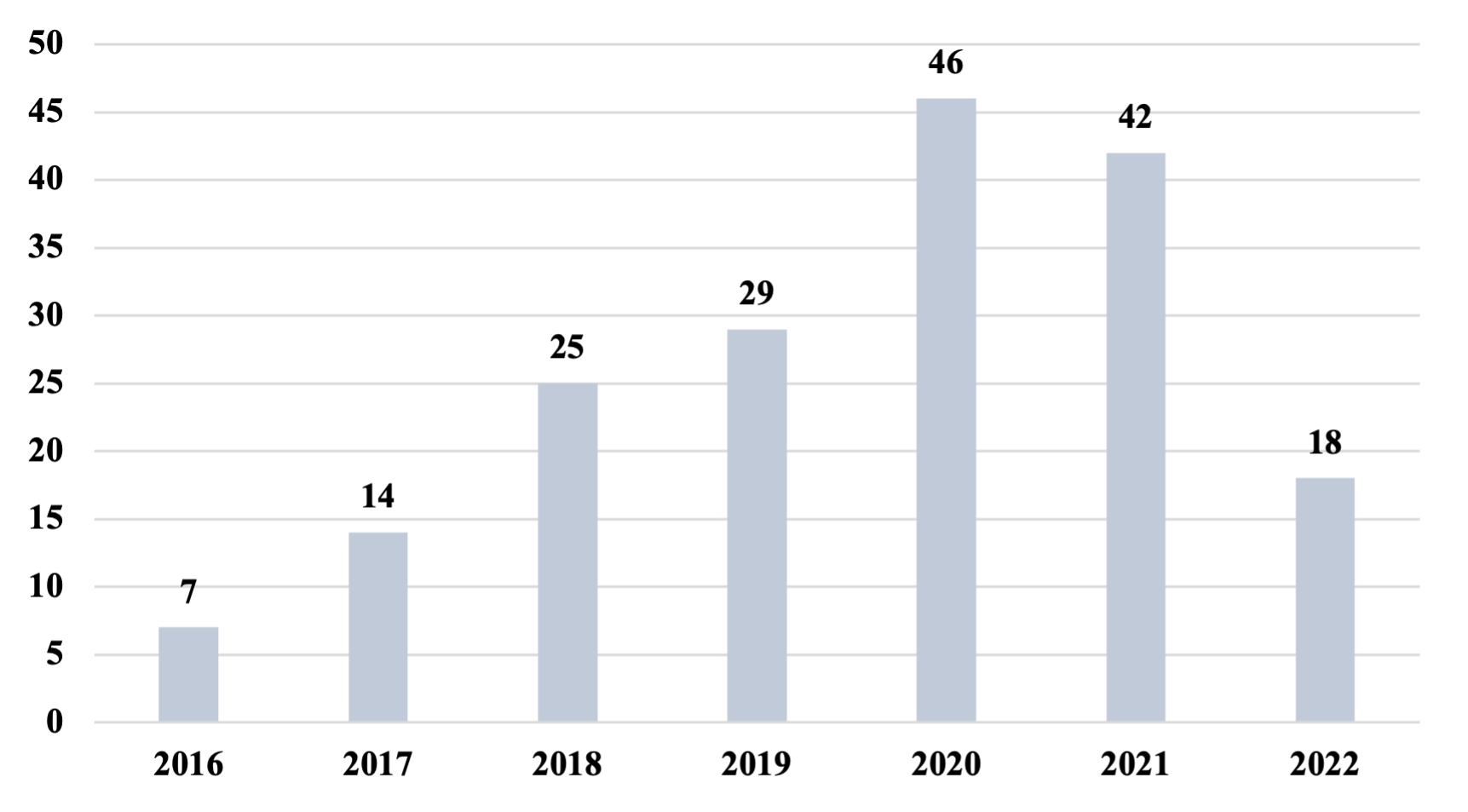}
    \caption{The number of the publications in each year}
    \label{fig:publicationYear}
\end{subfigure}
\caption{The statistical information of the publications in ADS testing}
\label{fig:statistics}
\end{figure}

In this section, we analyze the collected papers from three perspectives, namely, the publication venues, the targeted system modules, and the publication years. We show the distribution of the publication venues of all the papers in Fig.~\ref{fig:publicationVenue} and the distribution of the targeted modules/system in Fig.~\ref{fig:publicationModule}. Moreover, we present the number of papers published in different years in Fig.~\ref{fig:publicationYear}.

\myparagraph{Publication venues} In Fig.~\ref{fig:publicationVenue}, we can see that, 
\begin{inparaenum}[(i)]
\item many of the papers, up to 38$\%$, are published in transportation venues such as \emph{International Conference on Intelligent Transportation Systems} (\emph{ITSC}) and \emph{IEEE Intelligent Vehicles} (\emph{IV}); 
\item 25$\%$ of the papers are published in software engineering venues such as \emph{International Conference on Software Engineering} (\emph{ICSE}) and \emph{International Conference on Automated Software Engineering} (\emph{ASE}); 
\item the adversarial attack methods for vulnerability detection of ADS are related to the security of the systems, and hence 10$\%$ are published in the security venues, such as \emph{USENIX Security Symposium} and IEEE \emph{Symposium on Security and Privacy} (\emph{S\&P});
\item since the ADS and artificial intelligence are closely related, 7$\%$ of the papers are published in artificial intelligence venues such as \emph{Computer Vision and Pattern Recognition Conference} (\emph{CVPR}) and \emph{AAAI Conference on Artificial Intelligence} (\emph{AAAI}). 
\end{inparaenum}

\myparagraph{Target modules} In Fig.~\ref{fig:publicationModule}, we can see that, obviously, the papers on system-level testing dominate the largest percentage, up to 54$\%$. These papers involve testing techniques that span over both simulation-based testing and mixed-reality testing. Moreover, the number of the papers concerning with the perception module is the second largest, up to 22$\%$. The perception module takes charge of object detection and image semantic segmentation using deep learning, which is important but vulnerable to safety and security threats, and thus becomes a popular research direction. Compared to the perception module, there are fewer papers concerning other modules, such as the planning module or the control module.

\myparagraph{Publication year} In Fig.~\ref{fig:publicationYear}, we can see that the number of the papers related to ADS testing shows a general ascending trend, from 2015 to 2021. This trend indicates that the safety and security of ADS are attracting more and more research attention from researchers.
The reason for fewer relevant papers in 2022 is that only partial papers had been published by the time we collected the papers.

\section{Literature of Empirical Study on ADS Testing}\label{sec:static_analysis}
In this section, we provide an overview of the papers that perform empirical study in the field of ADS testing. By the term of empirical study, we mean that, instead of executing the systems in a simulated or real-world environment, these studies perform empirical analysis based on existing databases, such as project repositories and public crash reports. In general, empirical study is an essential step before the experimental ADS testing, 
since it provides experiences and insights in the distribution of potential safety risks.

We classify these studies into three categories, namely, \emph{system study}, \emph{bug/issue study} and \emph{public report study}. 
System study, shown in~\S{}\ref{subsec:systemStudy}, mainly analyzes the architectures of ADS and is thus beneficial for understanding the system behavior before running it.
Bug/issue study, shown in~\S{}\ref{subsec:issueStudy}, focuses on collecting and analyzing the bugs and issues of ADS, which are usually raised by users, developers, and researchers and published in project repositories. 
Public report study, shown in~\S{}\ref{subsec:accidentStudy}, refers to the analysis on those real-world disengagements and crashes reported in various databases (e.g., the crash reports released by \emph{California Department of Motor Vehicles} (\emph{CADMV})~\cite{CADMV}).
These reports target at real-world system failures, and they provide important references for understanding system reliability in the real world.

\subsection{System Study} \label{subsec:systemStudy}
Because of the high complexity of the system architectures of ADS, it is necessary to have a comprehensive understanding of the systems before performing their evaluation.
The system studies,
e.g., on \Apollo~\cite{peng2020first}, build the logical architectures that disclose the connections over different modules in ADS. As a result, these lines of work can bring insights into the potential vulnerabilities and suggest useful metrics for system testing. 

Peng et al.~\cite{peng2020first} investigate the collaboration between the code and the DNN models in \Apollo; specifically, they study
which roles are played by the code and the underlying DNN models, respectively.
They find that the 28 DNN models used in \Apollo interact with each other in diverse ways,
e.g., the output of one DNN can be used as the input of another DNN, and the outputs of multiple DNNs can be combined as the input of another DNN.
Moreover, the code also plays an important role in the system workflow, e.g., it can be used for filtering out invalid output of DNNs, and it can complement the imperfect outcome of DNNs.

\subsection{Bug/Issue Study} \label{subsec:issueStudy}
For those open-source ADS, issues and bugs reported in their public repositories (e.g., GitHub) reflect real problems encountered by users and developers during the development and deployment. Therefore, systematic analysis on these issues~\cite{wang2021exploratory, zampetti2022empirical} can provide insights into the root causes of system failures. In this section, we review two studies~\cite{garcia2020comprehensive, tang2021issue} in the field of ADS testing.

Garcia et al.~\cite{garcia2020comprehensive} present a comprehensive study of bugs in two ADS, namely \Apollo and \Autoware. Specifically, they collect bugs from the commits across the \Apollo and \Autoware repositories in GitHub and perform a manual analysis on these bugs and commits. As a result, they obtain 13 root causes (e.g., algorithm, data, memory) for system crashes, 20 symptoms (e.g., speed and velocity control, vehicle trajectory) and 18 bug-related components (e.g., perception, planning, control), based on their analysis of 499 bugs in the two ADS. 

Tang et al.~\cite{tang2021issue} perform a study on issue analysis for \OpenPilot. They collect 235 bugs from 1293 pull requests and 694 issues of the \OpenPilot project in GitHub and Discord\footnote{\url{https://discord.com/}}. These bugs are then classified into 5 categories, including (DNN) model bugs, plan/control bugs, car bugs, hardware bugs, and UI bugs. Among these different types of bugs, they find that the car bugs related to the interface with different car models dominate $31.48\%$, and plan/control bugs related to the control of car behaviors account for $25.95\%$.

\subsection{Public Report Study} \label{subsec:accidentStudy}
The following works all perform analysis on public reports, i.e., \emph{CADMV}~\cite{CADMV}, which is a database involving disengagement and crash records on public roads.
Specifically, a \emph{disengagement} refers to a failure that requires a human driver to take over control of the vehicle; a \emph{crash} refers to a collision with other traffic participants. These empirical studies investigate the relevant factors, such as the causes, the correlations, and the impacts of these system failures, and they also shed light on future system developments. 

\myparagraph{Analysis of disengagement reports}
In the works~\cite{lv2017analysis, boggs2020exploring, khattak2020exploratory}, the authors analyze the disengagements based on different metrics. Lv et al.~\cite{lv2017analysis} classify the disengagement events into two types, namely \emph{active disengagement} and \emph{passive disengagement}, and investigate the root causes of each group. Boggs et al.~\cite{boggs2020exploring} apply the binary logistic regression~\cite{train2009discrete} to categorize the cause of the disengagements in more details. The results show that the planning discrepancy (e.g., improper localization, motion planning) accounts for $41\%$ of ADS disengagements. Khattak et al.~\cite{khattak2020exploratory} investigate the relationship between disengagements and crashes, and find relevant factors that could increase the likelihood of a disengagement without a crash.

\myparagraph{Analysis of crash reports}
The following works~\cite{leilabadi2019depth, favaro2017examining,wang2019exploring, das2020automated, aziz2021data, song2021automated, esenturk2021analyzing} analyze the crash reports and identify the contributing factors.
Leilabadi et al.~\cite{leilabadi2019depth} apply text analysis to the crash reports, and they find that the crashes mostly occur when vehicles run in the automated mode, and the most frequent ADS crash type is the rear-end collision.
Favaro et al.~\cite{favaro2017examining} focus on the dynamics aspect and present the speed distribution of those crash vehicles. 
Wang et al.~\cite{wang2019exploring} adopt regression and classification tree (CART) to investigate the types and severity of these crashes. They find that the severity increases significantly when an automated vehicle is responsible for the event.
Das et al.~\cite{das2020automated} utilize Bayesian latent class model to perform the analysis and identify six collision patterns. Aziz et al.~\cite{aziz2021data} investigate both crash data involving ADS and without ADS, and build a spatial-temporal mapping of the contributing factors between them.
Song et al.~\cite{song2021automated} conclude that the most representative crash pattern is the ``collision following ADS stop'', i.e., an automated vehicle stops suddenly and gets hit by other vehicles on the road. Besides \emph{CADMV}, the crash data in other databases such as \emph{UK's STATS19}~\cite{UK19} are also analyzed with statistical approaches~\cite{esenturk2021analyzing, wang2017analysis}.

\begin{table}[!tb]
\footnotesize
\caption{Summary of the papers for empirical study on ADS testing}
\label{tab:empirical study}
\renewcommand{\arraystretch}{1.2}
\resizebox{\textwidth}{!}{
\begin{tabular}{p{0.2\textwidth}|p{0.7\textwidth}|p{0.1\textwidth}<{\centering} }
\toprule
\textbf{Category}
& \textbf{Description} &\textbf{ Literature}
\\ 
\hline
System study
& Introducing the interaction between the code and the DNN models in \Apollo
& \cite{peng2020first}\\
\hline
\multirow{3}{*}{Bug/issue study}
& Finding the root causes, symptoms, and bug-related components based on analysis on bugs of \Apollo and \Autoware
& \cite{garcia2020comprehensive}
\\
\cline{2-3}
& Performing categorization and analysis on bugs of \OpenPilot
&\cite{tang2021issue}
\\
\hline
\multirow{3}{*}{\begin{tabular}[c]{@{}l@{}}Public report study\end{tabular}}
& The analysis and classification of the disengagements based on different perspectives (e.g., modules)
& \cite{lv2017analysis, boggs2020exploring, khattak2020exploratory}
\\
\cline{2-3}
& The identification of the common crash types by different methods (e.g., text analysis)
& 
\cite{leilabadi2019depth, favaro2017examining,wang2019exploring, das2020automated, aziz2021data, song2021automated, esenturk2021analyzing, wang2017analysis}
\\
\bottomrule
\end{tabular}
}
\end{table}

\subsection{Discussion}
\label{sec:empirical study discussion}
Table~\ref{tab:empirical study} summarizes the collected papers that empirically study the issues in ADS testing. 
Several existing system studies focus on \Apollo,
and there are also studies that cover other open-source ADS, such as \Autoware and \OpenPilot.
Moreover, there are many works~\cite{lv2017analysis, boggs2020exploring, khattak2020exploratory,leilabadi2019depth, favaro2017examining, das2020automated, aziz2021data, song2021automated,leblanc2011driver, esenturk2021analyzing} that target the disengagement and crash reports for identifying the root causes or failure types, as these investigations are critical to understanding the ADS safety performances in the real world.

\smallskip
\setlength{\fboxsep}{8pt}
\begin{center}
\Ovalbox{
\begin{minipage}{0.9\textwidth}
\small \textbf{Summary:}
Many empirical studies focus on studying the systems themselves, e.g., \Apollo, to understand the characteristics of the systems or bugs/issues from their public project repositories. There are also many studies that analyze the public crash reports to understand the safety problems of ADS in the real world.
\end{minipage}
}
\end{center}
\smallskip

\section{Literature of Techniques on Module-Level ADS Testing}
\label{sec:module}
In this section, we introduce the works on module-specific testing for ADS with the  goal of answering RQ1 in \S{}\ref{sec:introduction}. These modules under test include the ones that have been introduced in~\S{}\ref{sec:ADS_architecture}, namely, the sensing module (in~\S{}\ref{sec:sensing}), the perception module (in~\S{}\ref{sec:perception}), the planning module (in~\S{}\ref{sec:planning}), the control module (in~\S{}\ref{sec:control}) and the end-to-end module (in~\S{}\ref{sec:end2end}). 

We introduce these studies from three perspectives, namely, test methodology, test oracle and test adequacy. Concretely, 
\begin{inparaenum}[(i)]
\item test methodology introduces various methods or technical innovations for testing; 
\item test oracle defines metrics that can be used to judge whether the module behaves correctly;
\item test adequacy proposes coverage criteria that tell if the test cases in a test suite are sufficient.
\end{inparaenum}
Note that, due to the different features of different modules, it can be the case that, for a specific module, not all of the three perspectives are identified as important scientific topics, so we may only introduce the related literature from only a part of the perspectives.

\subsection{Sensing Module} \label{sec:sensing}
The sensing module is the frontier module of an ADS and the performance of the physical sensors (e.g., camera, radar, LiDAR) in this module is critical to the safety and security of the whole ADS. Relevant studies on the test methodology of this module can be divided into \emph{physical testing} (shown in~\S{}\ref{subsec:physicalTesting}) and \emph{deliberate attack} (shown in~\S{}\ref{subsec:deliberateAttack}). Physical testing aims to test the performance of the sensors under different physical conditions, while deliberate attack interferes with the input signals of the sensors to diminish the sensing quality. 

\subsubsection{Physical Testing} \label{subsec:physicalTesting}
Physical testing~\cite{kutila2016automotive, kutila2018automotive} aims to assess the sensors’ capabilities of handling specific tasks under different physical environments, such as harsh weather conditions.
Kutila et al.~\cite{kutila2016automotive} perform a detection distance testing of LiDAR in the foggy and snowy conditions. The results show that the maximum measurable distance by the LiDAR decreases by $20-40m$ under harsh weather conditions.
They also compare the detection capability of LiDAR with different wavelengths in their follow-up work~\cite{kutila2018automotive}. Concretely, they test the detection accuracy of LiDAR at $905nm$ and $1550nm$ wavelengths in foggy and rainy weather, and the results indicate that the LiDAR with a larger wavelength can detect the environment more accurately when the visibility is low.

\subsubsection{Deliberate Attack}\label{subsec:deliberateAttack}
Unlike physical testing, deliberate attack refers to the intentional attacks launched by human attackers.
This type of attack on the sensors of ADS can be classified into \emph{jamming attack} and \emph{spoofing attack}, which are introduced below.

\myparagraph{Jamming attack} This is a basic type of attack on sensors by generating noises using specific tools to interfere with the sensors and damage their normal functionalities. 
Shin et al.~\cite{shin2017illusion} propose a blinding attack method against LiDAR by using intense light with the same wavelength as the target sensor.
Yan et al.~\cite{yan2016can} utilize a laser to cause irreversible damage to cameras and an ultrasonic jammer to interfere with ultrasonic sensors.
Another attack on ultrasonic sensors~\cite{lim2018autonomous} works by placing an ultrasonic sensor opposite to the target sensor. 

\myparagraph{Spoofing attack} Spoofing attack is performed by injecting fake data to deceive sensors. 
Meng et al.~\cite{meng2019gps} and Zeng et al.~\cite{zeng2018all} spoof the GPS receivers to a wrong destination by modifying the raw signals of these sensors.
Komissarov et al.~\cite{komissarov2021spoofing} utilize a \emph{Software Defined Radio} to fool the \emph{mmWave radar}, e.g., they make it produce the wrong measurement of vehicle speed.
Wang et al.~\cite{wang2021can} first utilize the features of \emph{infrared lights} to perform spoofing attack. Specifically, the proposed approach could create invisible objects with simple LEDs to fool the camera sensors, and thus introduce localization errors to the vehicles.

\begin{table}[!tb]
\caption{Summary of the papers for the sensing module testing}
\label{tab:sensing}
\footnotesize
\resizebox{\textwidth}{!}{
\renewcommand{\arraystretch}{1.2}
\begin{tabular}{p{0.1\textwidth} p{0.1\textwidth}|p{0.5\textwidth}|p{0.1\textwidth}|p{0.2\textwidth}}
\toprule
\multicolumn{2}{l|}{\textbf{Methodology}}   
& \textbf{Description}  
& \textbf{Literature} 
& \textbf{Test Sensor} \\ [2ex]
\hline
\multicolumn{2}{l|}{\multirow{2}{*}{Physical testing}}     
& Testing the detection distance of LiDAR under different weather conditions 
& \cite{kutila2016automotive,kutila2018automotive}  & LiDAR       \\ 
\hline
\multicolumn{1}{l|}{\multirow{7}{*}{\parbox{0.1\textwidth}{Deliberate attack}}}
& \multirow{4}{*}{\parbox{0.1\textwidth}{Jamming attack}}
& 
Using intense light to blind the sensors
& \cite{shin2017illusion}          & LiDAR                \\ 
\cline{3-5} 
\multicolumn{1}{l|}{}   &   & Utilizing a laser and a jammer to interfere with the sensors
& \cite{yan2016can}     & Camera and ultrasonic sensors    \\ 
\cline{3-5} 
\multicolumn{1}{l|}{}  &  
& Placing an opposite ultrasonic sensor
& \cite{lim2018autonomous}        & Ultrasonic sensors   \\ 
\cline{2-5} 
\multicolumn{1}{l|}{}   
& \multirow{3}{*}{\parbox{0.1\textwidth}{Spoofing attack}} 
& Modifying the raw data
& \cite{meng2019gps,zeng2018all} & GPS  \\ 
\cline{3-5} 
\multicolumn{1}{l|}{}& & Utilizing a Software Defined Radio & \cite{komissarov2021spoofing}
& Radar
\\
\cline{3-5}
\multicolumn{1}{l|}{}& & Creating invisible objects with simple LEDs & \cite{wang2021can}
& Camera \\
 \bottomrule
\end{tabular}}
\end{table}

\subsubsection{Discussion}
\label{subsubsec:sensingDiscussion}
Table~\ref{tab:sensing} shows the summary of the papers for the sensing module testing.
It can be seen that the existing physical testing works~\cite{kutila2016automotive, kutila2018automotive} mainly focus on testing LiDAR sensors under different weather conditions, e.g., the foggy weather, the snowy weather. This is because the LiDAR sensor has become a key component in ADS, and its robustness is of great significance to the vehicle's safety.
Besides, we find more works that perform deliberate attack including \emph{jamming attack}~\cite{shin2017illusion, yan2016can,lim2018autonomous} and \emph{spoofing attack}~\cite{meng2019gps,zeng2018all,komissarov2021spoofing,wang2021can} on other physical sensors. With the usage of specific devices, e.g., lasers~\cite{yan2016can} and LEDs~\cite{wang2021can}, these two types of attacks have been demonstrated to be effective for finding abnormal behaviors of the target sensors.

\smallskip
\setlength{\fboxsep}{8pt}
\begin{center}
\Ovalbox{
\begin{minipage}{0.9\textwidth}
\small \textbf{Summary:}
Physical testing focuses on testing sensors under different weather conditions. There are more works performing deliberate attack on the sensing module, e.g., jamming attack and spoofing attack, with the usage of specific hardware devices.
\end{minipage}
}
\end{center}
\smallskip

\subsection{Perception Module} \label{sec:perception}
The perception module receives and processes sensor data; based on that, it perceives external environments. The literature we collected includes the test methodologies (shown in~\S{}\ref{subsec:perceptionMethodology}), the test oracles (shown in~\S{}\ref{subsec:perceptionOracle}), and the test adequacy criteria (shown in~\S{}\ref{subsec:perceptionAdequacy}) for testing the DNN models used in the perception module of ADS.

\subsubsection{Testing Methodology}\label{subsec:perceptionMethodology}
Adversarial attack is the major approach for testing the DNN models used in the perception module, which attempts to generate \emph{adversarial examples} to trigger wrong inference results of perception. 
Based on the attacker's knowledge about the target model, adversarial attacks can be classified into \emph{white-box attacks}, in which the attackers have access to the training parameters of the target model, and \emph{black-box attacks}, in which the attackers have limited or no knowledge of the model. Based on the attackers' desired outcomes, there exist \emph{targeted attacks}, in which the prediction that the model makes is limited to specific classes, and \emph{non-targeted attacks}, in which the model can predict an arbitrary
wrong class~\cite{kurakin2018adversarial}.
In general, there are three basic methods for performing adversarial attack, namely, by solving an optimization problem, by leveraging the \emph{generative adversarial networks} (\emph{GAN})~\cite{goodfellow2014generative} and by poisoning the training data. In the following, we introduce the literature that adopts these methods. 

\myparagraph{Optimization-based attack}
We denote by $F$ a DNN model, which takes as input a picture $x$ and gives as output a label $y$. In general, an adversarial attack consists in solving the following optimization problem: 
\begin{equation}\label{eq:adversarialAttack}
    \min \delta \qquad s.t. \quad F(x + \delta) = y^{*}, \quad y^{*}\neq y^{o}
\end{equation}
where $\delta$ is a perturbation added to the picture $x$, and $y^{*}$ is a wrong label that is different from the correct label $y^{o}$. In other words, an adversarial attack involves finding the minimum perturbation that leads a DNN model to the wrong inference result. In most cases, the collected literature on adversarial attack follows this general framework; meanwhile, these papers also differ in their applications and motivations.

\smallskip
The following works~\cite{chen2018shapeshifter, zhao2019seeing,zhang2019camou,im2022adversarial,xu2020adversarial,li2020adaptive,kumar2020black} focus on performing adversarial attacks on camera-based perception tasks (e.g., object detection, traffic sign recognition, and semantic segmentation). 
Chen et al.~\cite{chen2018shapeshifter} propose an attack method, called \emph{ShapeShifter}, to generate perturbations against the object detector \emph{Faster R-CNN}~\cite{ren2015faster}. 
To make the perturbations more robust, they adopt the \emph{Expectation over Transformation} technique~\cite{athalye2018synthesizing} that adds random distortions iteratively, during the optimization process for generating perturbations. Zhao et al.~\cite{zhao2019seeing} propose two approaches for generating adversarial perturbations: one is called \emph{hiding attacks} that can make object detectors unable to recognize objects; and the other is \emph{appearing attack} that can lead the object detectors to make incorrect recognition.
Zhang et al.~\cite{zhang2019camou} propose an attack method for object detectors, which could generate camouflage on 3D objects, i.e., vehicles, and make it undetectable by target models.
Unlike the classification loss adopted by most studies, Choi et al.~\cite{im2022adversarial} consider the object loss defined as the detector's confidence on the existence of objects in an area. The adversarial perturbations generated by their approach could make the target object detector \emph{YOLOv4}~\cite{bochkovskiy2020yolov4} produce numerous false positives, i.e., those objects that do not exist in the clean images are unexpectedly detected. Xu et al.~\cite{xu2020adversarial} perform an adversarial attack on the popular segmentation model \emph{DeepLab-V3+}~\cite{chen2018encoder}. The perturbations generated are quite small and can be stealthily projected to an unnoticed area in the original image. 
Li et al.~\cite{li2020adaptive} propose the first black-box attack on traffic sign recognition models, which could generate adversarial perturbations efficiently.
Kumar et al.~\cite{kumar2020black} present another black-box attack method on traffic sign recognition models. Instead of maximizing the loss of the correct class, they accelerate the convergence through minimizing the loss of the class which is incorrectly predicted by target models.

In addition to attacking the camera-based object detectors, there are also works~\cite{cao2019adversarial,  cao2021invisible, sun2020towards,wang2021adversarial, chen2021camdar, zhu2021can,yang2021robust,zhu2021adversarial,li2021fooling} that focus on attacking the LiDAR-based 3D object detectors.
Cao et al.~\cite{cao2019adversarial} present a white-box attack method on a LiDAR-based perception module by adding the spoofed points into the original 3D point clouds. In the later work~\cite{cao2021invisible}, their generated adversarial perturbations could fool both the camera and the LiDAR-based perception algorithms.
Black-box attacks on the LiDAR-based object detectors are performed in~\cite{sun2020towards, wang2021adversarial, chen2021camdar, zhu2021can} and experimental results show that the target models are highly sensitive to those adversarial 3D perturbations.
Yang et al.~\cite{yang2021robust} consider both white-box and black-box scenarios and generate perturbations for roadside objects such that they can be misidentified as vehicles by the perception module.
Zhu et al.~\cite{zhu2021adversarial} generate perturbations for roadside objects but they target LiDAR-based semantic segmentation tasks.
Unlike existing works that generate perturbations for 3D objects, Li et al.~\cite{li2021fooling} add perturbations to the vehicle trajectories and their method can result in a significant drop in the precision of the object detector, to nearly zero.

While the adversarial attack framework in Eq.~\ref{eq:adversarialAttack} is effective in fooling DNN models, it does not consider the realism of the perturbed pictures.
There is literature that considers the adversarial attack problem under physical conditions.
Eykholt et al.~\cite{eykholt2018robust, song2018physical} propose an attacking method, called \emph{Robust Physical Perturbations} (\emph{$RP_{2}$}), that induces road sign classifiers to produce wrong classification results under real-world physical conditions, e.g., different viewpoint angles and different distances to the signs. 
Experimental results show that the attacked classifier misclassifies the traffic signs with a rate of 100$\%$ in the lab environment and 84.8$\%$ in the real world.

\myparagraph{GAN-based attack}
This type of attack~\cite{xiao2018generating} generates adversarial perturbations to fool a DNN model by training a GAN~\cite{goodfellow2014generative}. 
A GAN consists of two neural network models, namely, a generator $G$ and a discriminator $D$; specifically, $G$ is used to generate perturbations and add them to an input image, and $D$ is used to distinguish the generated image by $G$ and the original image.
The objective of training a generator $G$ is to make the perturbed image of $G$ indistinguishable by the discriminator $D$; this can be implemented by optimizing a loss function $L_{G}$. For fooling the target DNN, another loss function $L_{D}$ is needed to stimulate the adversarial images produced by the GAN to be misclassified.
As a result, the final objective function is formalized as follows:
\begin{align*}
    L \;=\;\gamma\cdot L_{G} + L_{D}
\end{align*}
where $\gamma$ is a parameter that controls the relative importance of $L_G$ and $L_D$.

Liu et al.~\cite{liu2019perceptual} propose a GAN-based attack framework called \emph{perceptual-sensitive GAN} (\emph{PS-GAN}), which generates adversarial patches with high visual fidelity. Experimental results show that the adversarial patches can significantly reduce the classification accuracy of the target DNNs.
Xiong et al.~\cite{xiong2021multi} propose a multi-source attack method based on GAN, which generates adversarial perturbations that can fool both camera-based and LiDAR-based perception models. 
Yu et al.~\cite{yu2018intelligent} utilize the \emph{cycle-consistent generative adversarial network} (\emph{CycleGAN})~\cite{zhu2017unpaired} to synthesize corner cases for testing traffic sign detection models.

\myparagraph{Trojan attack}
This type of attack~\cite{liu2017trojaning} is also  called \emph{poisoning attack} or \emph{backdoor attack}. Specifically, it works by injecting malicious samples with \emph{trigger patterns} into the training data of the target DNN models. Then the models can learn the malicious behaviors and make incorrect predictions when the inputs contain such triggers.
The following works~\cite{jiang2020poisoning,ding2019trojan} are all based on this idea.

Jiang et al.~\cite{jiang2020poisoning} utilize \emph{particle swarm optimization}~\cite{eberhart1995new} to perform this type of attack on traffic sign recognition models. Experimental results show that the classification accuracy could drop to 62$\%$ due to only 10$\%$ injected training data.
Ding et al.~\cite{ding2019trojan} propose the Trojan attack for deep generative models such as DeRaindrop Net~\cite{qian2018attentive}, which is a GAN-based network for raindrops removal. Experimental results show that the model could be triggered to misclassify the traffic light or the value on the speed limit sign when it normally removes the raindrops.

\subsubsection{Test Oracle} \label{subsec:perceptionOracle}
A test oracle defines a metric used to distinguish between the expected and unexpected behavior of the system under test.
Sometimes, an oracle is obviously identified; however, that is not always the case. In the perception testing, due to the huge input space (that involves all the possible input images) of the DNN models, it is a great challenge to specify the oracles for all the input images.
We collect several types of test oracles that have been adopted for perception testing, namely, ground-truth labeling~\cite{zhou2019automated,philipp2021automated}, metamorphic testing~\cite{shao2021testing,bai2021metamorphic,zhou2019metamorphic,woodlief2022semantic,wang2021object,ramanagopal2018failing} and formal specifications~\cite{dokhanchi2018evaluating,balakrishnan2021percemon} to judge whether a bug exists in the perception module.

\myparagraph{Ground-truth labeling}
The general approach of testing a DNN in the perception module is to match the inferred label by the DNN with the ground-truth label, given an image. Usually, these ground-truth labels are obtained by manual labeling. For instance, the ground-truth labels in~\cite{kondermann2016hci, sun2020scalability} are produced in this way.
However, manual labeling is notoriously expensive and laborious; to that end, automatically labeling methods are pursued by researchers. Zhou et al.~\cite{zhou2019automated} propose an automatic labeling method to detect the \emph{road} component in the camera sensor images. Their method identifies the road component in the 3D point cloud captured by a LiDAR for the same scene, and projects the identified area onto the corresponding image. The projected area labels the \emph{road} component in the camera sensor images, which can be used for the validation of semantic segmentation models.
Philipp et al.~\cite{philipp2021automated} propose another approach for automatically generating dimension and classification references for object detection. 
The dimension references are calculated by considering the occurred situations of each object and measuring the related features, e.g., projection angle, based on a given HD map.
The classification references are generated by a decision tree, which considers the features such as kinematic behavior and the interaction with infrastructure elements of each object.

\myparagraph{Metamorphic testing}
Metamorphic testing~\cite{chen2020metamorphic} was introduced by Chen et al. to tackle the problem when the test oracle is absent in traditional software testing. Consider the testing of a program $f$ that implements the trigonometric function $\sin$. Normally, for any input $x$, given the ground-truth value $\sin(x)$ as the oracle for $f(x)$, we can assess the correctness of $f$ by checking if $f(x) = \sin(x)$. However, assume that the ground-truth value $\sin(x)$ is unknown. In this case, testing $f$ by checking if $f(x) = \sin(x)$ is not possible; instead, we can use the \emph{metamorphic testing} that tests the program based on a metamorphic relation. For instance, in this case,
a metamorphic relation can be built as $f(x) = f(\pi - x)$, due to the property $\sin(x) = \sin(\pi - x)$ held by $\sin$. Hence, the correctness of $f$ can be assessed by metamorphic testing, which consists in checking if $f(x) = f(\pi-x)$, for any input $x$.

Metamorphic testing has been studied for testing the perception module of an ADS; various metamorphic relations have been proposed, over images~\cite{shao2021testing,bai2021metamorphic,zhou2019metamorphic,woodlief2022semantic,wang2021object} and frames in a scenario~\cite{ramanagopal2018failing}.

\begin{compactitem}[$\bullet$]
\item{\it Metamorphic relations over images.} 
Shao et al.~\cite{shao2021testing} introduce a metamorphic relation in object detection, that is, the detected object in the original images should also be detected in the synthetic images.
For testing traffic light recognition models, Bai et al.~\cite{bai2021metamorphic} propose another metamorphic relation, which states that, when traffic lights change from one color to another, the recognition results of the target models should change correspondingly.
Zhou et al.~\cite{zhou2019metamorphic} propose a metamorphic relation for LiDAR-based object detection, that is, the noise points outside the \emph{Region of Interest} (\emph{ROI}) should not affect the detection of objects within the \emph{ROI}.
Woodlief et al.~\cite{woodlief2022semantic, wang2021object} check the model inconsistencies between original images and mutated images, e.g., the images of a vehicle with changed color.

\item{\it Metamorphic relations over frames in a scenario.} Ramanagopal et al.~\cite{ramanagopal2018failing} propose two metamorphic relations, respectively for identifying temporal and stereo inconsistencies that exist in different frames of a scenario. The temporal  metamorphic relation says that an object detected in a previous frame should also be detected in a later frame; 
the stereo metamorphic relation is defined in a similar way, for regulating the spatial consistency of the objects in different frames of a scenario.
\end{compactitem}

\myparagraph{Formal specifications} 
Recently, temporal logics-based formal specifications have been adopted in the monitoring of the perception module of ADS. In general, temporal logics are a family of formalism used to express temporal properties of systems, e.g., an event should \emph{always} happen during a system execution; flagship temporal logics include \emph{linear temporal logic} (\emph{LTL})~\cite{pnueli1977temporal} and \emph{metric temporal logic} (\emph{MTL})~\cite{koymans1990specifying}.
Dokhanchi et al.~\cite{dokhanchi2018evaluating} propose an adaptation of temporal logic to express desired properties of perception; the new formalism is called \emph{Timed Quality Temporal Logic} (\emph{TQTL}). Specifically, TQTL can be used to express temporal properties that should be held by the perception module during object detection, e.g., ``\emph{whenever a lead car is detected at a frame, it should also be detected in the next frame}''. Conceptually, the properties expressed by TQTL are similar to the ones in~\cite{ramanagopal2018failing}; however, by adopting such a formal specification to express these properties, one can synthesize a monitor that automatically checks the satisfiability of the system execution. TQTL is later extended to \emph{Spatio-Temporal Quality Logic} (\emph{STQL})~\cite{balakrishnan2021percemon}, which has enriched syntax to express more refined properties over the bounding boxes used in object detection.  The authors also propose an online monitoring framework, named \emph{PerceMon}, for monitoring the perception module at runtime of the ADS.

\subsubsection{Test Adequacy} \label{subsec:perceptionAdequacy}
Measuring the adequacy of the testing for DNN models in the perception module is challenging, due to the complexity of DNN models. Compared to program execution, DNN inference involves a completely different logical process, which is deemed to be non-interpretable. In this domain, various metrics, analogous to the test adequacy criteria for programs, have been proposed; some of the metrics are for general DNN testing, while some are dedicated to ADS testing. Below, we introduce two typical lines of such adequacy criteria.

\myparagraph{Structural coverage} Neuron coverage is proposed in~\cite{pei2017deepxplore}, inspired by the structural coverage used in traditional software testing. Pei et al.~\cite{pei2017deepxplore} analogize DNN inference to program execution, and consider the \emph{neuron activation} as a symbol that indicates whether a neuron is ``covered''. Based on this analogy, they define \emph{neuron coverage} by what percentage of the neurons that are activated, as the counterpart of structural coverage in DNN. Inspired by~\cite{pei2017deepxplore}, a number of other neuron coverage criteria are proposed. For instance, \emph{k-multisection neuron coverage}~\cite{ma2018deepgauge1} is the refined version of neuron coverage that considers not only ``activated'' neurons but also ``not activated'' neurons; \emph{surprise adequacy}~\cite{kim2019guiding} pursues the novelty of an individual test case based on whether it is out of the distribution of the training data. 

\myparagraph{Combinatorial coverage}
Combinatorial testing~\cite{nie2011survey} utilizes \emph{combinatorial coverage} for test case generation, which measures the coverage of the combinations of different system parameters. The $t$-way combination coverage is a typical criterion, which is defined by the number of the $t$-wise combinations covered by the test suite, out of the total number of possible $t$-wise combinations. For instance, consider a system that has 3 binary parameters $a, b$ and $c$. Given a test suite $T=\{\langle0,0,1\rangle, \langle0,1,0\rangle, \langle1,0,0\rangle, \langle1,1,0\rangle\}$ that involves 4 test cases, the $2$-way combination of $T$ is computed by $\frac{1}{3}$, which indicates that one combination $ab$ is covered by $T$ (since $T$ involves all the possible cases $00,01,10,11$ of $ab$), over all the three possible combinations $ab, ac,$ and $bc$.

Combinatorial coverage has been used to solve the adequacy problem in the testing of the perception module. Gladisch et al.~\cite{gladisch2020leveraging} characterize the scenarios by using multiple parameters concerning different features, such as lane types and road types. 
They then apply combinatorial coverage as a guidance to generate test cases that can reveal system failures and achieve high coverage.
Cheng et al.~\cite{cheng2018quantitative} propose $k$-projection coverage that aims to reduce the combinatorial explosion during test case generation, by incorporating domain expertise.
Xia et al.~\cite{xia2018test} utilize the \emph{analytic hierarchy process} to identify the key factors and then generate test cases for a lane detection algorithm with combinatorial coverage guarantee.

\begin{table}[!tb]
\footnotesize
\caption{Summary of the papers for the perception module testing: Part \rom{1}}
\label{tab:perception}
\renewcommand{\arraystretch}{1.2}{
\begin{subtable}[t]{\textwidth}
\resizebox{\textwidth}{!}{
\begin{tabular}{p{0.12\textwidth}|p{0.34\textwidth}|p{0.1\textwidth}<{\centering}|p{0.34\textwidth}|p{0.1\textwidth}}
     \toprule
\textbf{Methodology} 
& \textbf{Description} 
& \textbf{Literature} 
& \textbf{Test Objective} 
& \textbf{Environment} \\
\hline 
\multirow{24}{*}{\parbox{0.12\textwidth}{Optimization-based attack}} 
& Replacing true traffic signs with generated adversarial traffic signs      
& \cite{chen2018shapeshifter}
& Object detector: Faster-RCNN\cite{ren2015faster}  
& Real world  \\
\cline{2-5}
& Generating transferable adversarial traffic signs and stickers
& \cite{zhao2019seeing}
& Object detectors: Faster-RCNN\cite{ren2015faster} and YOLOv3~\cite{redmon2018yolov3}                 
& Real world  \\
\cline{2-5}
& Generating camouflage on 3D objects
&\cite{zhang2019camou}
& Object detectors: Mask R-CNN~\cite{he2017mask} and YOLOv3-SPP~\cite{redmon2018yolov3}
& Simulation
\\
\cline{2-5}
& Focusing on the objectness loss
&\cite{im2022adversarial}
& Object detector: YOLOv4~\cite{bochkovskiy2020yolov4}
& Digital dataset
\\
\cline{2-5}

& Adding perturbations to the unnoticed area
&\cite{xu2020adversarial}
& Segmentation models: ResNet-101~\cite{he2016deep} and MobileNet~\cite{howard2017mobilenets}
& Digital dataset\\
\cline{2-5}
& Performing black-box attacks on traffic sign recognition
models
& \cite{li2020adaptive,kumar2020black}
& Models from the Kaggle Competition~\cite{Kaggle}
& Digital dataset
\\
\cline{2-5}
& Adding spoofed points into the original 3D point clouds
& \cite{cao2019adversarial}
& Perception module of \Apollo   
& Simulation  \\
\cline{2-5}
& Generating adversarial images against multi-sensor fusion based perception
&\cite{cao2021invisible}
& Perception module of \Apollo                             
& Simulation  \\
\cline{2-5}
& Performing the black-box attack by the occlusion information
&
\cite{sun2020towards, wang2021adversarial, chen2021camdar, zhu2021can}
& Perception module of \Apollo and LiDAR-based object detectors~\cite{shi2019pointrcnn, lang2019pointpillars,yang2018pixor}
& Digital dataset \\
\cline{2-5}

& Generating perturbations for roadside objects
& \cite{yang2021robust, zhu2021adversarial}
& LiDAR-based object detectors~\cite{shi2019pointrcnn, lang2019pointpillars, shi2020pv} and a segmentation model~\cite{qi2017pointnet}
& Real world
\\
\cline{2-5}
& Adding perturbations to the trajectories of vehicles 
& \cite{li2021fooling}
& Object detectors: PointRCNN~\cite{shi2019pointrcnn} and PointPillar++~\cite{hu2020you}
& Digital dataset
\\
\cline{2-5}
& Pasting generated adversarial stickers on traffic signs     
& \cite{eykholt2018robust, song2018physical}
& Classifiers: LISA-CNN~\cite{mogelmose2012vision}, GTSRB-CNN~\cite{Stallkamp2012} and Inception-v3~\cite{krizhevsky2012imagenet}          
& Digital dataset \\
\hline
\multirow{6}{*}{\parbox{0.12\textwidth}{GAN-based attack}} 
& Generating adversarial patches with high visual fidelity  
& \cite{liu2019perceptual}
& Classifiers: VGG16~\cite{simonyan2014very}, ResNet~\cite{he2016deep} and VY~\cite{VY}                      
& Digital dataset  \\
\cline{2-5}
& Proposing a multi-source attack method 
& \cite{xiong2021multi}
&Semantic segmentation model: VAE-GAN~\cite{larsen2016autoencoding} 
& Digital dataset \\
\cline{2-5}
& Synthesizing corner cases by utilizing \emph{CycleGANs}
& \cite{yu2018intelligent}
& Object detector: PatchGAN~\cite{johnson2016perceptual}
& Digital dataset
\\
\hline
\multirow{4}{*}{Trojan attack}
& Utilizing particle swarm optimization
&\cite{jiang2020poisoning}
&Classifier: LeNet-5~\cite{lecun1998gradient}
&Digital dataset
\\
\cline{2-5}
& Performing the Trojan attack for models used for raindrops removal
& \cite{ding2019trojan}
& Claffifiers: DeRaindrop Net~\cite{qian2018attentive} and RCAN~\cite{zhang2018image}
& Digital dataset
\\
 \bottomrule
\end{tabular}
}
\end{subtable}}
\end{table}

\begin{table}[!tb]\ContinuedFloat
\caption{Summary of the papers for the perception module testing: Part \rom{2}}
\footnotesize
\renewcommand{\arraystretch}{1.2}{
\begin{subtable}[t]{\textwidth}
\resizebox{\textwidth}{!}{
\begin{tabular}{p{0.21\textwidth}|p{0.69\textwidth}|p{0.1\textwidth}<{\centering}}
\toprule
\textbf{Oracle} 
& \textbf{Description} 
& \textbf{Literature} \\
\toprule
\multirow{2}{*}{Ground-truth labeling}
& Generating road label automatically by LiDAR
& \cite{zhou2019automated}
\\
\cline{2-3}
&- Generating dimension and classification references for object detection
& \cite{philipp2021automated}
\\
\hline
\multirow{6}{*}{Metamorphic testing}
& The detected object in the original images should also be detected in the synthetic images
& \cite{shao2021testing}
\\
\cline{2-3}

& The recognition results of target models should change when traffic lights change
& \cite{bai2021metamorphic}
\\
\cline{2-3}
& The detection of objects should be same with the affect of noise
& \cite{zhou2019metamorphic}
\\
\cline{2-3}
& Check the model inconsistencies between the original images and the mutated images
& \cite{woodlief2022semantic,wang2021object}
\\
\cline{2-3}
& The object detected in a previous frame should also be detected in a later frame 
& \cite{ramanagopal2018failing}
\\

\hline
\multirow{2}{*}{Formal specifications}
& Adapting TQTL to express desired properties of perception
& \cite{dokhanchi2018evaluating,ramanagopal2018failing}
\\
\cline{2-3}
& Adapting STQL to express more refined properties
& \cite{balakrishnan2021percemon}\\
\bottomrule
\end{tabular}}
\end{subtable}
 \hfill
\begin{subtable}{\textwidth}
\resizebox{\textwidth}{!}{
\begin{tabular}{p{0.21\textwidth}|p{0.69\textwidth}|p{0.1\textwidth}<{\centering}}
\toprule
\textbf{Adequacy} 
& \textbf{Description} 
& \textbf{Literature} \\
\toprule
\multirow{5}{*}{Structural coverage}
& Neuron coverage: the percentage of the neurons that are activated
& \cite{pei2017deepxplore}
\\
\cline{2-3}
& K-multisection neuron coverage: considering activated neurons and not activated neurons
& \cite{ma2018deepgauge1}
\\
\cline{2-3}
& Surprise adequacy: the novelty of a test case based on its range in the training data distribution
& \cite{kim2019guiding}
\\
\hline
\multirow{3}{*}{Combinatorial coverage}
& Characterizing the scenarios by multiple parameters 
& \cite{gladisch2020leveraging}
\\
\cline{2-3}
& Incorporating domain expertise to reduce the combinatorial explosion
& \cite{cheng2018quantitative}
\\
\cline{2-3}
& Utilizing the analytic hierarchy process to identify the key factors
& \cite{xia2018test}
\\
\bottomrule
\end{tabular}}
\end{subtable}
}
\end{table}

\subsubsection{Discussion} 
\label{subsubsec:perceptionDiscussion}
Table~\ref{tab:perception} summarizes the collected papers for testing the perception module. This module includes a number of DNN models for understanding the environmental information. It is important, since many crashes are caused due to the vulnerabilities of this module~\cite{garcia2020comprehensive}.
It can be seen that a large number of papers perform adversarial attacks on this module. These methods include three categories, namely, optimization-based attack, GAN-based attack, and Trojan attack. The first two types of methods could generate \emph{adversarial examples} to fool the DNN models, while the third type of method targets the training process of the models.
The DNN models take charge of various tasks of perception, including object classification~\cite{eykholt2018robust, liu2019perceptual}, semantic segmentation~\cite{xu2020adversarial, zhu2021adversarial}, and camera-/LiDAR-based object detection~\cite{chen2018shapeshifter, zhao2019seeing,zhang2019camou,im2022adversarial,xu2020adversarial,li2020adaptive,kumar2020black, cao2019adversarial,  cao2021invisible, sun2020towards,wang2021adversarial, chen2021camdar, zhu2021can,yang2021robust,li2021fooling}. Since the generated adversarial perturbations may not be effective in a noisy physical environment~\cite{lu2017no}, a number of methods (e.g., \emph{Robust Physical Perturbations}~\cite{eykholt2018robust}) are proposed to overcome this challenge. 

As mentioned in~\S{}\ref{subsec:perceptionOracle}, it is also a great challenge to judge the correctness of the output of the perception module. We collect three types of approaches, including ground-truth labeling~\cite{zhou2019automated, philipp2021automated}, metamorphic testing~\cite{shao2021testing,bai2021metamorphic,zhou2019metamorphic,woodlief2022semantic,wang2021object,ramanagopal2018failing}, and formal specifications~\cite{dokhanchi2018evaluating,balakrishnan2021percemon} for tackling this problem. In summary, the first approach focuses on automatically generating ground-truth labels for single images, while the other two methods tend to express the properties between continuous frames and are thus suitable for evaluating the perception module at runtime.

Traditional coverage metrics, e.g., code coverage, are typically not suitable for estimating the test adequacy of DNN-based models. Structural coverage metrics like neuron coverage~\cite{pei2017deepxplore} have become a mainstream substitute. 
Recently, there is a different  voice~\cite{yan2020correlations, harel2020neuron, ma2021test} saying that neuron coverage and its extensions may lack effectiveness in guiding ML testing.
In addition to neuron coverage, combinatorial testing~\cite{gladisch2020leveraging,cheng2018quantitative,xia2018test} is another approach for tackling the test adequacy problem of the perception module. 

\smallskip
\setlength{\fboxsep}{8pt}
\begin{center}
\Ovalbox{
\begin{minipage}{0.9\textwidth}
\small \textbf{Summary:}
A large number of papers perform adversarial attacks on the perception module, covering various perception tasks, e.g., camera-/LiDAR-based object detection and semantic segmentation. Testing oracle problem of this module has been studied through different approaches, e.g., metamorphic testing. Structural coverage metrics (e.g., neuron coverage) and combinatorial testing techniques are widely adopted for the guarantee of testing adequacy.
\end{minipage}
}
\end{center}
\smallskip

\subsection{Planning Module} \label{sec:planning}
The planning module takes the information from the perception module as input and produces a suitable driving trajectory as a reference for the control module to make decisions. In the planning module, we introduce the studies on test methodology (shown in~\S{}\ref{subsec:planningMethodology}), test oracle (shown in~\S{}\ref{subsec:planningOracle}) and test adequacy (shown in~\S{}\ref{subsec:planningAdequacy}). 

\subsubsection{Test Methodology}\label{subsec:planningMethodology} 
Testing of the planning module consists in providing traffic scenarios for an ADS, and checking if the planner module generates trajectories that satisfy properties such as safety, comfort, and low cost. Note that, the path planning module is usually integrated into the whole ADS, and highly coherent with other modules: the input of the module comes from the perception module, and the output trajectory is a reference for the control module, rather than the actual one observable from the system. Therefore, testing the planning module independently is a challenging task. 

Due to the above reasons, there exist no large numbers of studies on the testing dedicated to the planning module. The studies we collected are based either on dedicated path planning systems~\cite{laurent2019mutation,laurent2020achieving,arcaini2021targeting}, or on the assumption of the perfection of the perception and control modules~\cite{tang2021route}. 
In summary, \emph{search-based testing} is the major technique for testing of the planning module, and it is adopted in most of the works~\cite{laurent2019mutation,laurent2020achieving,arcaini2021targeting,arcaini2022less,althoff2018automatic,klischat2019generating,bak2022stress,kahn2022know} for this module.

\myparagraph{Search-based testing} 
According to~\cite{menzel2018scenarios}, scenarios are defined on three abstraction levels, namely, \emph{functional scenarios}, \emph{logical scenarios}, and \emph{concrete scenarios}. A functional scenario has the highest abstraction level and defines only the basic conditions and participants of a scenario; on top of a functional scenario, a logical scenario is defined by a set of parameters and their ranges; with the parameter values fixed in a logical scenario, a concrete scenario is generated. 
In the context of scenario generation, search-based testing usually consists in searching in the parameter space of a logical scenario for a concrete scenario, with specific objectives. Below are some examples of applying search-based testing to generate concrete scenarios for the testing of the planning module. 

\smallskip
The works~\cite{laurent2019mutation,laurent2020achieving, arcaini2021targeting,arcaini2022less} use a dedicated path planning system from their industry collaborator, which computes the trajectories of the ADS based on several constraints, e.g., safety and traffic regulations. The aggressiveness of the path planning strategy is decided by a system parameter, named \emph{weight}. 
Laurent et al.~\cite{laurent2019mutation} define a coverage criterion named \emph{weight coverage}, which is used to characterize the testing adequacy of the weight parameter. Later in~\cite{laurent2020achieving}, they propose two search-based techniques, named \emph{single-weight approach} and \emph{multi-weight approach}, that automatically generate testing scenarios guided by weight coverage. Specifically, the single-weight approach searches for the scenarios that cover one specific weight of the path planner, while the multi-weight approach generates scenarios that cover different weights simultaneously using the multi-objective search.
Arcaini et al.~\cite{arcaini2021targeting}  consider searching for the \emph{driving patterns} that are identified by the features appearing in the planned trajectory, such as longitudinal/lateral acceleration and curvature. The driving patterns that take place in a trajectory for a considerable duration are relevant to the characteristics of the path planner, and thus facilitate engineers in system assessment.
Since the testing scenarios in their previous works contain numerous irrelevant elements and are thus hard to debug, in their latest work~\cite{arcaini2022less}, they target the simplification of testing scenarios---they remove all the irrelevant traffic participants, but the failures can still be triggered.

\smallskip
Althoff et al.~\cite{althoff2018automatic} propose the notion of \emph{drivable area} for motion planning algorithms, which represents a safe solution space in which the ADS can avoid collision. Then they adopt a search method to generate scenarios that are highly critical in the sense that the drivable area is limited.
In their follow-up work~\cite{klischat2019generating}, the authors consider the interference of other traffic participants in the drivable area, in order to increase the complexity of the scenarios. In their experiment, the evolutionary algorithms~\cite{back1993overview} are demonstrated to be advantageous in finding a local optimum over these complex and diverse scenarios.
Bak et al.~\cite{bak2022stress} apply \emph{random exploring trees} (\emph{RRT}) to search for the \emph{adversarial agent perturbations}, which indicate that the behaviors of other vehicles are only  modified slightly. 
Kahn et al.~\cite{kahn2022know} generate occlusion scenarios for testing the behavior planning module of an ADS. To be specific, they apply an occlusion-guided search method to inject vehicles into the scenarios extracted from naturalistic data. Experimental results show that the number of occlusion-caused collisions generated by their approach is 40 times higher than that from the naturalistic data.

\subsubsection{Test Oracle}\label{subsec:planningOracle}
Assessing the correctness of the output of the planning module, i.e., a planned trajectory, is a challenging problem, due to the lack of an oracle that represents the ``correct'' trajectory. In this section, we introduce the studies~\cite{ calo2020generating, calo2020simultaneously} that define different metrics
as the oracles to evaluate the correct functionality of the planning module. 

Cal{\`o} et al.~\cite{calo2020generating, calo2020simultaneously} define the notion of \emph{avoidable collisions}, to distinguish them from \emph{unavoidable collisions}, in a dedicated path planner. By their definition, a collision is \emph{avoidable} if it can be avoided from happening in the same scenario by using a different system configuration of the ADS. Compared to the unavoidable ones, the avoidable collisions are considered critical, since these collisions require system reengineering to rectify the unsafe behavior.

\subsubsection{Test Adequacy}\label{subsec:planningAdequacy}
As mentioned in~\S{}\ref{subsec:planningMethodology}, the inputs to the planning module involve both external parameters that identify a scenario and internal parameters of the ADS. Because there are infinitely many possible combinations of these parameters, generating test cases that are sufficiently diverse remains a great challenge. In this section, we collect the studies~\cite{laurent2019mutation, tang2021route} that propose coverage measurements on the space of the parameters. Namely, the \emph{weight coverage} criterion~\cite{laurent2019mutation} refers to the coverage of  the possible configurations of the path planner under test; the \emph{route coverage} criterion~\cite{tang2021route} is proposed to measure whether different features of a map have been explored by the test suite.

Laurent et al.~\cite{laurent2019mutation} propose a coverage criterion, named \emph{weight coverage}, to test a dedicated path planning system. In their path planner, there is a \emph{weight} function that consists of six weight parameters, which affect the path planning decisions from different aspects, such as safety and comfort. In order to cover diverse planning decisions made by the system,  the authors use the weight coverage to guide the exploration of the weight parameter space. Thereby, they manage to generate scenarios that cover more diverse combinations of the weight parameters.

Tang et al.~\cite{tang2021route} propose another coverage criterion called \emph{route coverage} for testing the route planning functionality of \Apollo. Based on a \emph{Petri net} abstracted from the map, they quantify the route diversity based on the \emph{junction topology feature} and \emph{route feature}. The junction topology feature describes the relative position and connection relationship of the roads at a junction, while the route feature describes the action of \Apollo to track a selected road.
By mutating the test cases, they achieve a high route coverage ratio and thus obtain a diverse test suite that covers various features of the map.

\begin{table}[!tb]
\caption{Summary of the papers for the planning module testing}
\label{tab:planning}
\footnotesize
\renewcommand{\arraystretch}{1.2}
\begin{subtable}[t]{\textwidth}
\resizebox{\textwidth}{!}{
\begin{tabular}{p{0.12\textwidth}|p{0.38\textwidth}|p{0.1\textwidth}<{\centering}|p{0.25\textwidth}|p{0.15\textwidth}}
\toprule
\textbf{Methodology} 
& \textbf{Description} 
& \textbf{Literature} 
& \textbf{Test Objective} 
& \textbf{Environment} \\
\hline 
\multirow{9}{*}{\parbox{0.12\textwidth}{Search-based testing}}
& Searching for testing scenarios with weight coverage guarantee
&\cite{laurent2019mutation,laurent2020achieving}
& \multirow{3}{*}{\parbox{0.25\textwidth}{A dedicated path planner}}
&\multirow{2}{*}{Simulation} \\ 
\cline{2-3}
& Searching for the specific driving patterns
&\cite{arcaini2021targeting,arcaini2022less}
&
&\\
\cline{2-5}
& Searching for the scenarios in which the drivable area is limited 
&\cite{althoff2018automatic,klischat2019generating}
& Motion planners
& Digital dataset\\
\cline{2-5}
& Applying RRT to search for adversarial agent perturbations
&\cite{bak2022stress}
& Five path planners, e.g., Frenet Planner~\cite{werling2010optimal}
& Simulation
\\
\cline{2-5}
& Injecting vehicles into the scenarios extracted from naturalistic data
&\cite{kahn2022know}
& Strategic planners
& Digital dataset
\\
\bottomrule
\end{tabular}
}
\end{subtable}
\hfill
\begin{subtable}[t]{\textwidth}
\resizebox{\textwidth}{!}{
\begin{tabular}{p{0.2\textwidth}|p{0.6\textwidth}|p{0.2\textwidth}<{\centering}}
\toprule
\textbf{Oracle} 
& \textbf{Description} 
& \textbf{Literature} 
\\
\hline 
Avoidable collisions
& A collision can be avoided from happening in the same scenario by using a different system configuration of the ADS.
& \cite{calo2020generating, calo2020simultaneously}
\\
\bottomrule
\end{tabular}
}
\end{subtable}
\hfill
\begin{subtable}[t]{\textwidth}
\resizebox{\textwidth}{!}{
\begin{tabular}{p{0.2\textwidth}|p{0.6\textwidth}|p{0.2\textwidth}<{\centering}}
\toprule
\textbf{Adequacy} 
& \textbf{Description} 
& \textbf{Literature} 
\\
\hline 
Weight coverage
& Based on a weight function consisting of weight parameters which affect planning decisions
&\cite{laurent2019mutation}
\\
\hline
Route coverage
& Based on the junction topology feature and route feature
& \cite{tang2021route}
\\
\bottomrule
\end{tabular}
}
\end{subtable}
\end{table}

\subsubsection{Discussion}
\label{subsubsec:planningDiscussion}
The summary of the collected papers for testing the planning module is shown in Table~\ref{tab:planning}. We find that search-based testing is a dominant technique that has been demonstrated to be effective in revealing faults in the planning module~\cite{laurent2019mutation,laurent2020achieving, arcaini2021targeting,arcaini2022less, klischat2019generating, althoff2018automatic,bak2022stress,kahn2022know}.
In addition, several metrics are proposed for facilitating the testing on the planning module, e.g., \emph{avoidable collision}~\cite{calo2020generating} for tackling the oracle problem, \emph{weight coverage}~\cite{laurent2019mutation} and \emph{route coverage}~\cite{tang2021route} for evaluating the sufficiency of the test suite. 
However, most of these metrics are dedicated to specific path planning systems, and it needs to be further explored whether they could be generalized to the planning modules of other systems.

\smallskip
\setlength{\fboxsep}{8pt}
\begin{center}
\Ovalbox{
\begin{minipage}{0.9\textwidth}
\small \textbf{Summary:}
Search-based testing is an effective technique for revealing faults in the planning module. Several metrics have been proposed for testing particular path planning systems, and it needs further exploration on how to generalize these metrics to other systems.
\end{minipage}
}
\end{center}
\smallskip

\subsection{Control Module} \label{sec:control}
Based on the trajectories produced by the planning module, the control module takes charge of the lateral and longitudinal control of the ADS. By using various control algorithms, such as \emph{model predictive control} (\emph{MPC}) and \emph{proportional integral derivative} (\emph{PID}) \emph{control}, it generates control signals, e.g., acceleration, deceleration, and steering angle, to the CAN bus for the control of the whole system. In this module, we introduce the works on the testing of the control module, from the perspectives of test methodology (shown in~\S{}\ref{subsec:controlMethodology}) and test oracle (shown in~\S{}\ref{subsec:controlOracle}).

\subsubsection{Test Methodology} \label{subsec:controlMethodology}
The control module takes charge of multiple functionalities, such as the longitudinal control and  the lateral control of the ADS. Hence, the testing of this module focuses on detecting vulnerabilities in the control mechanisms.

\myparagraph{Fault injection}
Fault injection is a method that deliberately introduces faults into a system, in order to assess the fault tolerance of the system.
Uriagereka et al.~\cite{uriagereka2017fault} adopt this technique for testing the fault tolerance ability of the control module of an ADS. Specifically, they inject faulty GPS signals into the lateral control function of the ADS, which makes it produce wrong steering commands. By calculating the \emph{fault tolerant time interval}, which denotes the duration from the activation of the fault to the occurrence of unsafe behavior, they find the lateral control system can tolerate this type of fault for as long as $177ms$.
Zhou et al.~\cite{zhou2022strategic} inject faulty control signals through \emph{CAN} bus to cause collisions without being detected by ADS safety mechanisms, e.g., \emph{forward collision warning}. They evaluate the method with \OpenPilot and find that the lateral control of the system is the typically vulnerable part with a high attack success rate.

\myparagraph{Sampling}
We refer to \emph{sampling} as the statistical method that samples values from a probability distribution.
Wang et al.~\cite{wang2020behavioral} sample the relevant parameters, e.g., speeds of the \emph{non-player characters} (\emph{NPCs}) and the ego vehicle, for generating scenarios of different challenge levels. The method is evaluated on the unprotected left-turn scenario and experimental results demonstrate the robustness of the \emph{MPC} controller.
In their later work~\cite{wang2021interaction}, \emph{game theory} is applied to characterize the interactive behaviors of NPCs in highway merging scenarios.

\myparagraph{Falsification} Temporal logic-based falsification~\cite{zhang2018two, zhang2021effective, annpureddy2011s, donze2010breach, zhang2022falsifai, song2022cyber, ErnstABCDFFG0KM21} is applied to ADS testing in ~\cite{tuncali2016utilizing, tuncali2019rapidly}. Originally, falsification refers to a technique for testing of the general \emph{cyber-physical systems}, guided by the quantitative semantics of temporal logic specifications, which indicates how far is the system from being unsafe.
Tuncali et al.~\cite{tuncali2016utilizing} propose a falsification-based automatic test generation framework for testing collision avoidance controllers. They utilize a cost function, i.e., the quantitative semantics of the temporal logic specification, as a guidance in searching for the critical scenarios in which the relative speed of the two vehicles in the collision is minimal. The obtained  scenarios can be taken as the behavioral boundary that divides the safe and unsafe behaviors.
As a follow-up work, Tuncali et al.~\cite{tuncali2019rapidly} utilize the \emph{Rapidly-exploring Random Trees} (\emph{RRT}) algorithm  for ADS falsification. They incorporate a new cost function that applies \emph{time-to-collision} to measure the seriousness of the collision. As a result, the new method achieves better effectiveness in searching for safety-critical scenarios, thanks to the exploration brought by the RRT algorithm.

\subsubsection{Test oracle}\label{subsec:controlOracle}
Like the planning module, the control module also faces the oracle problem in its testing---indeed, it is usually not straightforward to determine whether a control decision is ``correct''. 
In~\cite{djoudi2020simulation}, Djoudi et al. propose a framework to determine whether the control module makes ``the correct decision''. They design a model to generate an oracle area in the given scenario, which is the closest safe position ahead of the vehicle. A control decision is then considered as ``the correct decision'', if it could drive the vehicle close to the oracle area.

\begin{table}[!tb]
\footnotesize
\caption{Summary of the papers for the control module testing}
\label{tab:control}
\renewcommand{\arraystretch}{1.2}
\begin{subtable}[t]{\textwidth}
\resizebox{\textwidth}{!}{
\begin{tabular}{p{0.16\textwidth}|p{0.32\textwidth}|p{0.1\textwidth}<{\centering}|p{0.3\textwidth}|p{0.12\textwidth}}
\toprule
\textbf{Methodology}   
&\textbf{Description}
&\textbf{Literature}
&\textbf{Test Objective}      
& \textbf{Environment} \\ 
\hline
\multirow{4}{*}{Fault injection} & Injecting faulty GPS signals 
&\cite{uriagereka2017fault}     
& Lateral control module of an urban vehicle
& Simulation          \\ 
\cline{2-5}
& Injecting faulty control signals through the CAN bus
&\cite{zhou2022strategic}
& The control module of \OpenPilot 
& Simulation
\\
\hline
Sampling
& Sampling the relevant parameters for generating different challenge level scenarios
&\cite{wang2020behavioral,wang2021interaction}
&
MPC controller~\cite{mpc}
&
Simulation
\\
\hline
Falsification & Utilizing temporal logic-based falsification to search for critical scenarios  
& \cite{tuncali2016utilizing,tuncali2019rapidly}      
& Collision avoidance controller
& Simulation          \\ 
\bottomrule
\end{tabular}
}
\end{subtable}
\hfill
\begin{subtable}[t]{\textwidth}
\resizebox{\textwidth}{!}{
\begin{tabular}{p{0.3\textwidth}|p{0.6\textwidth}|p{0.1\textwidth}<{\centering}}
\toprule
\textbf{Oracle} 
& \textbf{Description} 
& \textbf{Literature} 
\\
\hline 
Optimization-based oracle model
& The model can generate an oracle area in a given scenario
& \cite{djoudi2020simulation}
\\
\bottomrule
\end{tabular}
}
\end{subtable}
\end{table}

\subsubsection{Discussion}
\label{subsubsec:controlDiscussion}
Table~\ref{tab:control} summarizes the collected papers for the testing of the control module, where the number of studies is not too large. Note that currently most of the control modules of ADS adopt mature control techniques directly, such as \emph{PID}~\cite{pid} and \emph{MPC}~\cite{mpc}, which partially explains why this module is not extensively studied.
The collected studies adopt three major techniques for testing the control module, including fault injection~\cite{uriagereka2017fault,zhou2022strategic}, sampling~\cite{wang2020behavioral, wang2021interaction}, and falsification~\cite{tuncali2016utilizing, tuncali2019rapidly}.
To tackle the oracle problem in control module testing, the framework proposed in~\cite{djoudi2020simulation} could generate an oracle area for judging whether the control decision is ``correct''.
However, we do not find much work that handles the test adequacy problem for this module. In general, since the control module deals with continuous dynamics, it is challenging and thus requires further exploration to define adequacy criteria for test cases in the future.

\smallskip
\setlength{\fboxsep}{8pt}
\begin{center}
\Ovalbox{
\begin{minipage}{0.9\textwidth}
\small \textbf{Summary:}
Since most of the control modules of the ADS adopt those mature control techniques, e.g., \emph{MPC} and \emph{PID}, there are not many works studying the testing of the control module. Existing  techniques mainly include fault injection, sampling, and falsification. There are some works that study the oracle problem of control modules. There is few work that studies the adequacy criteria for testing control modules.
\end{minipage}
}
\end{center}
\smallskip
 
\subsection{End-to-End Module} \label{sec:end2end}
The end-to-end (e2e) module is a special design adopted by many modern ADS, which integrates the functionalities of perception, planning and control in a single DNN-based model. 
The DNN model is often developed by \emph{supervised learning}, which is trained by using a training dataset consisting of realistic driving data. Each element of the dataset is a pair $\langle I, c\rangle$, which maps the information $I$ at the end of the sensor to a label $c$ that indicates the desired control decision at the end of the controller. After training, a model can infer control decisions based on the driving environment at runtime in order to drive the ADS properly.
For instance, in some modern ADS that perform steering angle control, the end-to-end DNN model takes as input the sensing information including road conditions and the status of other cars, and outputs a series of predicted steering angles for controlling the ADS. In this section, we introduce the collected studies on the testing of the end-to-end module from the perspectives of test methodology (shown in~\S{}\ref{subsec:e2eMethodology}), test oracle (shown in~\S{}\ref{subsec:e2eOracle}), and test adequacy (shown in~\S{}\ref{subsec:e2eAdequacy}).

\subsubsection{Test Methodology}\label{subsec:e2eMethodology}
As mentioned before, an end-to-end DNN model integrates three functionalities, namely perception, planning and control, in a single module. Among these three functionalities, perception is the most vital part as it provides input information to other modules; meanwhile, it is also the most vulnerable to external environments, as it essentially involves image recognition tasks that rely on deep learning. Compared to a DNN just for perception, although an end-to-end DNN does not directly output the perception information, the control decisions it makes still depend on the perception information. Therefore, like the case in the perception module, generating adversarial images or scenarios that fool the end-to-end DNN is still the major testing methodology for testing the end-to-end modules.

We introduce three approaches, namely, search-based testing, optimization-based adversarial attack, and GAN-based attack. The first approach has been introduced in~\S{}\ref{subsec:planningMethodology}; the last two approaches have been introduced  in~\S{}\ref{subsec:perceptionMethodology}.

\myparagraph{Search-based testing}
Search-based testing searches for a target test case in the input space, guided by certain objectives. One commonly used objective is the coverage of the test suite---maximizing the cumulative coverage of a test suite can expose more diverse behavior of the system, and thus allow a better chance of detecting the target test case.
In the context of DNN testing, neuron coverage is proposed by Pei et al.~\cite{pei2017deepxplore} to analogize the structural coverage in traditional programs. In their follow-up work, Tian et al.~\cite{tian2018deeptest} propose a coverage-guided testing framework called \emph{DeepTest} for DNN testing. They propose various image operations, e.g., scaling, shearing, and rotating, as the test input (image) mutation methods; then they generate test cases by applying these operations to seed images, and keep only those mutants that enlarge the cumulative neuron coverage of a test suite. Experiments are conducted on three end-to-end models, and the results show the effectiveness of their method in test case generation. 

In addition to coverage, the seriousness of the unsafe behavior is another factor that can be used as the search objective, and this has been considered by Li et al.~\cite{li2021testing}.
In their work, the seriousness of the unsafe behavior of the end-to-end module is formulated as the deviation of the actual steering angle made in the test scenario from the expected steering angle. The authors design an objective function that takes into account both the coverage and the seriousness, such that they can detect not only diverse but also serious unsafe test cases.

\myparagraph{Optimization-based attack}
The optimization-based adversarial attacking framework has been introduced in~\S{}\ref{subsec:perceptionMethodology}. 
Zhou et al.~\cite{zhou2020deepbillboard} introduce a framework called \emph{DeepBillboard} that can generate adversarial perturbations which are added to billboard. The perturbations they generate can mislead the steering angles in a series of frames captured by camera sensors during the driving process, in spite of the physical conditions, such as different distances and different angles to the billboard.
Later Pavlitskaya et al.~\cite{pavlitskaya2020feasibility} extend \emph{DeepBillboard} with the \emph{projected gradient sign} method~\cite{madry2017towards}, and experimental results show that the curved and rainy scenes are more vulnerable to these adversarial attacks.
In another line of work, adversarial black lines are utilized to attack the end-to-end driving models~\cite{boloor2019simple, boloor2020attacking}. These black lines are easy to paint on the public road and can lead to a deviation of an ADS from the original path.

\myparagraph{GAN-based attack}
GAN has been introduced in~\S{}\ref{subsec:perceptionMethodology}, and it has been considered as a major approach for adversarial attacking.
Kong et al.~\cite{kong2020physgan} propose a GAN-based approach called \emph{PhysGAN} which utilizes 3D tensors, i.e., a slice of video containing hundreds of frames, to generate adversarial roadside signs that can continuously mislead the end-to-end driving models with  high efficacy and robustness.
In another work, to generate realistic adversarial images, 
Zhang et al.~\cite{zhang2018deeproad} propose a GAN-based approach called \emph{DeepRoad}. They demonstrate that their generated adversarial images are realistic under various weather conditions, and effective in detecting unsafe system behaviors.

\subsubsection{Test Oracle}\label{subsec:e2eOracle}
An oracle of the end-to-end module indicates which is the correct control decision at each moment of a scenario. Although this can be done with the help of human drivers, it is too expensive and prone to errors. Existing works propose various automatic methods to solve the oracle problem of the end-to-end module, including metamorphic testing~\cite{tian2018deeptest, zhang2018deeproad,pan2021metamorphic}, differential testing~\cite{pei2017deepxplore}, and model-based oracle~\cite{stocco2020misbehaviour,hussain2022deepguard}.

\myparagraph{Metamorphic testing}
As introduced in~\S{}\ref{subsec:perceptionOracle}, metamorphic testing is a viable way to solve the oracle problem. In the testing of end-to-end models, there are a few works that leverage metamorphic relations to define the test oracles, e.g., \emph{DeepTest}~\cite{tian2018deeptest} and \emph{DeepRoad}~\cite{zhang2018deeproad}.
The metamorphic relation introduced by DeepTest~\cite{tian2018deeptest} is that, the steering angle should not change significantly for the same scenes under different weather and lighting conditions.
Similarly, DeepRoad~\cite{zhang2018deeproad} aims to detect \emph{model consistency}, which means, for a synthetic image and the original image, the difference between two predicted steering angles is smaller than a threshold.
Pan et al.~\cite{pan2021metamorphic} introduce a metamorphic relation for testing end-to-end models in a foggy environment; the relation requires that the density and direction of fog not affect the output steering angle of the target models.

\myparagraph{Differential testing}
Pei et al.~\cite{pei2017deepxplore} apply differential testing to generate scenarios which reveal the inconsistencies between different DNN models. For the same scenario, they expect that the DNNs under test should give the same inference result. The violation of this property is considered as an unexpected behavior.

\myparagraph{Model-based oracle}
Stocco et al.~\cite{stocco2020misbehaviour} propose a so-called \emph{self-assessment oracle} for the potential risk prediction of ADS. The self-assessment oracle involves training a probabilistic model that characterizes the distribution of the potential risks under various real scenarios. This model can be used to monitor the real environment during the execution of the ADS and predict situations that are probably not  handled by the ADS. 
This novel idea is also studied by Hussain et al.~\cite{hussain2022deepguard}.

\subsubsection{Test Adequacy}\label{subsec:e2eAdequacy}
Combinatorial coverage is also adopted in end-to-end module testing, e.g., the 2-way combinatorial testing based on image transformations~\cite{chandrasekaran2021combinatorial}.
In~\S{}\ref{subsec:perceptionAdequacy}, we introduce the structural coverage for DNN testing, which analogizes the structural coverage in traditional program testing. Since the end-to-end module also relies on DNN models, these structural coverage criteria are also used in the testing of the end-to-end module. 
Neuron coverage, which has been introduced in~\S{}\ref{subsec:perceptionAdequacy}, is used by its authors for a coverage-guided testing~\cite{tian2018deeptest}, as mentioned in~\S{}\ref{subsec:e2eMethodology}. The refined structural coverage criteria for DNNs, such as \emph{k-multisection neuron coverage} (\emph{KMNC}) and \emph{neuron boundary coverage} (\emph{NBC}) ~\cite{ma2018deepgauge1,li2021testing}, which is also elaborated on in~\S{}\ref{subsec:e2eMethodology}.

\begin{table}[!tb]
\footnotesize
\caption{Summary of the papers for the end-to-end module testing}
\renewcommand{\arraystretch}{1.2}
\begin{subtable}[t]{\textwidth}
\resizebox{\textwidth}{!}{
\begin{tabular}{p{0.12\textwidth}|p{0.38\textwidth}|p{0.08\textwidth}<{\centering}|p{0.32\textwidth}|p{0.1\textwidth}}
\toprule
\textbf{Methodology} 
& \textbf{Description}  
& \textbf{Literature}   
& \textbf{Test Objective} 
& \textbf{Environment}\\
\hline
\multirow{4}{*}{\parbox{0.12\textwidth}{Search-based testing}}
& Generating transformed images with high neuron coverage
& \cite{tian2018deeptest}           
& Three DNN models: Rambo~\cite{rambo}, Chauffeur~\cite{chauffeur} and Epoch~\cite{epoch}
& Digital dataset \\
\cline{2-5}
& Designing an objective function to search for the diverse and serious unsafe test cases
& \cite{li2021testing}
& Three DNN models: Dave-1~\cite{0bserver07}, Dave-3~\cite{Dave-3} and Chauffeur
& Digital dataset
\\
\hline
\multirow{6}{*}{\parbox{0.12\textwidth}{Optimization-based attack}}
& Replacing the original billboard with an adversarial billboard 
& \cite{zhou2020deepbillboard}     
& Four DNN models: Dave-1, Dave-2~\cite{Dave-2}, Dave-3 and Epoch
& Digital dataset \\
\cline{2-5}
& Extending attack methods in~\cite{zhou2020deepbillboard} to generate adversarial patches
&\cite{pavlitskaya2020feasibility}
& An end-to-end driving model called DriveNet~\cite{hubschneider2017adding}
&Simulation \\
\cline{2-5}
&Generating adversarial black lines on the road
&\cite{boloor2019simple,boloor2020attacking}
& Two end-to-end driving models in \carla
& Simulation \\
\hline
\multirow{4}{*}{\parbox{0.12\textwidth}{GAN-based attack}}
& Generating adversarial roadside sign
& \cite{kong2020physgan}
& Three DNN models: Dave-2, Epoch and Rambo
& Digital dataset \\
\cline{2-5}
& Generating realistic adversarial images
&\cite{zhang2018deeproad}
& Three DNN models: Autumn~\cite{autumn}, Chauffeur and Rwightman~\cite{Rwightman}
& Digital dataset\\
 \bottomrule
\end{tabular}
}
\end{subtable}
\hfill
\begin{subtable}[t]{\textwidth}
\resizebox{\textwidth}{!}{
\begin{tabular}{p{0.2\textwidth}|p{0.65\textwidth}|p{0.15\textwidth}<{\centering}}
\toprule
\textbf{Oracle} 
& \textbf{Description} 
& \textbf{Literature} 
\\
\hline 
\multirow{3}{*}{Metamorphic testing}
& The steering angle should not change significantly under different conditions
& \cite{tian2018deeptest,zhang2018deeproad}
\\
\cline{2-3}
& The density and direction of fog should not affect the output steering angle of the target models
&\cite{pan2021metamorphic}
\\
\hline
Differential testing
& The DNNs under test should give the same inference result for the same scenario 
& \cite{pei2017deepxplore}
\\
\hline
Model-based oracle
& Predicting the situation that the ADS is probably not able to handle  
& \cite{stocco2020misbehaviour,hussain2022deepguard}
\\
\bottomrule
\end{tabular}
}
\end{subtable}
\hfill
\begin{subtable}[t]{\textwidth}
\resizebox{\textwidth}{!}{
\begin{tabular}{p{0.2\textwidth}|p{0.65\textwidth}|p{0.15\textwidth}<{\centering}}
\toprule
\textbf{Adequacy} 
& \textbf{Description} 
& \textbf{Literature} 
\\
\hline 
Combinatorial coverage
& 2-way combinatorial testing based on image transformations
& \cite{chandrasekaran2021combinatorial}
\\
\bottomrule
\end{tabular}
}
\end{subtable}
\end{table}

\subsubsection{Discussion}
\label{subsubsec:end2endDiscussion}
As with the perception module, the end-to-end module also contains many DNN-based models; however, these models are not only used for perception, but also for the control of the vehicles. Consequently, adversarial attack methods used in the perception module testing, including optimization-based method~\cite{zhou2020deepbillboard, pavlitskaya2020feasibility, boloor2019simple, boloor2020attacking} and GAN-based method~\cite{kong2020physgan, zhang2018deeproad}, are also adopted as the testing methodologies for this module. One observation is that, compared to perception module testing that tests DNN models using single images, the work~\cite{zhou2020deepbillboard, kong2020physgan} for end-to-end module testing often use a series of images, i.e., the frames captured by cameras in a system execution. Another major testing
approach is the coverage-based testing~\cite{pei2017deepxplore, tian2018deeptest, zhang2018deeproad, li2021testing}, in which the testing is guided by coverage criteria proposed for measuring whether the system behavior has been sufficiently explored. 

Since it is hard to evaluate the correctness of the output steering angle for an input image, metamorphic testing~\cite{tian2018deeptest, zhang2018deeproad,pan2021metamorphic} and differential testing~\cite{pei2017deepxplore} are adopted for tackling this problem. In addition, we find that other oracle techniques, e.g., model-based oracles~\cite{stocco2020misbehaviour,hussain2022deepguard}, can be used to solve the oracle problem for this module.

\setlength{\fboxsep}{8pt}
\begin{center}
\Ovalbox{
\begin{minipage}{0.9\textwidth}
\small \textbf{Summary:}
Many of the testing techniques used for testing the perception module can also be used for testing the end-to-end module, such as adversarial attack. One notable difference from the perception module is that, in the end-to-end module, these techniques are applied in a driving context involving a series of continuously-changing images, rather than a single image. Besides those metrics that have been used in the perception module, e.g., metamorphic testing, new techniques, e.g., differential testing, are employed to solve the oracle problem.
In terms of test adequacy metrics, this module is very similar to the perception module.
\end{minipage}
}
\end{center}
\smallskip

\subsection{Answer to RQ1}
In total, we survey over 80 papers that study the testing of different modules of ADS. Various testing techniques have been proposed for testing different modules. Based on our survey, we can draw the following conclusions: 
\begin{inparaenum}
\item for the sensing module, physical testing and deliberate attack on the sensors could effectively find their abnormal behaviors;
\item for the perception module and the end-to-end module, adversarial attack is the most widely-used approach, since the two modules mainly rely on the use of DNN-based models;
\item for the planning module, though the relevant studies are not so many, search-based testing has been extensively adopted;
\item for the control module, main testing techniques include fault injection, sampling, and falsification.
\end{inparaenum}

Despite the numerous techniques dedicated to different modules, we also find some open challenges for the testing of these modules. For example, the neuron coverage in~\cite{pei2017deepxplore} may not be effective  for testing the perception module and the end-to-end module. More details about these open challenges are also discussed in \S{}\ref{sec:challenges&opportunities}.

\section{Literature of Techniques on System-level ADS Testing}
\label{sec:system}
In this section, we introduce the research works on system-level ADS testing with the  goal of answering RQ2 in~\S{}\ref{sec:introduction}. Different from module-level testing, system-level testing focuses on the failures that threaten the safety of the whole vehicle due to the collaborations between modules.
In~\S{}\ref{sec:module}, most of the testing works are done in simulated environments, implemented by various software simulators. In this section, we introduce not only simulation-based testing in~\S{}\ref{sec:systemSimulator}, but also introduce the \emph{mixed-reality testing} in~\S{}\ref{sec:mixed_testing} that introduces real hardware in the testing loop. 

\subsection{System-Level Testing with Simulators}\label{sec:systemSimulator}
We first introduce system-level testing conducted with the help of software simulators. Similar to the module-level works, we also present these studies from three perspectives, namely, test methodology (shown in~\S{}\ref{subsec:systemMethodology}), test oracle (shown in~\S{}\ref{subsec:systemOracle}) and test adequacy (shown in~\S{}\ref{subsec:systemAdequacy}).

\subsubsection{Test Methodology} \label{subsec:systemMethodology}
In the literature, we find various testing methods for the system-level testing of ADS, including search-based testing, adaptive stress testing, sampling-based methods, and adversarial attack. In this section, we introduce these testing methods.

\myparagraph{Search-based testing} 
Search-based testing (or a similar concept named \emph{fuzzing}\footnote{Search-based testing and Fuzzing are similar concepts coming from different communities. The former emphasizes on the testing methodology via search, which relies on well-defined fitness functions and applies search heuristics, e.g., evolutionary algorithms, to find the target test cases. Fuzzing comes from the security community and its methodological essence lies at its randomness. Similar to search-based testing, fuzzing also comes with an objective function as a guidance that helps it achieve the target more efficiently.}), is one of the most widely-adopted methodologies in ADS testing. As introduced in~\S{}\ref{subsec:planningMethodology}, it consists in searching in the parameter space for specific parameter values that achieve a testing objective. In this section, we introduce the works~\cite{dreossi2019compositional, abdessalem2018testing,gambi2019automatically, riccio2020model,zohdinasab2021deephyperion,zohdinasab2022efficient,castellano2021analysis,goss2022eagle,zheng2020rapid,li2020av,hauer2019fitness,kolb2021fitness,luo2021targeting,birchler2022single, birchler2022cost, ben2016testing,batsch2021scenario, beglerovic2017testing, sun2020adaptive} to illustrate the ideas.

Dreossi et al.~\cite{dreossi2019compositional} propose a compositional search-based testing framework, and apply it for the testing of ADS with machine learning components (i.e., mostly perception). The basic idea in their work is the cooperative use of the perception input space and the whole system input space: the constraints on one space can reduce the search efforts in the other space. In this way, they improve the efficiency of searching for counterexamples. 
Abdessalem et al.~\cite{abdessalem2018testing} propose a multi-objective search algorithm for detecting errors caused by \emph{feature interaction}. A \emph{feature interaction} describes the interaction between different ADS functionalities, e.g., an AEB command could be overridden by an ACC command since the two functionalities both control the braking actuator.
In practice, search-based testing has also proved to be effective for industry-level ADS. Li et al.~\cite{li2020av} propose \emph{AV-Fuzzer} used for testing of \Apollo, and they show the effectiveness of this framework in finding dangerous scenarios.

There is a line of work~\cite{riccio2020model, zohdinasab2021deephyperion, zohdinasab2022efficient} that studies the relationship between test input and system behavior.
Riccio et al.~\cite{riccio2020model} propose the notion of \emph{frontier of behaviors}, which represents the boundary of inputs where the system starts to behave abnormally.  
In their later work~\cite{zohdinasab2021deephyperion}, they firstly provide an \emph{interpretable feature map} that explains the correlations between test inputs and system behaviors, and leverage \emph{Illumination Search}~\cite{mouret2015illuminating} to explore the feature space. This approach is enhanced in their follow-up work~\cite{zohdinasab2022efficient} for finding those test inputs that contribute to the exploration of the feature map.

Lane keeping system is an important target in ADS testing. When search-based testing is applied, different road representations can affect the effectiveness of the approach. Castellano et al.~\cite{castellano2021analysis} compare six road representations for testing lane keeping systems, and found that curvatures and orientation are essential factors which affect the behaviors of these systems.
Gambi et al.~\cite{gambi2019automatically} propose a novel approach called \emph{ASFAULT} to generate virtual roads for testing lane keeping systems. 
Experiments on \BeamNGAI~\cite{BeamNG.AI} and \DeepDriving~\cite{Deepdrive} demonstrate that the proposed approach could generate effective testing roads that cause vehicles to deviate from the correct lane.
Open-source search-based tools, e.g., \emph{Frenetic}~\cite{castellano2021frenetic}, are also developed and the comparison of these tools are reported in~\cite{panichella2021sbst, gambi2022sbst}.

There are other works that aim to improve the search efficiency by designing better search algorithm. Abdessalem et al.~\cite{abdessalem2018testing, abdessalem2018testing_v} combine multi-objective search with decision tree classification for test generation of ADS. In their framework, the classification checks whether the scenario is a critical one, and accelerates the search process. 
Goss et al.~\cite{goss2022eagle} apply \emph{Rapidly-exploring Random Trees} (\emph{RRT}) based on an \emph{Eagle Strategy} to estimate the critical scenario boundaries. Zheng et al.~\cite{zheng2020rapid} propose a quantum genetic algorithm that allows lower population size. 
Luo et al.~\cite{luo2021targeting} study the test case prioritization techniques and employ multi-objective search algorithms to find violations with a higher probability of occurrence. 
Test case prioritization techniques are also utilized to accelerate the regression testing of ADS ~\cite{birchler2022single, birchler2022cost} and achieve remarkable results.

Search-based testing is usually based on system simulations; however, even with software simulators, the simulations of ADS can still be expensive and slow. 
There is another line of work~\cite{ben2016testing, batsch2021scenario, beglerovic2017testing, sun2020adaptive} that trains surrogate models as the substitute for testing acceleration.
Abdessalem et al.~\cite{ben2016testing} train a surrogate model that maps the scenario parameters to fitness functions, and use the surrogate to detect the non-critical parameters for search space reduction.
\emph{Gaussian Process} is also leveraged for training a surrogate model in~\cite{batsch2021scenario}.
To search for more collision scenarios, Beglerovic et al.~\cite{beglerovic2017testing} train a surrogate model by utilizing the critical scenarios that already exist. 
For finding an optimal surrogate model for ADS testing, Sun et al.~\cite{sun2020adaptive} compare six types of surrogate models, e.g., \emph{Extreme Gradient Boosting} (\emph{EGB}) and \emph{Kriging} (\emph{KRG}) surrogates, in two logical scenarios.

\myparagraph{Adaptive stress testing} 
Stress testing has been widely adopted in various domains of the industry, which performs testing by providing test cases beyond the capability of the system under test. \emph{Adaptive stress testing}, literally, performs stress testing in an adaptive manner; namely, it prioritizes the test cases and allocates different testing resources to them accordingly. Therefore, specifying the policy of priority assignment is the key to adaptive stress testing. 
Koren et al.~\cite{koren2018adaptive} apply adaptive stress testing for ADS, and design a priority assignment policy based on the difference between the expected behavior and the actual behavior. In a later work~\cite{corso2019adaptive}, they propose a new priority assignment policy based on \emph{Responsibility-Sensitive Safety} (\emph{RSS})~\cite{shalev2017formal}, which defines the utopian behavior of the cars by which no collision will happen in a scenario. The new policy is thus defined according to the distance of the ADS behavior compared to the utopian cases in the RSS rules~\cite{corso2019adaptive}.
Baumann et al.~\cite{baumann2021automatic} adopt reinforcement learning, namely, \emph{Q-learning}~\cite{watkins1992q}, for exposing more critical scenarios in the overtaking scenario. 
Reinforcement learning is also combined with RSS rules for generating edge cases in ~\cite{karunakaran2020efficient}.

\myparagraph{Sampling}
One use case in ADS testing is to generate scenarios by sampling from a natural scenario distribution, in order to make the generated scenario realistic. This has been studied in~\cite{akagi2019risk}.
Nitsche et al.~\cite{nitsche2018novel} propose a sampling-based framework for validating ADS at road junctions. Specifically, they first cluster the junction scenarios along with the representative variations from the real-world accident data, and then these relevant parameters are sampled by the \emph{Latin Hybercube Sampling} (\emph{LHS}) method and used to compose concrete scenarios for simulation testing.

Sampling is also used to help the identification of the failure features.
Corso et al.~\cite{corso2020interpretable} combine \emph{signal temporal logic} (\emph{STL}) with sampling method to generate disturbance trajectories for testing. Those trajectories are interpretable and easier for debugging due to the features of STL, i.e., the description of logical relationships over time. In another work~\cite{corso2020scalable} of them, dynamic programming is applied during the sampling process to discover more failure scenarios.

Batsch et al.~\cite{batsch2019performance} sample the simulation data in a traffic jam scenario with the \CarMaker~\cite{Carmaker} simulation platform. The obtained data sets are then used to train a \emph{Gaussian Process Classification} model, which could probabilistically estimate the boundary between safe scenarios and unsafe scenarios.
Sch\"{u}tt et al.~\cite{DBLP:conf/vehits/SchuttHMZS22} utilize Bayesian optimization and Gaussian process to identify the relevant parameters of a logical scenario, i.e., they find the vehicle speed has no influence in one \emph{vulnerable road user} (\emph{VRU}) testing scenario.
Birkemeyer et al.~\cite{birkemeyer2022feature} leverage a \emph{Feature Model}, i.e., features are organized as nodes in a tree structure, to represent a scenario space for sampling. Experimental results show that the FM-based sampling method is suitable for scenario selection for ADS testing. 

Moreover, advanced sampling techniques can be applied to achieve specific goals; for example, \emph{importance sampling}~\cite{tokdar2010importance} is a technique used to sample rare events. In normal occasions, unsafe scenarios are indeed rare to happen, so detecting those scenarios is hard and costly. In that case, importance sampling can be applied to accelerate the testing~\cite{o2018scalable, norden2019efficient}. 
Zhao et al.~\cite{zhao2016accelerated, huang2017accelerated, huang2017anaccelerated, zhao2017accelerated, huang2018versatile, huang2017evaluation,wang2021combining} work extensively in this direction. The main aim of their work is to spend less simulations to detect more system failures, under various scenarios. Specifically, in~\cite{zhao2016accelerated, huang2017accelerated, huang2017anaccelerated, huang2018versatile, huang2017evaluation}, they investigate the cut-in/lane change scenarios; in~\cite{zhao2017accelerated} and~\cite{wang2021combining}, they focus on the car-following scenario and the unprotected pedestrian crossing scenario, respectively.

\myparagraph{Adversarial attack} Adversarial attack has been introduced in~\S{}\ref{subsec:perceptionMethodology}, in which it is used for testing the perception module. Here, we introduce several works~\cite{sato2021dirty, rubaiyat2018experimental,nassi2022b,wang2021advsim} that also attack the perception module, but they assess the influence of the attack on the whole system. Sato et al.~\cite{sato2021dirty} generate attack patches, as a camouflage for dirty roads, that mislead the lateral control functionality of the victim ADS to deviate from the lane. 
Rubaiyat et al.~\cite{rubaiyat2018experimental} generate perturbations to camera-captured images, based on a system-level safety risk analysis, to assess the reliability of \OpenPilot under real-world environmental conditions. 
Nassi et al.~\cite{nassi2022b} leverage the print advertisement to perform the attack, e.g., they embed an adversarial traffic sign on the back of other vehicles, and mislead the system to wrong behaviors.
Wang et al.~\cite{wang2021advsim} perform an attack that adds perturbations to the trajectories of \emph{NPCs}, and modifies the corresponding LiDAR sensor data.

\subsubsection{Test Oracle}\label{subsec:systemOracle}
The oracles of the system-level testing of ADS are usually defined by safety metrics, such as \emph{time-to-collision}, which measures how far the ADS under test is from dangerous situations. These metrics can be directly computed by monitoring the system behavior in the simulator, or expressed as formal specifications, such as \emph{signal temporal logic} (\emph{STL}), which can automatically monitor the system behavior and compute the metric values. Besides, metamorphic relations are also used in some works for defining the oracle of ADS.

\myparagraph{Safety metrics} 
In system-level testing, a suitable safety metric, or called criticality metric, can be leveraged to find more system violations. There have been studies~\cite{mahmud2017application, jahangirova2021quality, westhofen2022criticality} that comprehensively investigate these safety metrics and here part of commonly-used metrics are listed in Table~\ref{tab:safetyMetrics}.
These safety metrics can be categorized into temporal metrics and non-temporal metrics. Temporal metrics describe the temporal requirements to moving objects, and the most popular ones are \emph{Time-to-Collision} (\emph{TTC})~\cite{tuncali2019rapidly} and its extensions, e.g., \emph{Worst-Time-to-Collision} (\emph{WTTC})~\cite{wakabayashi2003traffic}, that measure the closeness of the ego car to collision in the scenario. Weng et al.~\cite{weng2020model} propose the \emph{Model Predictive Instantaneous Safety Metric} (\emph{MPrISM}), which considers the interaction between moving vehicles. Another metrics include \emph{Time Headway} (\emph{THW}) and \emph{Time-to-React} (\emph{TTR})~\cite{tamke2011flexible}. The former calculates the time of the ego vehicle to reach the position of the lead vehicle, and the latter estimates the remaining time for a required reaction, e.g., a braking action.

Non-temporal metrics concern different aspects, such as distance, deceleration, and steering. One distance metric called \emph{Stop Distance} (\emph{SD}) ~\cite{li2020av} calculates the distance for a vehicle to stop with a maximum comfortable deceleration. Another distance metric is called \emph{Lateral Position} (\emph{LP})~\cite{zohdinasab2021deephyperion}, which defines the distance between the center of the vehicle and the center of the driving lane. Deceleration metrics, such as \emph{Deceleration Rate to Avoid a Crash} (\emph{DRAC})~\cite{almqvist1991use}, consider the deceleration rate during emergency. Steering metrics, such as \emph{Steering Angle Reversal Rate} (\emph{SARR})~\cite{mclean1971analysis},
focus on the steering angle of a vehicle during the driving process.

There are works~\cite{hauer2019fitness, kolb2021fitness} that propose to organize and utilize these safety metrics in an elegant manner. Also, Li et al.~\cite{li2016intelligence} propose to design metrics that involve more factors such as the relationship between scenarios, tasks, and functionalities of an ADS.

\begin{table}[!tb]
\caption{Commonly-used Safety Metrics}
\label{tab:safetyMetrics}
\scriptsize
\scalebox{1.0} {
\renewcommand\arraystretch{1.5}
\begin{tabular}{c|c|l}
\hline
\textbf{Category} & \textbf{Name} & \multicolumn{1}{c}{\textbf{Description}} \\ \hline
\multirow{5}{*}{\begin{tabular}[c]{@{}c@{}}Temporal\\ metrics\end{tabular}} & TTC~\cite{tuncali2019rapidly} & The time until two objects collide with the current speed and path\\ \cline{2-3} 
 & WTTC~\cite{wakabayashi2003traffic} &The time of the collision in the most likely accident scenario\\ \cline{2-3} 
 & MprISM~\cite{weng2020model}  & \begin{tabular}[c]{@{}l@{}}Estimating the TTC with the consideration of game interaction between vehicles\end{tabular} \\ \cline{2-3} 
 & THW ~\cite{tamke2011flexible} &The time between two objects reaching the same location  \\ \cline{2-3} 
 & TTR ~\cite{tamke2011flexible} & \multicolumn{1}{l}{\begin{tabular}[c]{@{}l@{}}The remaining time until the start of the last driving maneuver that can avoid\\ collisions with all objects in the scenario\end{tabular}} \\ \hline
\multirow{4}{*}{\begin{tabular}[c]{@{}c@{}}Non-Temporal\\ metrics\end{tabular}} & SD~\cite{li2020av} & The stopping distance of the vehicle at the maximum comfortable deceleration \\ \cline{2-3} 
 & LP~\cite{zohdinasab2021deephyperion} & \begin{tabular}[c]{@{}l@{}}The distance between vehicle center and lane center\end{tabular} \\ \cline{2-3} 
 & DRAC~\cite{almqvist1991use} & \multicolumn{1}{l}{\begin{tabular}[c]{@{}l@{}}The minimum deceleration rate required by a vehicle to avoid a crash\end{tabular}} \\ \cline{2-3}
 & SARR~\cite{mclean1971analysis}  & The number of steering angle reversals larger than a certain value \\ \hline
\end{tabular}
}
\end{table}

\myparagraph{Formal specifications} As introduced in~\S{}\ref{subsec:perceptionOracle}, formal specification uses temporal logic languages to express the properties which the system should hold during the running; then by specification-based monitoring, the satisfaction of the system behavior can be automatically decided. On the system-level testing of ADS, \emph{signal temporal logic} (\emph{STL}), which can express the properties over real-time continuous variables, is the proper selection of specification language. There are a few works that adopt STL as the specification language~\cite{dreossi2019compositional, tuncali2018simulation, tuncali2019requirements}, in which STL monitors are synthesized to decide whether the behavior of the ADS satisfy the desired safety properties.
Zhang et al.~\cite{zhang2021systematic} utilize formal specifications to represent driving rules and ADS behaviors to check the consistency between them.

\myparagraph{Metamorphic testing} Metamorphic testing has been discussed for the module-level testing in~\S{}\ref{subsec:perceptionOracle} and \S{}\ref{subsec:e2eOracle}. On the system-level testing, Han et al.~\cite{han2020metamorphic} utilize metamorphic relations to distinguish between real failures and false alarms. The metamorphic relation regulates that the behavior of the ADS should be similar in slightly different scenarios; otherwise, the collision in one of such scenarios is considered \emph{avoidable}, and thus a real failure. 

\subsubsection{Test Adequacy}\label{subsec:systemAdequacy}
In system-level testing, the adequacy of testing is embodied by the diversity of the testing scenarios for the ADS. 
In this section, we introduce two lines of work that define various metrics to characterize the diversity of scenarios. 

\myparagraph{Scenario coverage} 
There is a line of work that defines \emph{coverage} for scenarios. The intuition is that the testing is sufficient if all different types of scenarios are covered~~\cite{hauer2019did}.
Tang et al.~\cite{tang2021collision} classify the scenarios based on the topological structure of the map.
Kerber et al.~\cite{kerber2020clustering} define a distance measure over scenarios based on their spatiotemporal features, which enable scenario clustering.  
Besides, the temporal, spatial, and causal information of the simulation data can be further abstracted into \emph{situations} for covering more test scenarios~\cite{majzik2019towards, tahir2022intersection}.

\myparagraph{Combinatorial coverage}
Combinatorial coverage has been introduced in~\S{}\ref{subsec:perceptionAdequacy}. Unlike the above coverage criteria defined directly on the features of the scenarios, combinatorial coverage considers the coverage of the combinations of different parameters that identify different scenarios. Tuncali et al.~\cite{tuncali2018simulation, tuncali2019requirements} propose to use \emph{covering array} for scenario generation in ADS testing.
Covering array is a specific mechanism in software testing that guarantees the satisfaction of the $t$-way combination coverage of the parameters. See~\S{}\ref{subsec:perceptionAdequacy} for more details about $t$-way combination coverage.
Guo et al.~\cite{guo2020automated} propose the definition of \emph{scenario complexity} and apply combinatorial testing techniques to generate more complex testing scenarios.
Shu et al.~\cite{shu2021test} adopt the three-way combinatorial testing method on lane-changing scenarios, which ensures a high coverage of the generated critical scenarios. 
Li et al.~\cite{li2020ontology} utilize the ontology concept, i.e., formulations of entities and their relationships, to describe the driving environment of an ADS. Then the constructed ontologies are combined with combinatorial testing techniques for generating concrete scenarios with coverage guarantee.
Another work~\cite{li2022comopt} proposes a scenario generation framework called \emph{ComOpT}, based on t-way combinatorial testing, and finds numerous system failures of \Apollo.
Moreover, combinatorial testing is also used in~\cite{rocklage2017automated} to tackle the regression testing problem of ADS.

\begin{table}[!tb]
\footnotesize
\caption{Summary of the papers for simulation-based system-level testing: Part \rom{1}}
\label{tab:systemLevelTesting}
\renewcommand{\arraystretch}{1.2}
\begin{subtable}[t]{\textwidth}
\resizebox{\textwidth}{!}{
\begin{tabular}{p{0.11\textwidth}|p{0.48\textwidth}|p{0.1\textwidth}<{\centering}|p{0.19\textwidth}|p{0.12\textwidth}}
\toprule
\textbf{Methodology} & \textbf{Description} & \textbf{Literature} & \textbf{Test Objective}  & \textbf{\begin{tabular}[c]{@{}l@{}}Environment\end{tabular}}   \\ 
\hline
\multirow{13}{*}{\parbox{0.11\textwidth}{Search-based testing}}
& Incorporating the perception input space and the whole system input space to accelerate search
&\cite{dreossi2019compositional}
& AEB systems~\cite{lee2011development,AEBS1}
& Simulation\\
\cline{2-5}
& Searching for unsafe feature interactions with decision trees
& \cite{abdessalem2018testing, abdessalem2018testing_v}
& Systems from IEE~\cite{IEE}  & Simulation \\ 
\cline{2-5}
& Generating virtual roads for testing lane keeping systems
&\cite{gambi2019automatically,riccio2020model,zohdinasab2021deephyperion,zohdinasab2022efficient,castellano2021analysis}
&Lane keeping system in \BeamNG~\cite{BeamNG.tech}
&Simulation
\\
\cline{2-5}

& Searching for critical scenario boundaries
& \cite{goss2022eagle,zheng2020rapid}
& Systems built based on simulators
&Simulation
\\
\cline{2-5}
& Finding safety violations of an ADS in the dynamic environment
&\cite{li2020av}
& \Apollo      & Simulation \\ 
\cline{2-5}
& Finding violations with a higher probability of occurrence
& \cite{luo2021targeting,birchler2022single,birchler2022cost}
& Systems such as \BeamNGAI~\cite{BeamNG.AI}
& Simulation
\\
\cline{2-5}
& Training surrogate models to accelerate testing
&\cite{ben2016testing,batsch2021scenario,beglerovic2017testing,sun2020adaptive}
& Systems like AEB system
& Simulation
\\

\hline
Adaptive stress testing
& Assigning different priorities to the test cases
&\cite{koren2018adaptive,corso2019adaptive,baumann2021automatic,karunakaran2020efficient}
& Systems like the Intelligent Driver Model~\cite{treiber2000congested}
& Simulation\\
\hline
\multirow{10}{*}{\parbox{0.11\textwidth}{Sampling}}
& Sampling junction scenarios
&\cite{nitsche2018novel}
& Collision avoidance system
& Simulation
\\
\cline{2-5}
& Sampling the simulation data in a traffic jam scenario
&\cite{batsch2019performance}
& System in \CarMaker
& Simulation
\\
\cline{2-5}
& Combining STL with the sampling method
&\cite{corso2020interpretable,corso2020scalable}
& Intelligent Driver Model
& Simulation
\\
\cline{2-5}
&  Identifying the relevant parameters of a logical scenario
&\cite{DBLP:conf/vehits/SchuttHMZS22}
&
Intelligent Driver Model
& Simulation
\\
\cline{2-5}
& Sampling scenario space represented by feature model
&\cite{birkemeyer2022feature}
& AEB system
& Simulation
\\
\cline{2-5}
& Adopt Importance Sampling to sample rare events

&\cite{o2018scalable,norden2019efficient,zhao2016accelerated,huang2017accelerated,huang2017anaccelerated,zhao2017accelerated,huang2018versatile, huang2017evaluation,wang2021combining}
& Systems like the ACC system~\cite{10.5555/556515} & Simulation \\ 
\cline{3-5}
\hline
\multirow{6}{*}{\parbox{0.11\textwidth}{Adversarial attack}}
& Generating attack patches as a camouflage of dirty roads
&\cite{sato2021dirty}
& \OpenPilot 
& Simulation and real world \\
\cline{2-5}
& Generating adversarial perturbations under different weather
&\cite{rubaiyat2018experimental}
& \OpenPilot  & Simulation \\
\cline{2-5}
& Adding perturbations on the trajectories of NPCs
&\cite{wang2021advsim}
& Driving models~\cite{sadat2019jointly, zeng2019end, sadat2020perceive}
& Digital dataset
\\
 \bottomrule
\end{tabular}
}
\end{subtable}
\end{table}

\subsubsection{Discussion}
\label{subsubsec:systemDiscussion}
As shown by Table~\ref{tab:systemLevelTesting}, search-based testing~\cite{dreossi2019compositional, abdessalem2018testing,gambi2019automatically, riccio2020model,zohdinasab2021deephyperion,zohdinasab2022efficient,castellano2021analysis,goss2022eagle,zheng2020rapid,li2020av,hauer2019fitness,kolb2021fitness,luo2021targeting,birchler2022single, birchler2022cost, ben2016testing,batsch2021scenario, beglerovic2017testing, sun2020adaptive} is the most widely-used technique for testing the whole ADS, with different focuses, e.g., studying the relations between test input and system behavior~\cite{riccio2020model, zohdinasab2021deephyperion, zohdinasab2022efficient}, testing lane keeping systems~\cite{castellano2021analysis, gambi2019automatically, panichella2021sbst, gambi2022sbst} and test case prioritization ~\cite{luo2021targeting, birchler2022single, birchler2022cost}. Although simulation-based testing aims to solve the high cost problem of real-world testing, it may repeatedly simulate the same type of scenarios, which is also a time-consuming process.
Consequently, adaptive stress testing~\cite{koren2018adaptive,corso2019adaptive,baumann2021automatic,karunakaran2020efficient} and sampling-based techniques~\cite{akagi2019risk,nitsche2018novel,batsch2019performance,corso2020interpretable,DBLP:conf/vehits/SchuttHMZS22,birkemeyer2022feature, zhao2016accelerated, huang2017accelerated, huang2017anaccelerated} are applied for accelerating the testing process. 
As in the cases of the perception and end-to-end modules, adversarial attack~\cite{sato2021dirty, rubaiyat2018experimental, nassi2022b, wang2021advsim} has also been adopted for system-level testing, which aims to detect the vulnerabilities of the perception that affect the safety of the whole system.
Note that among these testing techniques, adaptive stress testing has not been studied extensively, but it has a high potential for future ADS testing since it is effective in various domains of the industry~\cite{ellestad2003stress}.

System-level testing usually relies on safety metrics, e.g., temporal and non-temporal metrics (as shown in Table~\ref{tab:safetyMetrics}) and metamorphic relations~\cite{han2020metamorphic}, as the oracles that measure the occurrences of safety violations during the testing process.
For ensuring the adequacy of system-level testing, there are two lines of work, namely, scenario coverage~\cite{tang2021collision, majzik2019towards, kerber2020clustering} and combinatorial testing~\cite{tuncali2018simulation, tuncali2019requirements, guo2020automated, shu2021test, li2020ontology, li2022comopt, rocklage2017automated}, which propose metrics to characterize the diversity of testing scenarios.

Overall, there exist more works in system-level testing than that in module-level testing. Moreover, there are many other works that study the differences between simulation-based testing and real-world testing. More details can be found in~\S{}\ref{sec:simRealWorld}.

\begin{table}[!tb]\ContinuedFloat
\footnotesize
\caption{Summary of the papers for simulation-based system-level testing: Part \rom{2}}
\renewcommand{\arraystretch}{1.2}
\begin{subtable}[t]{\textwidth}
\resizebox{\textwidth}{!}{
\begin{tabular}{p{0.2\textwidth}|p{0.7\textwidth}|p{0.1\textwidth}<{\centering}}
\toprule
\textbf{Oracle} 
& \textbf{Description} 
& \textbf{Literature} 
\\
\hline 
\multirow{2}{*}{Safety metrics}
& Temporal metrics and non-temporal metrics (more details in table~\ref{tab:safetyMetrics}).
& -
\\
\cline{2-3}
& Providing a set of fitness function templates for different testing goals
& \cite{hauer2019fitness,kolb2021fitness}
\\
\hline

\multirow{3}{*}{Formal specifications}
& Adopting STL as the specification language to express the properties over real-time continuous variables
& \cite{dreossi2019compositional, tuncali2018simulation, tuncali2019requirements}
\\
\cline{2-3}
& Utilizing formal specifications to represent driving rules and ADS behaviors 
& \cite{zhang2021systematic}
\\
\hline
Metamorphic testing
& The behavior of the ADS should be similar in slightly different scenarios
&\cite{han2020metamorphic}
\\
\bottomrule
\end{tabular}
}
\end{subtable}
\hfill
\begin{subtable}[t]{\textwidth}
\resizebox{\textwidth}{!}{
\begin{tabular}{p{0.2\textwidth}|p{0.65\textwidth}|p{0.15\textwidth}<{\centering}}
\toprule
\textbf{Adequacy} 
& \textbf{Description} 
& \textbf{Literature} 
\\
\hline 
\multirow{3}{*}{Scenario coverage}
& Based on the topological structure of the map
& \cite{tang2021collision}
\\
\cline{2-3}
& Based on spatiotemporal features
& \cite{kerber2020clustering}
\\
\cline{2-3}
& Based on the temporal, spatial, and causal information of the simulation data
& \cite{majzik2019towards, tahir2022intersection}
\\
\hline
\multirow{3}{*}{Combinatorial coverage}
& 
Applying t-way combinatorial testing techniques for scenario generation
& \cite{tuncali2018simulation,tuncali2019requirements,guo2020automated,shu2021test,li2022comopt,rocklage2017automated}
\\
\cline{2-3}
& Utilizing the ontology concept to describe the driving environment
&\cite{li2020ontology}
\\
\bottomrule
\end{tabular}
}
\end{subtable}
\hfill
\end{table}

\smallskip
\setlength{\fboxsep}{8pt}
\begin{center}
\Ovalbox{
\begin{minipage}{0.9\textwidth}
\small \textbf{Summary:}
There are a large number of studies that leverage different techniques, e.g., search-based testing, adaptive stress testing, and sampling-based techniques, for testing the ADS at the system level. Besides, numerous metrics have been proposed for different usages in the testing process, e.g., they are used to measure the occurrences of safety violations, and they are used to characterize the diversity of testing scenarios.
\end{minipage}
}
\end{center}
\smallskip

\subsection{Mixed-Reality Testing}\label{sec:mixed_testing}
\begin{figure}[!tb]
  \centering
  \includegraphics[width=0.8\textwidth]{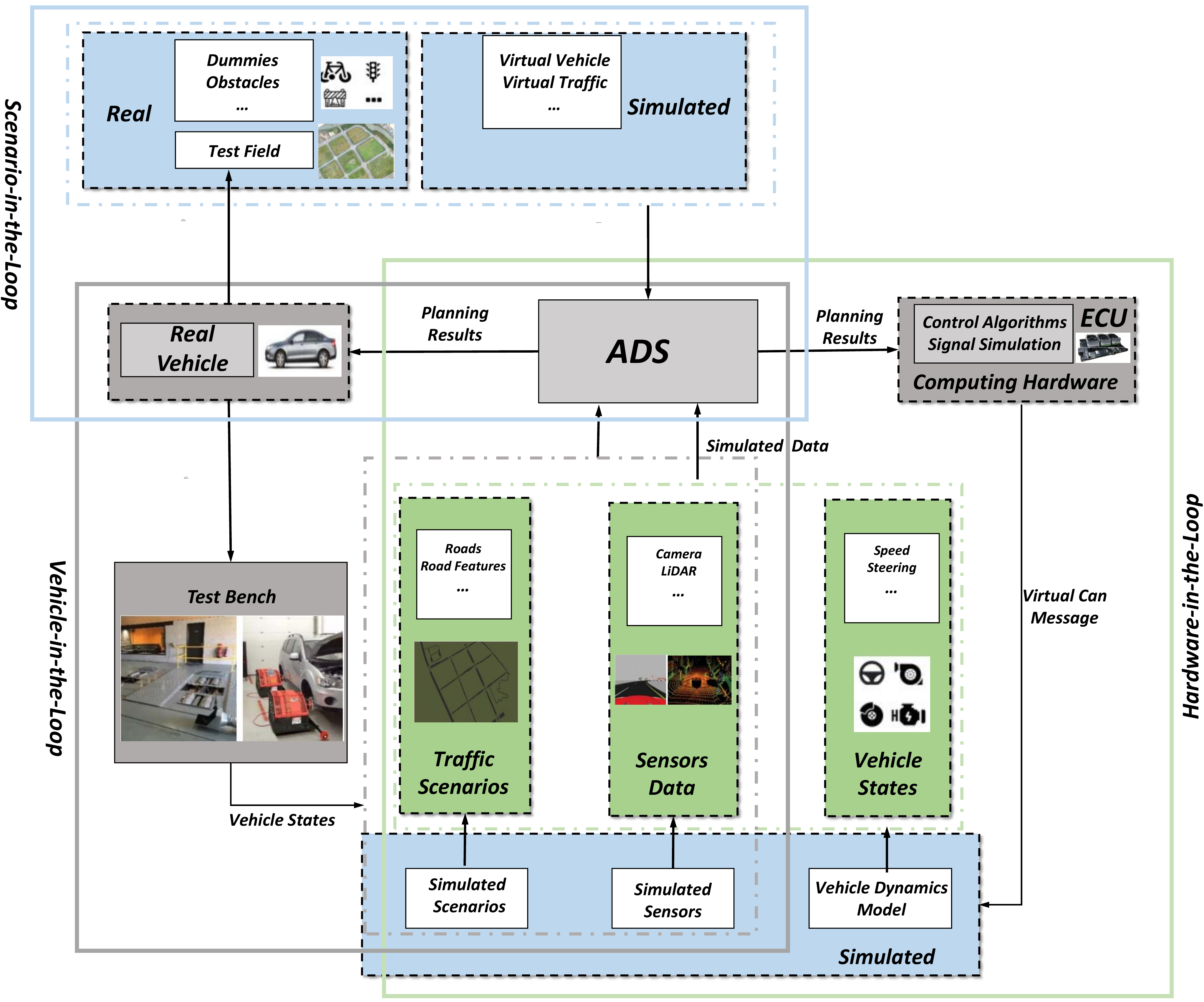}
  \caption{Illustration of the HiL, ViL, and SciL approaches}
  \label{fig:H-V-S_workflow}
\end{figure}

Due to the expensiveness of the real-world testing of ADS, most approaches in~\S{}\ref{sec:module} and in~\S{}\ref{sec:systemSimulator} test ADS in software simulators. 
Although modern simulators can be powerful and high-fidelity, simulation-based testing is not sufficient to reveal all the problems of ADS, due to the gap between simulators and the real world.
As a trade-off, mixed-reality testing combines simulation-based testing with real-world testing. In this section, we introduce several special testing schemes, which replace certain parts of the components in the testing loop, with physical components. Specifically, these schemes include \emph{hardware-in-the-loop} (\emph{HiL}), \emph{vehicle-in-the-loop} (\emph{ViL}), and \emph{Scenario-in-the-Loop} (\emph{SciL}); their mechanisms are illustrated in Fig.~\ref{fig:H-V-S_workflow}. 

\subsubsection{Hardware-in-the-Loop}
HiL testing usually introduces the real ECU hardware into the testing loop, as shown in the green box in Fig.~\ref{fig:H-V-S_workflow}. There is a line of work~\cite{chen2018autonomous, chen2019novel, brogle2019hardware, gao2020hardware} that adopts this testing method.
Chen et al.~\cite{chen2018autonomous, chen2019novel} propose an HIL testing platform that could simulate multi-agent interaction on large-scaled scenarios with the usage of \emph{OpenStreetMap}~\cite{osm}.
Brogle et al.~\cite{brogle2019hardware} build their HiL platform based on \carla and \emph{robot operating system} (\emph{ROS}), which achieves high fidelity in vehicle dynamics and sensor data output. 
Gao et al.~\cite{gao2020hardware} design another HiL platform for AEB testing and find that the performance of the AEB functions in HiL tests is close to that in real road tests.

\subsubsection{Vehicle-in-the-Loop}
Different from HiL testing, ViL testing works by integrating a synchronized virtual scenario into a real vehicle, as shown in the gray box in Fig.~\ref{fig:H-V-S_workflow}. The following works~\cite{chen2020mixed, tettamanti2018vehicle, solmaz2020vehicle, li2021real, stocco2022mind} we collected are all based on this idea.
Chen et al.~\cite{chen2020mixed} propose a ViL testing platform which can reconstruct scenarios based on the corresponding HD map.
For simulating more realistic scenarios, these works~\cite{tettamanti2018vehicle, solmaz2020vehicle, li2021real} integrate popular traffic simulators, such as \SUMO~\cite{krajzewicz2002sumo} and \VISSIM~\cite{vissim}, into the ViL testing loop.
Stocco et al.~\cite{stocco2022mind}
utilize the Donkey Car platform ~\cite{DonkeyCar} to build a 1:16 scale car which is controlled by end-to-end driving models. They test these driving models in a closed-track environment and study the transferability of failures between simulation and the real world. 

\subsubsection{Scenario-in-the-Loop}
SciL testing narrows the gap between simulator and the real world by integrating more real components like the pedestrian dummies into the loop, as shown in the blue box in Fig.~\ref{fig:H-V-S_workflow}.
Szalay et al. first propose the concept of SciL testing in~\cite{szalay2019proof} and they develop a SciL testing platform based on \SUMO and \Unity~\cite{Unity} in a later work~\cite{szalai2020mixed}.
Horvath et al.~\cite{horvath2019vehicle} study the SciL testing by comparing the implementation process of this method with that of ViL testing. The authors find that the two testing methods have the same basis, but SciL testing is still at an early stage.

\subsection{Simulation-Based Testing vs. Real-World Testing}\label{sec:simRealWorld}
Regarding the efforts in simulation-based testing, a natural question arises that, how far is the simulation-based testing still from the real-world testing. Moreover, Kalra et al.~\cite{kalra2016driving} find that the ADS should be driven hundreds millions of miles to demonstrate their reliability.
As an emerging issue, this topic has attracted increasing research attention; here, we introduce the latest progress from two perspectives, namely, the realism of test cases and the realism of simulators.

\myparagraph{Realism of test cases}
One question arises in the simulation-based testing that the virtual scenarios generated by testing algorithms that lead to system failures may never happen in the real world. Indeed, simulators give a high liberty to create traffic participants, of which, nevertheless, only a subset can really happen.  

There is a line of work that aims to bridge this gap and thus generate natural scenarios for ADS testing. Nalic et al.~\cite{nalic2019development} propose a co-simulation framework using two simulation tools \CarMaker (for vehicle dynamics) and \VISSIM (for traffic simulation); their framework can generate scenarios based on calibrated traffic models derived from real-world data. 
In their later work~\cite{nalic2020stress}, stress testing method, which has been introduced in~\S{}\ref{subsec:systemMethodology}, is applied for increasing the number of detected critical scenarios under the co-simulation environment.
Klischat et al.~\cite{klischat2020scenario} utilize \emph{OpenStreetMap} to extract real-world road intersections, and combine with \SUMO to generate realistic traffic scenarios.
Wen et al.~\cite{wen2020scenario} focus on triggering the events in a specific area near the ego vehicle, and a CNN-based selector is utilized to choose those scenario agents which could achieve more realistic results.

The following works~\cite{bashetty2020deepcrashtest, gambi2019generating, huynh2019ac3r,el2022virtual,xinxin2020csg} focus on reconstructing scenarios from public crash reports.
Mostadi et al.~\cite{el2022virtual} utilize a distance metric, i.e., \emph{Manhattan distance}, to align the virtual scenarios to real-world scenarios.
Computer vision algorithms, i.e., object detection and tracking, are adopted in~\cite{bashetty2020deepcrashtest, xinxin2020csg} to extract the trajectories of the vehicles from the crash videos.
Gambi et al.~\cite{gambi2019generating, huynh2019ac3r, gambi2019automatically} utilize \emph{natural language processing} (\emph{NLP}) techniques to extract the relevant information and then calculate the abstract trajectories for recreating the crash. Experimental results show that the method could accurately reconstruct the crashes in public reports, and the generated test cases are able to expose faults in open-source ADS, i.e., \DeepDriving~\cite{Deepdrive}.

There is also a line of work~\cite{zhou2021deep, balaji2019deepracer, o2020f1tenth, lehner20223d} that focuses on narrowing the reality gap in the training process. By including components such as augmented data, small scale cars, and real-world tracks, they could generate more realistic cases to train the perception model or reinforcement learning algorithms for automated driving. 

\myparagraph{Realism of simulators}
A comparative study on the assessment of testing in different levels of simulation is performed by Antkiewicz et al.~\cite{antkiewicz2020modes}. In their work, the authors study simulation-based testing, mixed-reality testing, and real-world testing on two scenarios, i.e., car following and surrogate actor pedestrian crossing. They propose various metrics, e.g., realism, costs, agility, scalability, and controllability, and based on these metrics, they compare the different testing schemes under evaluation. As their conclusion, they quantitatively show the performance difference among the testing schemes: although real-world testing is better in terms of realism, it is more costly, and less agile, scalable, and controllable, compared to simulation-based testing; the performance of mixed-reality testing is in the middle of them. 
Testing ADS in different simulators is studied by Borg et al.~\cite{borg2021digital}, in which they utilize search-based testing techniques to generate scenarios in two simulators, i.e., \PreScan~\cite{prescan} and \ProSiVIC~\cite{belbachir2012simulation}. They find notable differences of the test outputs, e.g., they detect different safety violations. Consequently, they recommend involving multiple simulators for more robust simulation-based testing in the future.

Although simulation-based testing cannot achieve the same realism as real-world testing, to what extent can the results of simulation-based testing benefit real-world testing? This question is investigated in~\cite{fremont2020formal}, where the authors perform simulation-based testing to identify critical scenarios, and map them to a real-world environment.
Their key insights involve that, $62.5\%$ of the unsafe  scenarios detected by the simulators translate to real collisions; and $93.3\%$ of the safe scenarios with the simulators are also safe in the real world.
Another question is whether the simulator-generated dataset can substitute real-world dataset for DNN-based ADS testing, which have been studied in~\cite{haq2020comparing, haq2021can}. Moreover, they also compare offline testing, e.g., module-level testing, and online testing, e.g., system-level testing, in terms of their pros and cons. Experiments on DNN-based ADS show that: the average prediction error difference on two datasets is less than 0.1, which means the simulator-generated dataset can serve as an alternative to the real-world dataset; online testing is more suitable than offline testing for DNN-based ADS testing, since online testing could detect more errors, i.e., those errors caused by accumulation over time, than offline testing.
Reway et al.~\cite{reway2020test} evaluate the simulation-to-reality gap by testing an object detection algorithm under three different environments, namely, a real proving ground and two simulation software, considering four weather conditions. The gap is quantitatively calculated by considering metrics such as \emph{precision} and \emph{recall} on each platform. One of their experimental results is that the gap between real and simulation domains under nighttime and rainy conditions is larger than that under daytime conditions. 
\subsection{Answer to RQ2}
Overall, we have surveyed more than 90 papers dedicated to the system-level testing of the ADS. We find that those module-level testing techniques, such as search-based testing, sampling, and adversarial attack, are also widely adopted for finding failures arising from collaborations over different modules at the system level. Besides, more metrics, which can be found in \S{}\ref{subsec:systemOracle} and \S{}\ref{subsec:systemAdequacy}, are proposed or utilized for facilitating the testing process. Another observation is that more than 30 papers focus on bridging the gap between the simulation and the real-world environments, e.g., by introducing real components into the testing loop or by making a comparison between the simulation-based testing and the real-world testing.

Similar to module-level testing, there still remain several open challenges for system-level testing of the ADS. For example, since the system executions during testing are expensive and time-consuming, it needs future exploration on how to accelerate the testing process. 
More discussions about the challenges and future research directions can be found in \S{}\ref{sec:challenges&opportunities}.

\section{Statistics and Analysis of Literature}
\label{sec:analysis}
In this section, based on the survey results in~\S{}\ref{sec:static_analysis}, \S{}\ref{sec:module} and \S{}\ref{sec:system}, we perform a statistical analysis. Specifically, we provide the threat model for general ADS in~\S{}\ref{subsec:threatModel}, and we collect popular datasets, tool stacks, and programming languages for ADS testing in~\S{}\ref{subsec:datasetToolset}.

\subsection{The Threat Model for General ADS}\label{subsec:threatModel}
\begin{figure}[!tb]
  \centering
  \includegraphics[width=\textwidth]{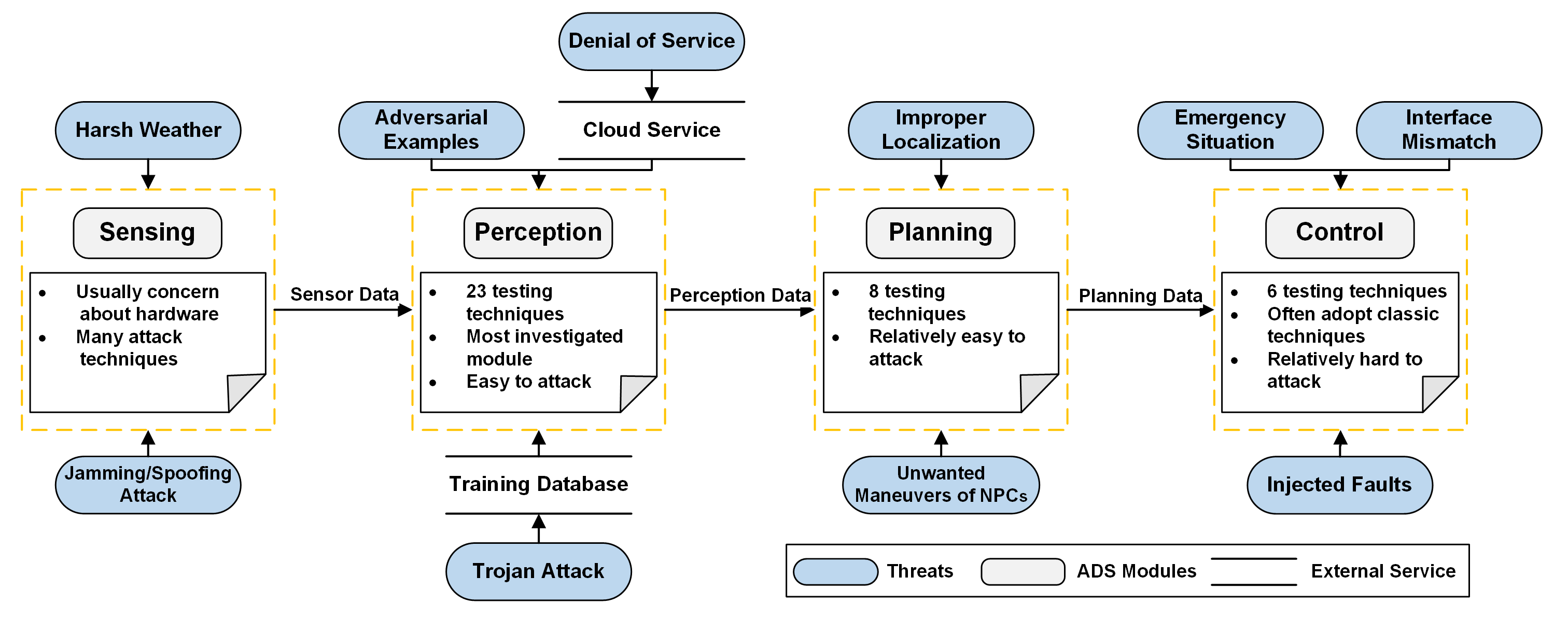}
  \caption{Threat model of ADS}
  \label{fig:threat_model}
\end{figure}

In this section, we construct a threat model in order to summarize the safety and security threats that each module may confront, based on our survey results. 
To build the threat model, we first summarize the threats discovered in the papers which we survey in the module-level testing in~\S{}\ref{sec:module}; then, as a complement, we review the bugs shown in the empirical studies~\cite{garcia2020comprehensive, tang2021issue} on open-source ADS, to understand the concrete issues encountered in each module during system development. 
Our threat model is shown in Fig.~\ref{fig:threat_model}.

\myparagraph{Threats to sensing} In this module, existing studies mainly concern about the hardware aspect, e.g., those physical sensors which are critical hardware used in an ADS for collecting the information of the external environments.
A common threat such as harsh weather conditions could reduce the capabilities of the intelligent sensors.
There are also many deliberate attack techniques, such as \emph{jamming attack}~\cite{ shin2017illusion, yan2016can,lim2018autonomous} and \emph{spoofing attack}~\cite{meng2019gps,zeng2018all,komissarov2021spoofing,wang2021can} (see details in~\S{}\ref{subsec:deliberateAttack}) that target this module, and could interfere with these sensors and harm their normal functionalities.

\myparagraph{Threats to perception}
The perception module is the most investigated and we collect 23 testing techniques dedicated to this module. Common threat comes from \emph{adversarial examples} that are generated by adding perturbations to normal images, which can fool the deep learning models in the perception module to make incorrect predictions, as shown by~\cite{eykholt2018robust, liu2019perceptual, xu2020adversarial, zhu2021adversarial, chen2018shapeshifter, zhao2019seeing,zhang2019camou,im2022adversarial,xu2020adversarial,li2020adaptive,kumar2020black, cao2019adversarial,  cao2021invisible, sun2020towards,wang2021adversarial, chen2021camdar, zhu2021can,yang2021robust,li2021fooling}. Another type of threat is called \emph{Trojan attack}~\cite{ jiang2020poisoning,ding2019trojan}, in which malicious data are injected into the training data of the deep learning models. Moreover, in the case when the ADS requires an HD map from the cloud service, \emph{Denial of service} (\emph{DoS})~\cite{long2005denial} or fake HD map data~\cite{sinai2014exploiting} can interfere with perception tasks such as localization.

\myparagraph{Threats to planning}
With the data produced by the perception module, the planning module takes charge of several tasks, e.g., object trajectory prediction, path planning, and we collect 8 testing techniques for this module. A common threat comes from the unwanted maneuvers of \emph{non-player characters} (\emph{NPCs})~\cite{laurent2019mutation,laurent2020achieving, arcaini2021targeting,arcaini2022less, klischat2019generating, althoff2018automatic,bak2022stress,kahn2022know}, which can interfere with the prediction for moving objects and thus lead to an unsafe trajectory plan. Moreover, improper localization from the perception module can also threaten the accuracy of output trajectories.

\myparagraph{Threats to control}
This module mostly adopts those mature control
techniques, e.g., \emph{MPC} and \emph{PID}, and thus are relatively hard to attack. One threat comes from the injected faults~\cite{uriagereka2017fault, zhou2022strategic}, which could affect the longitudinal/lateral control of the vehicle. Another threat emerges when an emergency situation is encountered, e.g., when an emergency braking is needed, and the control module may fail to handle these cases. Moreover, the output signals are sent via the CAN bus to the ECU for controlling the vehicle. Since this process involves a data transmission between software and hardware, a potential threat is the interface mismatch~\cite{opIssue}, e.g., an inappropriate steering angle rate, in practical usage.

\subsection{Datasets and Tool Stacks for ADS Testing}\label{subsec:datasetToolset}
\begin{table}[!tb]
\centering
\scriptsize
\caption{Scenario driven datasets for ADS testing}
\label{tab:datasets} 
\begin{subtable}[t]{\textwidth}
\renewcommand{\arraystretch}{1.2}
\resizebox{\textwidth}{!}{
\begin{tabular}{p{0.05\textwidth}|p{0.2\textwidth}|p{0.45\textwidth}|p{0.15\textwidth}|p{0.15\textwidth}}
\toprule
    \textbf{Time} & \textbf{Dataset} & \textbf{Description} &\textbf{Size} &\textbf{Reference} \\
    \hline
     2004&NGSIM~\cite{alexiadis2004next}&Vehicle trajectory data at four different locations & -
    & \cite{klischat2019generating}\\
    \hline
    2012 & KITTI~\cite{geiger2013vision}
    & Driving scenes captured by a standard station wagon & 12,919 images & \cite{tuncali2018simulation,cao2021invisible,xiong2021multi,sun2020towards,dokhanchi2018evaluating,ramanagopal2018failing,bashetty2020deepcrashtest,bashetty2020deepcrashtest,zhou2020deepbillboard,kong2020physgan,wang2021can} \\
    \hline
    2013 & GTSRB~\cite{Stallkamp2012} & Containing 43 classes of traffic signs in Germany&50,000+ images &\cite{eykholt2018robust,liu2019perceptual,li2020adaptive,jiang2020poisoning}\\
    \hline
    2014 & BelgiumTS~\cite{belgium} & A large dataset with traffic sign annotations & 10,000+ images &\cite{jiang2020poisoning}\\
    \hline
    2015 & LISA~\cite{philipsen2015traffic} & A traffic sign dataset containing US traffic signs& 43,000+ images &\cite{eykholt2018robust}\\
    \hline
    2016 & Cityscapes ~\cite{cordts2016cityscapes}& A diverse set of stereo video sequences recorded in street scenes & 25,000 images & \cite{zhou2019automated,xu2020adversarial}\\
    \hline
    2016 & Udacity~\cite{Udacity}
    & Video frames taken from urban road &410,530 images & \cite{pei2017deepxplore,zhang2018deeproad,kong2020physgan,li2021testing,zhou2020deepbillboard,chandrasekaran2021combinatorial,haq2020comparing,pan2021metamorphic}\\
    \hline
    2016 & SYNTHIA~\cite{ros2016synthia}& Multiple categories of virtual city rendering pictures&220,000+ images & -\\
    \hline
     2016 & Stanford Drone\cite{robicquet2016learning}
    & The movement and dynamics of pedestrians across the university campus
    & 69GB videos and images & -\\
    \hline
    2017& RobotCar~\cite{maddern20171}& Various combinations of weather, traffic and pedestrians, as well as long-term changes such as road engineering & - & -\\
    \hline
    2017 & CityPersons~\cite{zhang2017citypersons}& A dataset with a large proportion of blocked pedestrians images & 5,050 images & -\\
    \hline
    2017 & Mapillary Vistas~\cite{neuhold2017mapillary} & Street view of multiple cities under multiple seasons and weather conditions&25,000 images & - \\\hline
    2018 & GTA5~\cite{richter2016playing} & Synthetic images of urban traffic scenes collected using the game engine& 24,966 images & \cite{ramanagopal2018failing} \\
    \hline
    2018 & BDD100K~\cite{yu2018bdd100k}& Various scene types and weather conditions at different times of the day & 100,000 videos & -\\
    \hline
    2018 & comma2k19~\cite{schafer2018commute}& Over 33 hours of commute in California's 280 highway &33h videos & \cite{sato2021dirty}\\
    \hline
    2018 & highD~\cite{krajewski2018highd} & Traffic conditions of six different locations obtained by drone & 147h videos & -\\
    \hline
    2018 & ApolloScape~\cite{ApolloScape} & Images under different conditions and traffic density & 146,997 images &
    \cite{shao2021testing}
    \\
    \hline
    2019 & ACFR~\cite{zyner2019acfr} & Vehicle traces at 5 Roundabouts & 23,000 images & -\\
    \hline
    2019 & nuScenes~\cite{caesar2020nuscenes}
    & Images under different times of day and weather conditions & 1,400,000 images & -\\
    \hline
     2019 & INTERACTION~\cite{zhan2019interaction} & A dataset collected under interactive driving scenes with semantic maps & - & -\\
    \hline
    2019 & Waymo~\cite{sun2020scalability} & Including a perception dataset with high-resolution sensor data and labels, and a motion dataset with object trajectory and corresponding 3D map & 493,354 images & - \\
    \hline
    2020 & inD ~\cite{bock2020ind} & Naturalistic trajectories of vehicles and vulnerable road users recorded at German intersections & 10h videos &-  \\
    \hline
    2020 & Ford~\cite{agarwal2020ford} & Multiple seasons, traffic conditions, and driving environments & - & -\\
    \hline
    2020 & rounD ~\cite{rounDdataset} & Naturalistic trajectories of vehicles and vulnerable road users recorded at German roundabouts & - & -\\
    \hline
    2020 & openDD~\cite{breuer2020opendd} & A trajectory dataset covering seven roundabouts & 62h videos & -\\
    \hline
     2021 & Bosch Small Traffic Light~\cite{BoschNight} & An accurate dataset for vision-based traffic light detection & 13,427 images & \cite{wang2021can,bai2021metamorphic}\\
     \hline
     2022 & CrashD~\cite{crashD} & A synthetic LiDAR dataset to quantify the generality of 3D object detectors on out-of-domain samples & - & \cite{lehner20223d}\\
    \bottomrule
    \end{tabular}
    }
    \end{subtable}
\vfill
\end{table}
\myparagraph{Datasets} In the context of ADS, deep learning components handle safety-critical tasks, e.g., perception and end-to-end control, so it is necessary to validate their robustness under various scenarios. This process typically relies on data from the real world, which is, however, hard to obtain in general. Fortunately, there are a collection of publicly available datasets to solve the problem, which involve large quantities of real-world pictures and videos recorded by onboard sensors. For example, the KITTI dataset~\cite{geiger2013vision} contains over 10,000 images of traffic scenarios, collected by a variety of sensors including high-resolution \emph{RGB/grayscale stereo cameras} and a \emph{3D laser scanner}.

In this section, we summarize the scenario-driven datasets for ADS testing in Table~\ref{tab:datasets}. The first column shows the time when each dataset was released. The next three columns give the name, brief description, and the size of each dataset, and the last column indicates the related works that adopt these datasets.
Note that the datasets for other machine learning testing tasks~\cite{zhang2020machine} that have nothing to do with ADS testing are not listed here; in other words, all the datasets listed here are dedicated to ADS testing.

As shown in Table~\ref{tab:datasets}, we collect 27 datasets released from 2004 to 2022, including popular ones like the KITTI dataset~\cite{geiger2013vision} and emerging ones like the CrashD~\cite{crashD} dataset. One observation is that these datasets span over various physical conditions, e.g., different times of the day~\cite{Udacity}, different weather conditions~\cite{caesar2020nuscenes,agarwal2020ford} and different traffic density~\cite{ApolloScape}.
They also span over various application scenarios, such as urban street~\cite{cordts2016cityscapes, neuhold2017mapillary,maddern20171}, highway~\cite{schafer2018commute, krajewski2018highd,santana2016learning}, and intersection~\cite{bock2020ind}. In addition, we find that some of these datasets are specific to a certain task, e.g., pedestrian detection~\cite{zhang2017citypersons, robicquet2016learning}, and traffic sign detection~\cite{Stallkamp2012,philipsen2015traffic,belgium}.

As the column of reference in Table~\ref{tab:datasets} shows, several datasets such as the KITTI dataset~\cite{geiger2013vision} and the Udacity dataset~\cite{Udacity} are frequently used in ADS testing due to the diverse tasks they support, such as object detection and semantic segmentation. However, we also find that a number of datasets have not been widely used, due to their own limitations. For example, the rounD~\cite{rounDdataset}  and openDD~\cite{breuer2020opendd} can only be used for validating the behavior planning of ADS in the scenario of roundabouts; SYNTHIA~\cite{ros2016synthia} and GTA5~\cite{manikand} contain synthetic images from virtual environments, which may be not realistic enough for ADS testing. 

\begin{table}[!tb]
\scriptsize
\centering
\caption{Simulation platforms for ADS testing}
\label{tab:toolsets_2}
\resizebox{\textwidth}{!}{
\renewcommand\arraystretch{1.5}
\begin{tabular}{c|c|cc|cccc|c|c}
\hline
\multirow{2}{*}{\textbf{Simulator}} & \multirow{2}{*}{\textbf{Open-source}} & \multicolumn{2}{c|}{\textbf{Vehicle dynamic}} & \multicolumn{4}{c|}{\textbf{X-in-the-loop}} &\multirow{2}{*}{\begin{tabular}[c]{@{}c@{}}\textbf{Interface to other} \\ \textbf{ simulators}\end{tabular}} & \multirow{2}{*}{\textbf{Reference}} \\ \cline{3-8}
& & \multicolumn{1}{c|}{\textbf{customization}} & \textbf{soft/rigid}  & \multicolumn{1}{c|}{\textbf{MiL}} & \multicolumn{1}{c|}{\textbf{SiL}} & \multicolumn{1}{c|}{\textbf{HiL}} & \textbf{ViL} &  \\ \hline
\matlabSimulink~\cite{MatlabSimulink} & $\times$  & \multicolumn{1}{c|}{\checkmark} & - & \multicolumn{1}{l|}{\checkmark} & \multicolumn{1}{l|}{\checkmark} & \multicolumn{1}{l|}{\checkmark} &-  & \begin{tabular}[c]{@{}l@{}}\CarSim, \CarMaker, \PreScan, \\ \Gazebo, \carla, \rFpro, \\ \VTD, \Cognata, \ADAMS\\ \ProSiVIC\end{tabular} & \begin{tabular}[c]{@{}l@{}}~\cite{tuncali2019requirements,tuncali2016utilizing,tuncali2019rapidly}\\\cite{abdessalem2018testing,beglerovic2017testing,solmaz2020vehicle}\\\cite{shu2021test,nitsche2018novel,wang2020behavioral}\\ \cite{kolb2021fitness,zheng2020rapid}\end{tabular} \\ \hline
\CarSim~\cite{Carsim} & $\times$  &\multicolumn{1}{c|}{\checkmark} &rigid   & \multicolumn{1}{l|}{\checkmark} & \multicolumn{1}{l|}{\checkmark} & \multicolumn{1}{l|}{\checkmark} &-  & \begin{tabular}[c]{@{}l@{}}\matlabSimulink, \rFpro,\\
\NVIDIADriveSim, \VTD,\\ \ProSiVIC, \DonkeyCar\end{tabular} &\cite{shu2021test,zheng2020rapid}\\ \hline
\VISSIM~\cite{vissim} &$\times$  &\multicolumn{1}{c|}{$\times$} &- & \multicolumn{1}{l|}{-} & \multicolumn{1}{l|}{\checkmark} & \multicolumn{1}{l|}{\checkmark} &\checkmark & \begin{tabular}[c]{@{}l@{}}\carla, \VTD, \PreScan,\\\CarMaker, \rFpro, SUMO\end{tabular}&\cite{nalic2019development,nalic2020stress} \\ \hline
\SUMO~\cite{krajzewicz2002sumo} &\checkmark  &\multicolumn{1}{c|}{$\times$} &- & \multicolumn{1}{l|}{-} & \multicolumn{1}{l|}{\checkmark} & \multicolumn{1}{l|}{\checkmark} &\checkmark  & \begin{tabular}[c]{@{}l@{}}\carla, \VISSIM, \Cognata,\\ \rFpro\end{tabular}&\begin{tabular}[c]{@{}l@{}}\cite{weng2020model,solmaz2020vehicle,goss2022eagle}\\\cite{szalay2019proof}\end{tabular}\\ \hline
\Webots~\cite{Webots} & \checkmark  & \multicolumn{1}{c|}{-} &rigid  & \multicolumn{1}{l|}{-} & \multicolumn{1}{l|}{\checkmark} & \multicolumn{1}{l|}{-} & - &- &\cite{bashetty2020deepcrashtest,tuncali2018simulation} \\ \hline
\VTD~\cite{Vires} & $\times$  &\multicolumn{1}{c|}{\checkmark} &- & \multicolumn{1}{l|}{\checkmark} & \multicolumn{1}{l|}{\checkmark} & \multicolumn{1}{l|}{\checkmark} & \checkmark  &\begin{tabular}[c]{@{}l@{}}\CarSim, \matlabSimulink, \\\ADAMS, \VISSIM, \rFpro\end{tabular}&\cite{rocklage2017automated,reway2020test} \\ \hline
\Gazebo~\cite{gazebo} &\checkmark  & \multicolumn{1}{c|}{-} &rigid &\multicolumn{1}{l|}{-} & \multicolumn{1}{l|}{\checkmark} & \multicolumn{1}{l|}{\checkmark} &-  & \matlabSimulink, \ADAMS &\cite{majzik2019towards,chen2019novel,chen2020mixed}\\ \hline
\PreScan~\cite{prescan} &$\times$  & \multicolumn{1}{c|}{-} &rigid  &\multicolumn{1}{l|}{\checkmark} & \multicolumn{1}{l|}{\checkmark} & \multicolumn{1}{l|}{\checkmark} &\checkmark  & \begin{tabular}[c]{@{}l@{}} \matlabSimulink, \VISSIM,\\ \ProSiVIC\end{tabular} &\begin{tabular}[c]{@{}l@{}}\cite{li2021testing,abdessalem2018testing,haq2020comparing}\\\cite{borg2021digital,zheng2020rapid} \end{tabular}\\ \hline
\BeamNG~\cite{beamng} &\checkmark  &\multicolumn{1}{c|}{\checkmark} &soft  & \multicolumn{1}{l|}{\checkmark} & \multicolumn{1}{l|}{\checkmark} & \multicolumn{1}{l|}{\checkmark} &\checkmark  & - &\begin{tabular}[c]{@{}l@{}}\cite{gambi2019generating,castellano2021analysis,gambi2019automatically}\\\cite{gambi2019generating,birchler2022single,riccio2020model}\\\cite{zohdinasab2021deephyperion,zohdinasab2022efficient}\end{tabular}\\ \hline
\carla~\cite{dosovitskiy2017carla} &\checkmark  &\multicolumn{1}{c|}{$\times$} &rigid  & \multicolumn{1}{l|}{\checkmark} & \multicolumn{1}{l|}{\checkmark} & \multicolumn{1}{l|}{\checkmark} &\checkmark  &\begin{tabular}[c]{@{}l@{}} \CarSim, \VISSIM, \SUMO, \\ \matlabSimulink  \end{tabular}&\begin{tabular}[c]{@{}l@{}} \cite{balakrishnan2021percemon,boloor2020attacking,zhong2021detecting} \end{tabular} \\ \hline
\AirSim~\cite{shah2018airsim} &\checkmark  &\multicolumn{1}{c|}{$\times$} &rigid & \multicolumn{1}{l|}{-} & \multicolumn{1}{l|}{\checkmark} & \multicolumn{1}{l|}{\checkmark} &-  & - &- \\ \hline
\rFpro~\cite{rFpro} &$\times$  &\multicolumn{1}{c|}{\checkmark}  & rigid & \multicolumn{1}{l|}{-} & \multicolumn{1}{l|}{\checkmark} & \multicolumn{1}{l|}{\checkmark} &-  & \begin{tabular}[c]{@{}l@{}}\CarSim, \matlabSimulink, \\ \CarMaker, \VISSIM, \VTD, \\ \SUMO\end{tabular} & -\\ \hline
\Cognata~\cite{Cognata} &$\times$  &\multicolumn{1}{c|}{\checkmark} & -& \multicolumn{1}{l|}{\checkmark} & \multicolumn{1}{l|}{\checkmark} & \multicolumn{1}{l|}{\checkmark} &-  & \matlabSimulink, \SUMO & -\\ \hline
\NVIDIADriveSim~\cite{NVIDIA} &\checkmark  &\multicolumn{1}{c|}{\checkmark} &- & \multicolumn{1}{l|}{-} & \multicolumn{1}{l|}{\checkmark} & \multicolumn{1}{l|}{\checkmark} &\checkmark  & \CarMaker, \CarSim & -\\ \hline
\LGSVL~\cite{rong2020lgsvl} &\checkmark  &\multicolumn{1}{c|}{$\times$} &- & \multicolumn{1}{l|}{\checkmark} & \multicolumn{1}{l|}{\checkmark} & \multicolumn{1}{l|}{\checkmark} &-  & - &\begin{tabular}[c]{@{}l@{}}\cite{li2020av,tang2021route,cao2021invisible}\\\cite{sato2021dirty,tang2021collision,fremont2020formal}\end{tabular}\\ \hline
\SCANeRStudio~\cite{avsimulation} &$\times$  &\multicolumn{1}{c|}{\checkmark} &soft/rigid & \multicolumn{1}{l|}{\checkmark} & \multicolumn{1}{l|}{\checkmark} & \multicolumn{1}{l|}{\checkmark} &\checkmark  & - &\cite{djoudi2020simulation}\\ \hline
\ADAMS~\cite{ADAMS} &$\times$  &\multicolumn{1}{c|}{\checkmark} &rigid & \multicolumn{1}{l|}{-} & \multicolumn{1}{l|}{\checkmark} & \multicolumn{1}{l|}{\checkmark} &-  &\begin{tabular}[c]{@{}l@{}}\Gazebo, \matlabSimulink,\\ \VTD\end{tabular} & -\\ \hline
\CarMaker~\cite{Carmaker} &$\times$  &\multicolumn{1}{c|}{\checkmark} &rigid & \multicolumn{1}{l|}{\checkmark} & \multicolumn{1}{l|}{\checkmark} & \multicolumn{1}{l|}{\checkmark} &\checkmark  &\begin{tabular}[c]{@{}l@{}}\matlabSimulink, \VISSIM, \\\rFpro, \NVIDIADriveSim \end{tabular} &\begin{tabular}[c]{@{}l@{}}\cite{nalic2019development,solmaz2020vehicle,baumann2021automatic}\\\cite{nitsche2018novel,batsch2019performance,nalic2020stress}\\ \cite{kolb2021fitness,batsch2019performance,reway2020test}\end{tabular}\\ \hline
\ProSiVIC~\cite{hiblot2010pro}&$\times$&\multicolumn{1}{c|}{\checkmark}& - &\multicolumn{1}{l|}{\checkmark}&\multicolumn{1}{l|}{\checkmark}&\multicolumn{1}{l|}{\checkmark} &\checkmark &\begin{tabular}[c]{@{}l@{}}\matlabSimulink, \CarSim, \\ \PreScan \end{tabular} &\begin{tabular}[c]{@{}l@{}}\cite{borg2021digital}\end{tabular}  \\ \hline
\DonkeyCar ~\cite{DonkeyCar} &\checkmark&\multicolumn{1}{c|}{$\times$}& -  &\multicolumn{1}{l|}{-}&\multicolumn{1}{l|}{\checkmark}&\multicolumn{1}{l|}{\checkmark} &\checkmark &\begin{tabular}[c]{@{}l@{}}
\matlabSimulink
\end{tabular} &\begin{tabular}[c]{@{}l@{}}\cite{stocco2022mind}\end{tabular}  \\ \hline
\end{tabular}
}
\end{table}
\myparagraph{Tool stacks} As mentioned before, simulation-based testing has become an important alternative approach for real-world testing. Simulators usually provide vehicle dynamics, e.g., longitudinal and lateral motion of the vehicle, and virtual traffic scenarios. Moreover, simulators can help generate those extreme scenarios for testing, e.g., harsh weather, which are rarely encountered in the real world. There have been many advanced simulation platforms developed for ADS testing in recent years. For example, \carla~\cite{dosovitskiy2017carla} is an open-source simulator for ADS training and testing, which supports various sensor models and environmental conditions. 

In this section, we summarize the simulation platforms, namely, the simulators usually used for ADS testing in Table~\ref{tab:toolsets_2}. As shown in the table, we collect 20 simulation platforms including classical platforms such as \matlabSimulink~\cite{MatlabSimulink} and \CarSim~\cite{Carsim}, and emerging popular simulators such as \carla and \LGSVL~\cite{rong2020lgsvl}. Since these simulators have their own pros and cons, we compare them in the table and  focus on several aspects of interest, e.g., their gap from real environment.
Specifically, the first column lists the name and the second column shows the accessibility of each simulator. The third column is relevant to physical aspects, that is, whether the simulator allows for customizing a dynamic model and whether it is a soft-body or rigid-body based simulator. The fourth column indicates the level of support for mixed-reality testing, including model-in-the-loop (MiL), software-in-the-loop (SiL), hardware-in-the-loop (HiL), and vehicle-in-the-loop (ViL). The fifth column presents the capability of these simulators to complement each other, e.g., whether they support co-simulation with other simulators. The last column indicates the related works that adopt these simulators in their research.

 Based on the table, we can draw the following conclusions: 
 \begin{compactitem}[$\bullet$]
 \item  First, there exist many commercial simulators which are not open-source, e.g., \CarMaker~\cite{Carmaker} and \PreScan~\cite{prescan}. These simulators could be expensive and difficult for researchers to satisfy their research goals. In comparison, open-source simulators like \carla could have broader prospects for future research;
 \item Second, accurate physical dynamic models are needed to bridge the gap between simulation-based testing and real-world testing, and satisfy different testing requirements, e.g., smooth road needs a lower friction coefficient.
 We find that there have been simulators, i.e., \BeamNG~\cite{BeamNG.tech} and \CarSim, dedicated to this aspect and allowing for dynamic model customization. In particular, \BeamNG is also a soft-body based simulator which supports more realistic collision effects (see more details in~\S{}\ref{subsec:open-source_toolsets});
 \item Third, we find that most simulators support software-in-the-loop and hardware-in-the-loop testing. Several simulators, e.g., \CarMaker and \VTD~\cite{Vires}, support vehicle-in-the-loop testing, which closes the gap between hardware-in-the-loop testing and real-world testing.
 \item  Lastly, we find that a number of simulators have built-in interfaces to other simulators. This is essential to perform co-simulation for ADS testing since these simulators have their own pros and cons, and co-simulation could 
complement each other for a more realistic testing environment. For example, \CarMaker (accurate vehicle dynamics) and \VISSIM (representational traffic flow) are combined into a co-simulation framework~\cite{nalic2019development} for generating more realistic testing scenarios. 
 \end{compactitem}
 
Overall, it can be seen from the reference column that \matlabSimulink and \BeamNG have been widely used for ADS testing. Simulators like \NVIDIADriveSim and \Gazebo also have great potential for future research since they cover multiple features we list in the table, e.g., whether they could perform ViL testing or co-simulation with other simulators.

\smallskip
\begin{table}[!tb]
\scriptsize
\caption{Open-source ADS}
\label{tab:openADS}
\renewcommand\arraystretch{1.5}
\begin{tabular}{l|l|l}
\hline
 \textbf{System Category}&\textbf{Name} & \textbf{Description} \\ \hline
Modular&\Apollo & A commercial-grade ADS developed by Baidu  \\ \cline{2-3}
& \Autoware~\cite{kato2018autoware} & A  L4 ADS developed by Nagoya University \\ \cline{2-3}
& \OpenPilot & A commercial-grade L2 ADAS developed by Comma.ai  \\ \cline{2-3}
& \Pylot~\cite{gog2021pylot} & A modular ADS with low-latency dataflow from academic  \\ \hline
End-to-end&\LBC~\cite{chen2020learning} & An end-to-end controller based on imitation learning  \\ \cline{2-3}
& \DeepDriving~\cite{chen2015deepdrivin} & A CNN based end-to-end system that provides ACC and ALC \\ \cline{2-3}
& \begin{tabular}[c]{@{}c@{}}\NvidiaCNNLaneFollower~\cite{bojarski2016end}\end{tabular} & An end-to-end lane following system based on CNN  \\ \cline{2-3}
&\begin{tabular}[c]{@{}c@{}}\UdacityDNNModels~\cite{udacitychallenge}\end{tabular} & \begin{tabular}[l]{@{}l@{}}DNN-based steering prediction models from the Udacity challenge, \\ e.g., \Chauffeur~\cite{chauffeur} and \Epoch~\cite{epoch}   \end{tabular} \\ \hline
Other & \BeamNGAI~\cite{BeamNG.tech} & An AI agent in \BeamNG, which can realize simple control of vehicles \\ \cline{2-3}
& \carlapid~\cite{dosovitskiy2017carla} & A specific controller built in \carla \\ \hline
\end{tabular}
\end{table}

Moreover, we also introduce several publicly available systems under test in Table~\ref{tab:openADS}. 
\OpenPilot, \Apollo, \Autoware and \Pylot are all modular systems and have already been described in~\S{}\ref{subsec:open-source_toolsets}, so we will not repeat them here. In addition to modular ADS systems, there also exist open-source end-to-end based systems: \LBC~\cite{chen2020learning} is an imitation learning controller, which uses camera images and direction commands as input to control the direction of the vehicle in the lane and intersection; 
\DeepDriving~\cite{chen2015deepdrivin} and \NvidiaCNNLaneFollower~\cite{bojarski2016end} are also widely used CNN-based end-to-end controllers;
Open-source DNN models from the Udacity self-driving challenge, such as \Chauffeur~\cite{chauffeur} and \Epoch~\cite{epoch}, are another line of end-to-end driving controllers;
Besides, there are also driving agents and controllers from simulators, such as \BeamNG and \carla. \BeamNGAI, which has been mentioned in~\S{}\ref{subsec:open-source_toolsets}, is an AI agent in \BeamNG simulator and could accept virtual images in the simulator as input for path planning and trajectory tracking. \carlapid is a specific module in \carla that performs calculations at the motion planning stage, and estimates the acceleration, braking, and steering inputs required to reach target positions.

\begin{table}[!tb]
\footnotesize
\caption{Programming languages for scenario generation}
\label{tab:progLangScenario}
\centering
\renewcommand{\arraystretch}{1.2}
\resizebox{\textwidth}{!}{
\begin{tabular}{p{0.15\textwidth}|p{0.15\textwidth}|p{0.2\textwidth}|p{0.4\textwidth}|p{0.1\textwidth}}
\toprule
\textbf{Language} & \textbf{Dependencies} & \textbf{Supported simulators} & \textbf{Other features} & \textbf{Reference}\\
\hline
OpenScenario & Unified Modeling Language (UML), XML & \carla, Matlab, \PreScan & A scenario is described in a ``Storyboard'' tag in XML, which includes a series of events & \cite{balakrishnan2021percemon}\\\hline
GeoScenario~\cite{queiroz2019geoscenario} & XML & An Unreal-based driving simulator & The language is based on open street map standard. Users can either program by dragging icons, or code in an XML editor. & -\\\hline
Scenic~\cite{fremont2019scenic} & Imperative, object oriented (Python-like) & \carla, \LGSVL, \Webots & It is a probabilistic programming language that can specify the input distributions of machine learning components, and use that information for testing and analysis. & \cite{fremont2020formal} \\\hline
stiEF~\cite{bock2019advantageous} & Domain specific language  & \VTD & It supports multilingual representations for scenario description. & - \\\hline
SceML~\cite{schutt2020sceml} & Graph-based modeling framework & \carla & It allows information modeling at different depths, to support scenarios at different abstraction levels. & - \\\hline
CommonRoad~\cite{althoff2017commonroad} & XML & \SUMO & It provides a benchmark set that contains scenarios for the study of motion planning. & \cite{klischat2019generating} \\\hline
SceGene~\cite{li2021scegene} & A hierarchical representation model  & - & It supports scenario generation via bio-inspired operations, such as crossover and mutation. & -\\\hline
paracosm~\cite{majumdar2021paracosm} & Reactive programming model & Udacity's self-driving simulator& It adopts reactive objects that allow to describe temporal reactive behavior of entities. It also defines coverage criteria for test case generation. & - \\ 
\bottomrule
\end{tabular}
}
\end{table}

\myparagraph{Programming languages}
In order to systematically generate test cases, it has become a trend to propose new programming languages for testing scenario description. In this way, the generation of a new test case boils down to writing a program that describes the scenario. Also, researchers can make use of existing coverage criteria for programs, such as the code coverage criteria, to assess the adequacy of the generated tests. 

To define such a programming language, researchers need to formally express the basic elements in an ADS scenario, e.g., the ego car, other cars, pedestrians, and static objects. Since these languages are usually dependent on existing formats, they vary in their ways of expressing those elements, based on their dependent formats. For instance, \emph{Scenic}~\cite{fremont2019scenic}, a python-like language, requires users to define those objects as variables; in contrast, \emph{GeoScenario}~\cite{queiroz2019geoscenario} provides users with a graphical interface where users can drag the icons to describe a scenario.
Moreover, these languages usually do not emerge independently; instead, they come with specific simulators, or even specific ADS. 

In this section, we summarize the state-of-the-art programming languages for test case generation in Table~\ref{tab:progLangScenario}, and introduce their dependent formalism, their bonded simulator, other features, and their adoption in ADS testing. There exists literature, e.g.,~\cite{ma2021traffic}, that surveys programming languages for the test generation of ADS. Compared to~\cite{ma2021traffic}, our main aim is to show the  ecosystems and the landscape of the use of these languages in ADS testing, as a reference for the readers to better understand the testing techniques in \S{}\ref{sec:module} and \S{}\ref{sec:system}. Also, our study includes some latest achievements, e.g., paracosm~\cite{majumdar2021paracosm} and SceGene~\cite{li2021scegene}, in this direction. 

As shown by Table~\ref{tab:progLangScenario}, we collect 8 representative programming languages, including the classic ones, such as OpenScenario, that have been widely used in different stages of the development of ADS, and emerging ones, such as paracosm~\cite{majumdar2021paracosm}. As our findings, first, different languages are designed for different purposes and attached with different features, e.g., Scenic~\cite{fremont2019scenic} allows probabilistic sampling for testing driving systems with machine learning components; SceGene~\cite{li2021scegene} designs bio-inspired operations, such as crossover and mutation, for scenario generation. Second, some of these languages provide more user-friendly features; for instance, some of the languages, e.g., SceML~\cite{schutt2020sceml} provide a GUI for users to define their scenarios. However, as the column of reference shows, most of these languages have not been widely adopted in practice. This can be due to several reasons: one possibility is that some languages are still too specialized for practitioners to adopt them in their work; also, since many of the languages, such as GeoScenario~\cite{queiroz2019geoscenario}, are designed for specific systems, they are still ad hoc and not easily extensible to be adopted in other systems. 

\smallskip
In conclusion, programming languages are increasingly deemed as powerful weapons for test case generation in ADS testing, but
currently they have not been widely adopted in practice yet.

\section{Challenges and Opportunities}
\label{sec:challenges&opportunities}
\begin{figure}[!tb]
    \centering
    \includegraphics[width=1\textwidth]{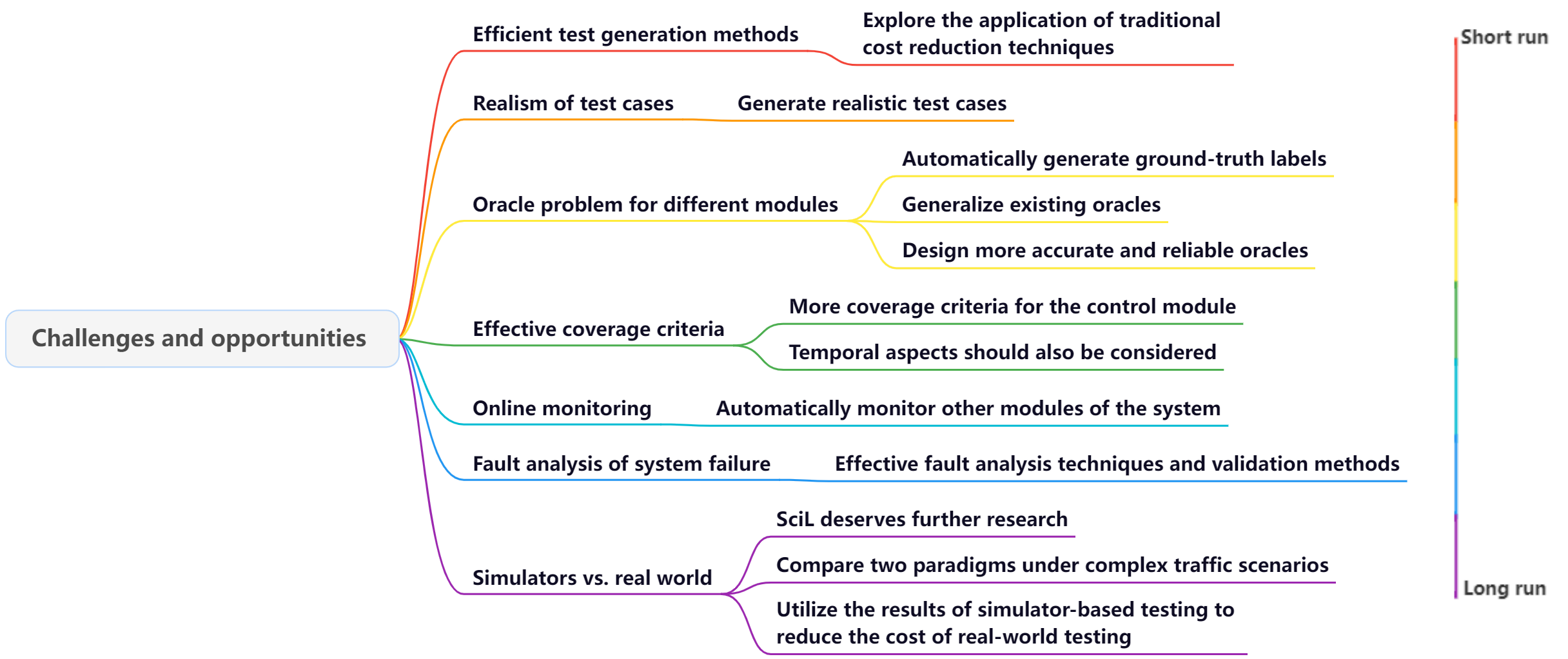}
    \caption{Illustration of challenges and opportunities}
    \label{fig:Challenges and opprtunities}
\end{figure}

As this survey reveals, ADS testing has experienced rapid growth in recent years. Nevertheless, there are still many challenges and open questions in its development and deployment.
Based on our analysis of the collected literature and our discussions in each section, we answer RQ3 by listing the challenges and opportunities in this direction , as shown in Fig.~\ref{fig:Challenges and opprtunities}. To account for it, there exist several solutions to the first four challenges that could be improved, while the last three challenges still need further exploration and require a long period of research.

\myparagraph{Efficient test generation methods} 
Efficiency is one of the most important objectives in ADS testing, since system executions, whether in simulator environments or the real world, are too expensive. There have been many methods that aim to reduce the number of system executions, e.g., training surrogate models~\cite{ben2016testing, batsch2021scenario, beglerovic2017testing, sun2020adaptive}, or adopting sampling-based methods~\cite{akagi2019risk,nitsche2018novel,batsch2019performance,corso2020interpretable,DBLP:conf/vehits/SchuttHMZS22,birkemeyer2022feature, zhao2016accelerated, huang2017accelerated, huang2017anaccelerated}, as discussed in~\S{}\ref{subsec:systemMethodology}. 
However, there are several limitations to these methods; for example, the process of preparing training data in~\cite{ben2016testing} for surrogate models is time-consuming. One potential future direction is to explore the application of traditional cost reduction techniques, such as test selection and test prioritization, to further accelerate the testing process.

\myparagraph{Realism of test cases} 
Generating realistic test cases that can really threaten the safety of ADS in the real world should be  another important goal of test case generation. Unrealistic test cases that cannot happen in the real world are meaningless and not worthy of being taken care of. However, compared to efficiency, this aspect is usually ignored. Generating realistic test cases is a demand over different modules, and some existing works have paid attention to this problem.  For example, in the perception module testing, RP$_2$~\cite{eykholt2018robust} is proposed to generate test cases under real physical conditions; in the planning module testing, \emph{avoidable collision}~\cite{calo2020generating} is proposed to filter out useless test cases; moreover, this is also a major issue in system-level testing, as discussed in~\S{}\ref{sec:simRealWorld}. In addition to these efforts, the problem is worthy of more attention, in order to find out those really useful test cases.

\myparagraph{Oracle problem for different modules} Although there have been many works that try to design suitable oracles for different modules of ADS, there still remain many open challenges in defining oracles regarding different characteristics of different modules. For the perception module, as discussed in~\S{}\ref{subsubsec:perceptionDiscussion}, the automatically labeling method in~\cite{zhou2019automated} targets only at semantic segmentation models, so one future direction is to explore how to automatically generate high-fidelity ground-truth labels for other types of models in the perception module. For the planning module, as discussed in~\S{}\ref{subsubsec:planningDiscussion},
the criteria such as \emph{avoidable collision}~\cite{calo2020generating} are ad hoc and may not be generalized to other systems. 
\emph{Metamorphic relations} are widely adopted by works~\cite{shao2021testing,bai2021metamorphic,zhou2019metamorphic,woodlief2022semantic,wang2021object,ramanagopal2018failing, tian2018deeptest, zhang2018deeproad} for different modules, but they may lack sufficiently accuracy and so lead to false positives. Hence, one potential future direction is to design more accurate and reliable oracles for the testing of different modules in ADS.

\myparagraph{Effective coverage criteria} 
Coverage criteria are used as guidance to generate diverse test cases for testing. As discussed in~\S{}\ref{sec:module} and~\S{}\ref{sec:system}, various coverage criteria have been proposed for testing different modules of the ADS, e.g., \emph{neuron coverage}~\cite{pei2017deepxplore} for perception and end-to-end modules, 
\emph{weight coverage}~\cite{laurent2019mutation} and \emph{route coverage}~\cite{tang2021route} for the planning module. 
Notably, few coverage criteria have been proposed for the control module, which indicates a future research direction. 
Moreover, one problem in the existing studies is that they mainly consider covering the spatial aspects of the test cases; for instance, \emph{neuron coverage}~\cite{pei2017deepxplore} is computed based on the activated neurons in a DNN model and used as a guidance to trigger diverse behavior of single DNNs.  However, in the testing of ADS which run over a time period, even though a strange behavior for a moment is triggered to happen, if it is immediately corrected, it may not affect the system behavior over the period. Therefore, in the testing of ADS, we need to trigger the diverse behavior of the DNNs over time. For instance, if a DNN keeps making wrong predictions for a period, it is likely to lead to a collision.
Besides, several studies~\cite{yan2020correlations, harel2020neuron, ma2021test} have demonstrated that neuron coverage may not be suitable for guiding the testing process. Whether these findings will effect the ADS testing or there exists more effective criteria dedicated to perception testing needs further exploration.
To sum up, coverage criteria dedicated to the control module are expected to be proposed in the future, and another research direction is to consider incorporating the temporal aspects into the existing coverage criteria.

\myparagraph{Online monitoring}
In this work, we mainly see testing techniques for ADS based on the posterior checking of the system execution; another effective quality assurance scheme is \emph{online monitoring}~\cite{bartocciBook, zhang2022online} that monitors the system behavior at runtime. As an advantage, online monitoring can detect unsafe behavior during the system execution, and thus warn drivers to take actions to avoid the safety risk.
As discussed in~\S{}\ref{subsubsec:perceptionDiscussion}, there have been some works, e.g.,~\cite{balakrishnan2021percemon}, that rely on formal temporal specifications to monitor the perception module at runtime. 
Besides, the model-based oracle proposed by Stocco et al.~\cite{stocco2020misbehaviour} is also a system-level online monitoring approach, as it predicts the misbehaviors of the system at runtime.
However, how to automatically monitor other modules of the system remains a great challenge. One potential future direction is to develop monitoring techniques for other modules. Meanwhile, more expressive specification languages should be provided to handle real-world system requirements.

\myparagraph{Fault analysis of system failure}
As this survey shows, the function of an ADS relies on the collaborative work of different modules; indeed, the wrong function of any module can cause a failure at the system level.
Therefore, one question arises that which module should be deemed as the main cause of the system failure.
Currently, as discussed in~\S{}\ref{subsubsec:systemDiscussion}, the research attention is mostly focused on failure detection, rather than fault analysis. 
Moreover, fault analysis of ADS is challenging in nature, because it requires defining the boundaries of each module properly and making the oracles of each module clear. Sometimes, the failure of the system is not due to single modules but to the interactions between different modules. Therefore, one future direction is to propose effective fault analysis techniques as well as their validation methods.

\myparagraph{Simulators vs. real world}
Because of the high cost of real-world testing, 
simulation-based testing is the most commonly used testing paradigm; however, even with modern high-fidelity simulators (e.g., \carla and \LGSVL), there is still a gap from real-world testing.
Recently, lightweight mixed-reality testing schemes, including \emph{hardware-in-the-loop} (\emph{HiL}), \emph{vehicle-in-the-loop} (\emph{ViL}), and \emph{scenario-in-the-loop} (\emph{SciL}) (more detail in~\S{}\ref{sec:mixed_testing}), that mix the simulation-based testing and the real-world testing, also emerge to achieve a trade-off between the two. 
While HiL and ViL testing have developed quickly over the years, SciL testing, which is closest to the real world, is still at a theoretical stage and has not yet been widely adopted.
As discussed in~\S{}\ref{subsec:datasetToolset}, existing simulators all have their pros and cons, and one future direction is to combine their distinguishing features, e.g., co-simulation, to enhance the realism of the simulation environment.
Moreover, there have been several works in~\S{}\ref{sec:simRealWorld} that try to estimate how far the simulation-based testing is from the real-world testing. Nevertheless, in the case of handling complex traffic scenarios in testing, there are still open questions, such as the selection between simulation-based testing and real-world testing, and how to mitigate the weaknesses of the selected testing paradigm, that are seeking for better answers.
To sum up, the gap between simulation-based testing and real-world testing still exists, and one research direction is to explore how to utilize the results of simulation-based testing to reduce the cost of real-world testing.

\myparagraph{Answer to RQ3}
Based on our survey results, we identify 7 major challenges for ADS testing and discuss the corresponding potential research opportunities. Moreover, as shown in Fig.~\ref{fig:Challenges and opprtunities}, we find that several challenges such as the efficiency of test generation could be improved in the short run; by contrast, some other challenges (for example, how to mitigate the gap between simulation and real-world environments) may require a long period of research.

\section{Conclusion}
\label{sec:conclusion}
This survey provides a comprehensive overview and analysis of the relevant studies on ADS testing. These testing works cover both module-level testing and system-level testing of ADS, and we also include the works on empirical study w.r.t. system testing, mixed-reality testing, and real-world testing. In the introduction to the testing of each module, we 
respectively unfold the landscape of the literature from three perspectives, namely, test methodology, test oracle and test adequacy. Based on the literature review, we also perform analysis on the landscape of ADS testing, and propose a number of challenges and research opportunities in this direction. 

Our work gives a specific emphasis on the technical differences in the testing of different modules, and also reveals the gap between simulation-based testing and real-world testing. Moreover, our analysis and discussion on the challenges and opportunities based on the literature review point out the future direction of research in this field. We hope that this work can inspire and motivate more contributions to the safety assurance of ADS, and we also hope that ADS can be sufficiently reliable to be adopted by more people as early as possible.

\begin{acks}
This work was supported in part by the Anhui Provincial Department of Science and Technology under Grant 202103a05020009, in part by National Natural Science Foundation of China under Grant 61972373, the Basic Research Program of Jiangsu Province under Grant BK20201192 and the National Research Foundation Singapore under its NSoE Programme (Award Number: NSOE-TSS2019-03). The research of Dr.~Xue is also supported by CAS Pioneer Hundred Talents Program of China. 
Lei Ma's research was supported in part by Canada CIFAR AI Chairs Program, NSERC (No.RGPIN-2021-02549, No.RGPAS-2021-00034, No.DGECR-2021-00019), as well as JSPS KAKENHI Grant No.JP20H04168, No.JP21H04877, JST-Mirai Program Grant No.JPMJMI20B8.

\end{acks}

\bibliographystyle{unsrtnat}
\bibliography{ACM}

\end{document}